\documentclass[]{gCFD2e}

\usepackage{subfigure}
\usepackage{graphicx}
\graphicspath{{figures/}}
\usepackage{amsmath,amsfonts,amssymb}
\usepackage{float}
\usepackage{morefloats}
\usepackage{color}
\usepackage{multirow}
\usepackage{algorithm}
\usepackage{algpseudocode}

\begin{document}


\title{A localized dynamic closure model for Euler turbulence}

\author{S. M. Rahman$^{\rm a}$ and O. San$^{\rm a}$$^{\ast}$\thanks{$^\ast$Corresponding author. Email: osan@okstate.edu
\vspace{6pt}}\\  $^{a}${School of Mechanical and Aerospace Engineering, Oklahoma State University, Stillwater, OK, USA}\\\received{\today} }

\maketitle

\begin{abstract}
In this work, we present a localized form of the dynamic eddy viscosity model for computationally efficient and accurate simulation of the turbulent flows governed by Euler equations. In our framework, we determine the dynamic model coefficient locally using the information from neighboring grid points through a test filtering process. We then develop an optimized Gaussian filtering kernel, using a consistent definition with respect to the test filtering ratio, which gives full attenuation at the grid cut-off wave number. A systematic a-posteriori analysis of our model is performed by solving two 3D test problems: (i) incompressible Taylor-Green vortex flow and (ii) compressible shear layer turbulence induced by Kelvin-Helmholtz instability to show the wide range of applicability of the proposed localized dynamic model. We demonstrate that the proposed dynamic model is robust and provides a better estimation of the inertial range turbulence dynamics than other numerical models tested in this study.


\end{abstract}

\section{Introduction}
\label{sec:1}

Euler turbulence refers to the turbulent flow phenomenon where the viscous forces are insignificant. The existence of the turbulence in inviscid conservative systems has been observed experimentally and numerically for both compressible \citep{abid1998experimental,mac2004control,porter1994kolmogorov} and incompressible flows \citep{araki2002energy,kawamura1986computation}. Also, examples of two-dimensional Euler turbulence are ubiquitous in the form of coherent structures of jets and vortices in geophysical flows \citep{kraichnan1980two,qi2014hyperviscosity}. The numerical simulation of turbulent flows at a very high Reynolds number $(Re)$ is often carried out by using Euler equations which can be considered as a form of Navier-Stokes equations \citep{cichowlas2005effective,bos2006dynamics,krstulovic2008two,zhou2014estimating,moura2017eddy}. However, an initial vorticity has to be defined to introduce vortex shedding or stretching in the flow and also, the addition of artificial viscosity is required to maintain the dissipative characteristics of Navier-Stokes equations \citep{frisch1995turbulence}.

Computational studies of these high speed turbulent flows governed by Euler equations have been investigated actively for past few decades because of their vast applications in a wide range of areas including astrophysical flows, atmospheric flows and aerospace simulations \citep{majda2006nonlinear,bertschinger1998simulations,banerjee2013exact,lee1999spurious,biskamp2001two,zhou2017rayleigh1,zhou2017rayleigh}. Designing accurate and computationally efficient schemes in a direct numerical simulation (DNS) or large eddy simulation (LES) of Euler equations is a challenging task since both discontinuities and turbulent features coexist in the flow. These kind of flows require some sort of regularization to prevent any oscillations in the solutions near the discontinuities. In order to do the regularization in simulations of shock dominated flows, a certain amount of artificial dissipation is added in the numerical scheme which should be sufficient enough to capture shocks but not too much dissipative to damp the small-scale turbulent eddies. For last few decades, lots of successful numerical techniques have been developed with a virtue of adding the artificial dissipation locally near the region of discontinuity while preserving the turbulent characteristics in smooth flow regions (e.g., see \cite{maulik2018adaptive} and references therein).

Among all the high order shock-capturing algorithms, upwind-biased and nonlinearly weighted approaches (e.g., Weighted Essentially Non-Oscillatory (WENO) scheme) are widely used in considerable amount of works because of their robustness to capture discontinuities in shock dominated flows and high order of accuracy in preserving turbulence features \citep{yamamoto1993higher,martin2006bandwidth,titarev2005weno,pirozzoli2011numerical,zhang2006effects}. The implicit large eddy simulation (ILES) framework utilizes shock-capturing schemes very frequently where the numerical dissipation is added through an upwinding scheme \citep{maulik2017resolution,zhou2014estimating,karaca2012implicit}. ILES has been successfully applied to complex flows in engineering and physical applications as an approach to simulate high $(Re)$ flows with computational efficiency \citep{zhao2014comparison,grinstein2007implicit,watanabe2016implicit,zhou2016comparison,domaradzki2003effective}. Many studies have been published on the development of WENO schemes to improve it's excessive dissipative nature \citep{wong2017high,hu2015efficient,san2014numerical,hu2010adaptive,henrick2005mapped}. Another approach named high-order discontinuous Galerkin (DG) method can be employed for simulation of Euler turbulence through the ILES approach \citep{beck2014high,vermeire2016implicit,bull2015simulation}. In the broad sense, DG works similarly as the ILES where truncation error of the discretization scheme contributes as the subgrid-scale (SGS) contribution and also shows a good agreement in treating the different physical waves in a consistent manner for simple solvers \citep{moura2017eddy,gassner2016split,liu2007central}. In our study, we utilize the WENO-Z \citep{borges2008improved} scheme coupled with a family of Riemann solvers \citep{san2015evaluation} to analyze the performance of our ILES framework to solve Euler turbulence equations.

We also investigate an explicit filtering approach in our test simulations where a low-pass filter is used to add dissipation on the truncated scales in LES \citep{mathew2006new}. We use a symmetric central linear reconstruction scheme combined with a relaxation filtering framework ($CS+RF$). The non-dissipative central scheme minimizes numerical dissipation \citep{hyman1992high} and  relaxation filter removes high-frequency content in the flow to prevent Gibbs oscillation as well as add the dissipative effect of the unresolved scales. Relaxation filtering has been appeared in numerous works over the years \citep{bose2010grid,fauconnier2013performance,lund2003use,berland2011filter}. Another explicit filtering strategy, called approximate deconvolution (AD) \citep{stolz1999approximate,germano2009new,germano2015similarity}, is based on the explicit filtering technique without using any preassumption of physics and can also be coupled with relaxation filters. A considerable amount of filters have been proposed as relaxation filter over the years in LES literature with a goal to find a consistent and efficient filtering framework for explicit filtering techniques\citep{de2002sharp,san2016analysis,najafi2015high,vasilyev1998general}. In this paper, we utilize the $6^{th}$ order symmetric central scheme along with discrete compact Pad\'{e} filter \citep{lele1992compact,visbal2002use} as relaxation filter to formulate an explicit filtering framework.

Based on the studies above, it is obvious by now that the addition of adequate artificial dissipation is the key to accurate and effective simulation of Euler equations. One might also consider the simplest form of LES eddy viscosity models \citep{meneveau2000scale,piomelli1999large}. In eddy viscosity models, the effective viscosity is increased to achieve sufficient numerical dissipation to keep the model consistent with Kolmogorov's hypothesis for three-dimensional isotropic turbulence \citep{frisch1995turbulence}. One of the most widely used eddy viscosity model, the Smagorinsky model \citep{smagorinsky1963general,deardorff1970numerical} utilizes the filter width as the characteristic length to compute eddy viscosity to account for the influence of the subgrid-scales in the inertial and dissipation range \citep{chapman1979computational}. One of the major characteristics of Smagorinsky model is that it contains the Smagorinsky constant $C_s$ which must be determined a priori in space and time. Later, $C_s$ is found to be dependent on physical problems and local flow information \citep{pope2001turbulent,vorobev2008smagorinsky,smagorinsky1993some}. One promising approach to overcome the limitations associated with the Smagorinsky model is to use dynamic models where $C_s$ will be problem dependent. Germano et al. \citep{germano1991dynamic} and Lilly \citep{lilly1992proposed} devised the dynamic Smagorinsky model where $C_s$ is dynamically calculated using the locally laminar or fully resolved flow.

There are a vast amount of literature available on dynamic LES approach for both compressible and incompressible flows \citep{moin1991dynamic,domaradzki2003effective,chai2012dynamic,kosovic2002subgrid,moin1991dynamic,park2006dynamic}. The issues with dynamic model of achieving stable computation lead to some extended dynamic models such as mixed models, Lagrangian model or dynamic localization model. Dynamic mixed model resolves the issue with the use of negative eddy viscosity in dynamic model but, it also can turn unstable \citep{zang1993dynamic,vreman1994formulation}. The performance of the dynamic localization model has been proved successfully in LES of complex flows \citep{ghosal1995dynamic,moin2002advances,carati1995representation,you2004computational}. There are other family of eddy viscosity models which do not use the Smagorinsky concept such as the variational multi-scale model \citep{hughes2000large,collis2001monitoring}, Vreman model \citep{vreman2004eddy} and regularization model \citep{stolz2001approximate}. All these multi-directional works prove that eddy viscosity model is one of the most widely utilized LES models in the computational fluid dynamics research. However, there are very few studies have been published on the application of eddy viscosity approaches in Euler turbulence, most probably, because of there is no physical viscosity present in Euler equations. Garnier et al. \citep{garnier1999use} showed several computations of the shock capturing schemes (ENO, Jameson, TVD-MUSCL) coupled with the Smagorinsky and the dynamic Smagorinsky models and concluded that the addition of a subgrid-scale model to the shock-capturing schemes is unnecessary and inconvenient. The present study also confirms this statement since the dissipation introduced by numerics due to upwinding dominates the effect of additional SGS models. However, their behavior using the underlying central schemes is largely unexplored, a topic we address in this study. In this paper, we introduce a localized dynamic eddy viscosity model coupled with non-dissipative symmetric central scheme in the simulation of Euler equations using an optimized test filter which is proved important for the success of dynamic eddy viscosity models. Based on the fact that the functional eddy viscosity models are often considered as explicit LES closures, we also analyze a localized dynamic approach with relaxation filtering (compact Pad\'{e} filter as relaxation filter in our test case) to investigate the resulting behavior of the flow field.

The numerical techniques discussed above are tested in the simulation of two classical numerical test problems, the inviscid Taylor-Green vortex (TGV) flow and the stratified Kelvin-Helmholtz instability (KHI) test case. The temporal and spatial evolution of the TGV flow is one of the popular approach for numerical models validation and LES characterization due to its simplicity as well as deterministic nature \citep{shu2005numerical,san2015posteriori,berselli2005large,drikakis2007simulation}. Also, the TGV problem exhibits vortex stretching and nonlinear interactions type of features like a real turbulent flow \citep{bull2015simulation}. The kinetic energy generated by the velocity shear is dissipated by the smallest scales which makes TGV a simple model for the development of energy cascade from larger to smaller scales. We have used the results obtained by WENO-Roe ILES simulations on $512^3$ grid points as a benchmark to test the performance of models since it shows a considerably larger inertial scale compared to our test simulations between $16^3$ and $128^3$ resolutions. The models are then applied to the KHI case which is an instability of the shear layer in stratified fluids observed in many atmospheric and ocean phenomena \citep{hwang2012first,sckopke1981structure,russell1979observation,norman1982structure}. Instabilities can occur when a wave draws energy from the system whether it is kinetic energy from a pre-existing motion or the potential energy from background stratification\citep{staquet2002internal,cocle2007investigation,zhou2017rayleigh,villard2004first,cohen2002three}. KHI is characterized by momentum exchange and mixing of the two media of different densities resulting in a growing wave from the unstable velocity difference at the interface of those two media flowing relative to each other \citep{thomson1871xlvi,matsumoto2004onset}. The KHI problem has already been used in various literature as a test problem to validate different numerical models in 2D \citep{san2015evaluation,san2014numerical} which inspired us to utilize a similar KHI test case in a 3D setting. The validation of the numerical schemes being studied here will be carried out through visual examination of field variables (in the form of absolute vorticity level for TGV and density contour for KHI problem) to observe the time evolution of the instability, and through the statistical quantities such as the total kinetic energy and the angle-averaged kinetic energy spectra which should be enough to draw a conclusion about the efficiency of the models in capturing the effect of modeled scales. The high fidelity higher resolution results achieved through a distributed multiprocessing environment using the message passing interface (MPI) approach \citep{gropp1999using} ensures that all the relevant scales of the flow are resolved and are used for the purpose of characterizing the performance of a numerical method on a much coarser grid.

The main objective of this work is to present a new form of modeling approach, localized dynamic model, as an extension to the dynamic eddy viscosity model which is shown to perform better for solving Euler equations than the other state-of-the-art modeling techniques. Other objectives are to provide a systematic assessment of the performances of different numerical schemes through a quantitative measure of the dissipation level and to validate the appropriateness of statistical test results by doing a-posteriori analysis on two well-known benchmark test cases. Furthermore, we present a thorough analysis of the ILES approach using improved WENO-Z scheme coupled with a family of Riemann solvers, and on the symmetric central reconstruction based explicit filtering approach with relaxation filtering to draw a comparative conclusion on their performances for resolving Euler equations. The analysis of the test problems shows that the proposed localized dynamic Euler model performs better than the other models discussed in this study.

The rest of the paper is organized as follows: Section $2$ details the governing equations for Euler turbulence in brief. The numerical formulations associated with different numerical techniques are illustrated in Section $3$. In Section $4$, the definition of the test cases are given, and the results obtained by the implementation of different numerical methods in the test problems are discussed. Finally, the summary of the results and some comments on the performance of these schemes are presented in Section $5$ as concluding remarks.

\section{Governing equations}
\label{sec:2}

\newcommand{\partfrac}[2]{\frac{\partial #1 }{\partial #2}}
\newcommand{\partfracsec}[2]{\frac{\partial^2 #1 }{\partial #2^2}}

In present work, the three-dimensional Euler system of equations has been considered as underlying governing laws for flow simulation. Euler equations are a set of nonlinear hyperbolic conservation laws without the effects of body forces, heat flux or viscous stresses. Explicitly, they can be expressed in their conservative dimensionless form as:
\begin{align}\label{EULERPDE}
  \frac{\partial \textbf{q}}{\partial t} + \frac{\partial \textbf{F}}{\partial x} + \frac{\partial \textbf{G}}{\partial y} + \frac{\partial \textbf{H}}{\partial z} = 0
\end{align}
with $\rho$, $P$, $u$, $v$, and $w$ are the density, pressure and velocity components of the flow field in the $x$, $y$, and $z$ Cartesian coordinates respectively, the quantities included in $\textbf{q}$, $\textbf{F}$, $\textbf{G}$, and $\textbf{H}$ are:
\begin{align}\label{InviscidMat}
\textbf{q} = \begin{bmatrix}
           \rho \\
           \rho u \\
           \rho v \\
           \rho w \\
           \rho e
         \end{bmatrix}, \ \ \
\textbf{F} = \begin{bmatrix}
           \rho u \\
           \rho u^2 + P \\
           \rho uv \\
           \rho uw \\
           \rho uH
         \end{bmatrix}, \ \ \
\textbf{G} = \begin{bmatrix}
           \rho v \\
           \rho uv \\
           \rho v^2 + P \\
           \rho vw \\
           \rho vH
         \end{bmatrix}, \ \ \
\textbf{H} = \begin{bmatrix}
           \rho w \\
           \rho uw \\
           \rho uv \\
           \rho w^2 + P \\
           \rho wH
         \end{bmatrix} \ \ \
\end{align}
Here, $e$ is the total energy and $H$ is the total enthalpy. With the assumption of ideal gas law, the total enthalpy, and total energy can be written as:
\begin{align}\label{Enthalpy}
  H = e + \frac{P}{\rho},
  \quad P = \rho (\gamma-1)\left(e - \frac{1}{2}(\textbf{u}\cdot\textbf{u}) \right)
\end{align}
where $\textbf{u}\cdot\textbf{u} = u^2 + v^2 + w^2$ and the ratio of specific heats $\gamma$ is set as $\frac{7}{5}$.
The convective flux Jacobian matrices for the conservation laws above can be shown as \citep{laney1998computational}:
\begin{align}\label{JacA}
  A = \frac{\partial \textbf{F}}{\partial \textbf{q}} =
  \begin{bmatrix}
    0 & 1 & 0 & 0 & 0 \\
    -u^2 + \frac{\gamma - 1}{2} (\textbf{u}\cdot\textbf{u}) & (3-\gamma) u & -(\gamma - 1)v & -(\gamma - 1)w & \gamma-1 \\
    -uv & v & u & 0 & 0 \\
    -uw & w & 0 & u & 0 \\
    -(\gamma e - (\gamma-1) \textbf{u}\cdot\textbf{u}) u~ & ~\gamma e - \frac{\gamma-1}{2} (2u^2 + \textbf{u}\cdot\textbf{u})~ & -(\gamma-1)uv~ & -(\gamma-1)uw & \gamma u
  \end{bmatrix},
\end{align}
\begin{align}\label{JacB}
  B = \frac{\partial \textbf{G}}{\partial \textbf{q}} =
  \begin{bmatrix}
    0 & 0 & 1 & 0 & 0 \\
    -uv & v & u & 0 & 0 \\
    -v^2 + \frac{\gamma - 1}{2} (\textbf{u}\cdot\textbf{u}) & -(\gamma-1) u & (3 - \gamma)v & -(\gamma - 1)w & \gamma-1 \\
    -vw & 0 & w & v & 0 \\
    -(\gamma e - (\gamma-1) \textbf{u}\cdot\textbf{u}) v~ & -(\gamma-1)uv~ & ~\gamma e - \frac{\gamma-1}{2} (2 v^2 + \textbf{u}\cdot\textbf{u})~ & -(\gamma-1)vw & \gamma v
  \end{bmatrix},
\end{align}
\begin{align}\label{JacC}
  C = \frac{\partial \textbf{H}}{\partial \textbf{q}} =
  \begin{bmatrix}
    0 & 0 & 0 & 1 & 0 \\
    -uw & w & 0 & u & 0 \\
    -vw & 0 & w & v & 0 \\
    -w^2 + \frac{\gamma - 1}{2} (\textbf{u}\cdot\textbf{u}) & -(\gamma-1) u & -(\gamma - 1)v & (3-\gamma)w & \gamma-1 \\
    -(\gamma e - (\gamma-1) \textbf{u}\cdot\textbf{u}) w~ & -(\gamma-1)uw~ & -(\gamma-1)vw~ & ~\gamma e - \frac{\gamma-1}{2} (2 w^2 + \textbf{u}\cdot\textbf{u})  & \gamma \omega
  \end{bmatrix}.
\end{align}
where $\textbf{u}$ is the three-dimensional velocity vector. A similar transform of each Jacobian matrix can be given by:
\begin{equation}\label{Eigensystem}
\left.\begin{aligned}
     L_A A R_A &= \Lambda_A \Rightarrow R_A \Lambda_A L_A = A, \\
     L_B B R_B &= \Lambda_B \Rightarrow R_B \Lambda_B L_B = B, \\
     L_C C R_C &= \Lambda_C \Rightarrow R_C \Lambda_C L_C = C. \\
\end{aligned}
\right\}
\end{equation}
where $\Lambda_A=\mbox{diag}[u-a,u,u,u,u+a]$, $\Lambda_B=\mbox{diag}[v-a,v,v,v,v+a]$, and $\Lambda_C=\mbox{diag}[w-a,w,w,w,w+a]$ are the diagonal matrices comprising of the real eigenvalues of $A, B$, and $C$ respectively. Here $a$ is the speed of the sound and defined as $a=\sqrt{\gamma P/\rho}$. In Eq.~(\ref{Eigensystem}), $L_A, L_B$, and $L_C$ are the matrices whose columns are the eigenvectors of this eigendecomposition (e.g., see \cite{parent2012positivity,bidadi2015investigation} for a detailed presentation of the left and right eigenvector matrices).

\section{Numerical Methods}
\label{sec:3}

\subsection{Finite Volume Framework Formulation}
We formulate a sixth-order finite volume approach (7-point stencil scheme) using a third-order Runge-Kutta time integrator (TVDRK3) in multi-blocks domains based on non-overlapping 3-point stencil MPI data streams. The semi-discrete form of the Euler equations can be written as:
\begin{align}\label{FVMEq}
\begin{gathered}
  \frac{d q_{i,j,k}}{dt} + \frac{ F_{i+\frac{1}{2},j,k} - F_{i-\frac{1}{2},j,k}}{\Delta x} + \frac{G_{i,j+\frac{1}{2},k} - G_{i,j-\frac{1}{2},k} }{\Delta y} + \frac{H_{i,j,k+\frac{1}{2}} - H_{i,j,k-\frac{1}{2}} }{\Delta z} = 0
\end{gathered}
\end{align}
where $q_{i,j,k}$  is the cell-averaged vector of dependant variables, $F_{\frac{i\pm1}{2},j,k}$,  $G_{i,\frac{j\pm1}{2},k}$, and $H_{i,j,\frac{k\pm1}{2}}$ are the approximation of the cell face flux reconstructions in $x$-, $y$-, and $z$-direction respectively. A TVDRK3 time evolution scheme \citep{shu1988efficient} is used to solve the ordinary differential equations resulting from a system of partial differential equations. The TVDRK3 scheme is selected because of its accuracy over other third-order Runge-Kutta schemes to compute hyperbolic conservation laws \citep{borges2008improved,san2012high}. To implement the TVDRK3 scheme, we use the method of lines to cast our system in the following form:
\begin{align}\label{ODE1}
  \frac{d q_{i,j,k}}{dt} = \pounds (q_{i,j,k})
\end{align}
Here the right-hand side of the equation represents the effect of various inviscid flux terms in the conservation equations. The time integration scheme is given as follows \citep{gottlieb1998total}:
\begin{equation}\label{RK3}
\left.\begin{aligned}
    q_{i,j,k}^{(1)}   &= q_{i,j,k}^{(n)} + \pounds (q_{i,j,k}^{(n)}) \Delta t, \\
    q_{i,j,k}^{(2)}   &= \frac{3q_{i,j,k}^{(n)} + q_{i,j,k}^{(1)} + \pounds (q_{i,j,k}^{(1)})\Delta t}{4}, \\
    q_{i,j,k}^{(n+1)} &= \frac{q_{i,j,k}^{(n)} + 2q_{i,j,k}^{(2)} + 2\pounds (q_{i,j,k}^{(2)}) \Delta t}{3},
\end{aligned}
\right\}
\end{equation}
where $\pounds (.)$ is the spatial operator, $q_{i,j,k}^{(n)}$ and $q_{i,j,k}^{(n+1)}$ are the data arrays of the solution field at the $n\textsuperscript{th}$ and $(n+1)\textsuperscript{th}$ timestep, respectively, and $q_{i,j,k}^{(1)}$ and $q_{i,j,k}^{(2)}$ are temporary arrays at the intermediate steps. A time step $\Delta t$ is prescribed through a CFL criterion as:
\begin{align}\label{timestep}
  \Delta t = \min \left(\eta \frac{\Delta x}{\max(|\Lambda_A|)},\eta \frac{\Delta y}{\max(|\Lambda_B|)},\eta \frac{\Delta z}{\max(|\Lambda_C|)}\right)
\end{align}
where $\max(|\Lambda_A|)$, $\max(|\Lambda_B|)$ and $\max(|\Lambda_C|)$ are the maximum absolute eigenvalues over the entire spatial domain at timestep $n$. The value of $\eta$ (i.e., should be $\eta \leq 1$) is set as $0.5$ to account for numerical stability.

\subsection{Upwind-Biased ILES Schemes: Nonlinear Interpolation}

We investigate the ILES scheme utilizing the damping characteristics of WENO schemes to act as an implicit filter preventing energy pile-up near the grid cut-off \citep{grinstein2007implicit,denaro2011does,boris1992new}. Since any upwind biased scheme has its own numerical dissipation, ILES framework uses this dissipative behavior of numerical discretizations to act similarly as the SGS model. Although ILES eliminates the computational cost of computing SGS term, there are a number of numerical errors associated with ILES which are difficult to control \citep{maulik2018explicit,thornber2007implicit}. In coming sections, we demonstrate the numerical formulation of the ILES approach implied in present study.

\subsubsection{Improved WENO ($Z$) Reconstructions} \hfill\\

We employ biased stencils for ILES schemes using the WENO nonlinear reconstructions to construct cell face quantities for our cell averaged variables. WENO scheme is widely used to solve the conservation laws accurately and efficiently in which the smoothness of the solutions is applied through adjustment of an adaptive stencil \citep{titarev2004finite,titarev2005weno}. The WENO approach is first introduced in \citep{liu1994weighted} for discontinuous problems to get an improvement over the essentially non-oscillatory (ENO) method \citep{shu1988efficient,shu1999high,harten1987uniformly}. For example, the WENO method is more accurate, and more efficiently parallelized \citep{jiang1996efficient,balsara2000monotonicity,shu1998essentially}. Both methods give especially good performance for problems with shocks and complicated smooth flow structures. In our study, the $5\textsuperscript{th}$ order accurate WENO scheme is used which is developed by the $3\textsuperscript{rd}$-order reconstructed fluxes from upwind $3$-point stencils. The order of accuracy depends on the length of the stencil chosen and has a substantial effect on the solution of the flow problem \citep{san2014numerical}. Considering reconstruction of the conserved variables, a modified implementation of the WENO reconstruction in the $x$-direction can be written as:
\begin{align}
\begin{split}
q_{i+1/2}^L = w_0 (\frac{1}{3} q_{i-2} - \frac{7}{6} q_{i-1} + \frac{11}{6} q_{i})
              +w_1 (-\frac{1}{6} q_{i-1} + \frac{5}{6} q_{i} + \frac{1}{3} q_{i+1})
              +w_2 (\frac{1}{3} q_{i} + \frac{5}{6} q_{i+1} - \frac{1}{6} q_{i+2})
\end{split}
\label{LeftFace}
\end{align}

\begin{align}
\begin{split}
q_{i-1/2}^R = w_0 (\frac{1}{3} q_{i+2} - \frac{7}{6} q_{i+1} + \frac{11}{6} q_{i})
              +w_1 (-\frac{1}{6} q_{i+1} + \frac{5}{6} q_{i} + \frac{1}{3} q_{i-1})
              +w_2 (\frac{1}{3} q_{i} + \frac{5}{6} q_{i-1} - \frac{1}{6} q_{i-2})
\end{split}
\label{RightFace}
\end{align}
Here, $q_{i+1/2}^L$ and $q_{i-1/2}^R$ are the left state (positive), $L$ and right state (negative), $R$ reconstructed fluxes, respectively, approximated at midpoints between cell nodes. The left ($L$) and right ($R$) states correspond to the possibility of advection from both directions. Since the procedures are similar in the $y$- and $z$-direction, we shall present stencil expressions only in the $x$-direction for the rest of this document. $w_k$ are the nonlinear WENO weights of the $k\textsuperscript{th}$ stencil where $k = 0, 1,..., r$ and $r$ is the number of stencils ($r = 2$ for the WENO5 scheme). The nonlinear weights are proposed by Jiang and Shu \citep{jiang1996efficient} in their classical WENO5-JS scheme as:
\begin{align}
w_k = \frac{\alpha_k}{\sum\limits_{k=0}^{2}{\alpha_k}}, \ \ \ \alpha_k = \frac{d_k}{(\beta_k + \epsilon)^p}
\end{align}
but the nonlinear weights defined by the WENO5-JS scheme are found to be more dissipative than many low-dissipation linear schemes in both smooth region and regions around discontinuities or shock waves \citep{henrick2005mapped}. There are several new or improved forms of WENO interpolation have been developed over the time to achieve improved computational efficiency with less numerical dissipation \citep{balsara2016efficient,ha2013improved,huang2016weno,kim2016modified,wong2017high,hu2015efficient,zhang2003numerical}. We refer to the literature above as examples for reader's interest to investigate more details about modified WENO schemes.
In our study, we have used an improved version of WENO approach proposed by \citep{borges2008improved}, often referred to as WENO5-Z scheme. One of the main reasons behind selecting WENO5-Z can be less dissipative behavior than classical WENO5-JS to capture shock waves. Also, there is a smaller loss in accuracy at critical points for improved nonlinear weights. The new nonlinear weights for the WENO5-Z scheme are defined by:
\begin{align}
w_k = \frac{\alpha_k}{\sum\limits_{k=0}^{2}{\alpha_k}}, \ \ \ \alpha_k = d_k\left(1 + \left(\frac{\tau_5}{\beta_k + \epsilon}\right)^p\right)
\end{align}
\begin{align}
\tau_5 = |\beta_2 - \beta_0|.
\end{align}
where $\beta_k$ and $p$ are the smoothness indicator of the $k\textsuperscript{th}$ stencil and a positive integer, respectively. Here $\tau_5$ is a fifth order reference smoothness indicator and $\epsilon = 1.0\times10^{-20}$, a small constant preventing zero division, and $p = 2$ is set in the present study to get the optimal fifth order accuracy at critical points. The expressions for $\beta_k$ in terms of cell values of $q$ are given by:
\begin{equation}
\left.\begin{aligned}
\beta_0 &= \frac{13}{12} (q_{i-2} - 2 q_{i-1} + q_i)^2 + \frac{1}{4} (q_{i-2} - 4 q_{i-1} + 3 q_i)^2, \\
\beta_1 &= \frac{13}{12} (q_{i-1} - 2 q_{i} + q_{i+1})^2 + \frac{1}{4}(q_{i-1} - q_{i+1})^2, \\
\beta_2 &= \frac{13}{12} (q_{i} - 2 q_{i+1} + q_{i+2})^2 + \frac{1}{4} (3 q_i - 4 q_{i+1} + q_{i+2})^2.
\end{aligned}
\right\}
\end{equation}
$d_k$ are the optimal weights for the linear high-order scheme which are given by:
\begin{align}
d_0 = \frac{1}{10},\ d_1 = \frac{3}{5},\ d_2 = \frac{3}{10}. \nonumber
\end{align}
There is another approach for WENO interpolation based on flux-splitting where positive and negative fluxes are obtained at the cell centers depending on the information propagation direction, and then the fluxes are computed at the cell boundaries using WENO interpolation \citep{jiang1996efficient,vcrnjaric2006different}.

\subsection{Riemann Solvers}

In the present study, we use the reconstruction based WENO interpolation procedure followed by a Riemann solver to determine and calculate the interfacial fluxes. It has been shown that Riemann solver selection has a significant effect on eddy resolving properties as well as turbulent statistics \citep{san2015evaluation}. We select three different Riemann solver in our investigation to evaluate their ability through numerical experiments.

\subsubsection{Rusanov Solver} \hfill\\

\cite{rusanov1961calculation} proposes the Riemann solver based on the information obtained from maximum local wave propagation speed, sometimes referred to as local Lax-Friedrichs flux \citep{lax1954weak,toro2013riemann}. The expression for Rusanov solver is as follows:
\begin{align}
  F_{i+1/2} = \frac{1}{2}(F^R + F^L) - \frac{c_{i+1/2}}{2} \left(q_{i+1/2}^R - q_{i+1/2}^L\right)
\end{align}
where $F^R = F(q^{R}_{i+1/2}) =$ The right constructed state flux component, $F^L = F(q^{L}_{i+1/2}) =$ The flux left constructed state component and $c_{i+1/2} = \max(r(A_i),r(A_{i+1})) =$ The local wave propagation speed. $r(A) = \max(|u|,|u-a|,|u+a|) =$ The spectral radius of convective Jacobian matrix $A$.

\subsubsection{Roe Solver} \hfill\\

\cite{roe1981approximate} introduces a Riemann solver based on the Godunov theorem \citep{godunov1959difference}. The Godunov scheme states that, for a hyperbolic system of equations, the exact values of fluxes at the interface can be computed by following relation (if the Jacobian matrix, $A$ is constant):
\begin{align}
  F_{i+1/2} = \frac{1}{2} (F^R + F^L) - \frac{1}{2} R_A |\Lambda_A| L_A \left(q_{i+1/2}^R - q_{i+1/2}^L\right)
\end{align}
where $|\Lambda_A|$, $R_A$, and $L_A$ are obtained from the eigendecomposition of Jacobian matrix $A$. Since $A$ is not constant for the systems with shocks, the Roe solver \citep{roe1981approximate} estimates the interfacial flux as follows:
\begin{align}
    F_{i+1/2} = \frac{1}{2} (F^R + F^L) - \frac{1}{2} \tilde{R}_A |\tilde{\Lambda}_A| \tilde{L}_A \left(q_{i+1/2}^R - q_{i+1/2}^L\right).
\end{align}
The eigensystem matrices can be constructed using the density weighted average values of the conserved variables by:
\begin{align}
  \tilde{u} = \frac{u_R \sqrt{\rho_R} + u_L \sqrt{\rho_L}}{\sqrt{\rho_R}+\sqrt{\rho_L}}
\end{align}
\begin{align}
  \tilde{v} = \frac{v_R \sqrt{\rho_R} + v_L \sqrt{\rho_L}}{\sqrt{\rho_R}+\sqrt{\rho_L}}
\end{align}
\begin{align}
  \tilde{H} = \frac{H_R \sqrt{\rho_R} + H_L \sqrt{\rho_L}}{\sqrt{\rho_R}+\sqrt{\rho_L}}
\end{align}
where the left and right states of the un-averaged conserved variables are available from the WENO5 reconstruction described earlier. To fix the entropy, Harten et al. proposed the following approach \citep{harten1983high} by replacing Roe averaged eigenvalues with the eigenvalues of the Roe averaged Jacobian matrix $\tilde{A}$, $\tilde{\lambda}_i$ which is given by
\begin{align}
  |\tilde{\lambda}_i| =
    \begin{cases}
        |\tilde{\lambda}_i|, & \text{if  } |\tilde{\lambda}_i| \geq 2\epsilon \tilde{a}\\ \tilde{\lambda}_i^2/(4\epsilon \tilde{a}), & \text{if  } |\tilde{\lambda}_i| < 2\epsilon \tilde{a}
    \end{cases}
\end{align}
Here, $\epsilon$, a small positive number, is set as $0.1$ in our computations.

\subsubsection{AUSM Solver} \hfill\\

The final Riemann solver we considered is Advection Upstream Splitting Method (AUSM) which was introduced by Liou et al. \citep{liou1993new}. The AUSM has been employed in a wide range of problems because of it's features like accurate capturing of shocks and contact discontinuities as well as entropy-satisfying solutions. The main idea behind the AUSM scheme is to split the inviscid flux into a convective flux part and a pressure flux part. The convective flow quantities are determined by a cell interface advection Mach number. There are many types of the AUSM solver which are developed to obtain a more accurate and robust result against shocks in all-speed regimes \citep{liou2001ten,zhang2017robust,kitamura2016reduced,matsuyama2014performance}. Here we use the AUSM as a low-diffusion solver to yield the interfacial numerical flux as:
\begin{align}
F_{i+1/2} = \frac{M_{i+1/2}}{2} \left[
\begin{pmatrix}
\rho a \\
\rho a u \\
\rho a v \\
\rho a w \\
\rho a H
\end{pmatrix}^L +
\begin{pmatrix}
\rho a \\
\rho a u \\
\rho a v \\
\rho a w \\
\rho a H
\end{pmatrix}^R
\right] - 
\frac{|M_{i+1/2}|}{2} \left[
\begin{pmatrix}
\rho a \\
\rho a u \\
\rho a v \\
\rho a w \\
\rho a H
\end{pmatrix}^R -
\begin{pmatrix}
\rho a \\
\rho a u \\
\rho a v \\
\rho a w \\
\rho a H
\end{pmatrix}^L
\right] +
\begin{pmatrix}
0 \\
P_L^{+} + P_R^{-} \\
0 \\
0   \\
0
\end{pmatrix}
\end{align}
where the mass flux is given by:
\begin{align}
M_{i+1/2} = M_L^{+} + M_R^{-}
\end{align}
in which the directional convective Mach number ($M = u/a$ in $x$-direction) is given as:
\begin{align}
M^{\pm} =
\begin{cases}
\pm \frac{1}{4} (M \pm 1)^2,\ \  \textnormal{if} \ \ |M| \leq 1 \\
\frac{1}{2} (M \pm |M|),\ \  \textnormal{otherwise}
\end{cases}
\end{align}
The following split formula is used for the pressure flux:
\begin{align}
P^{\pm} =
\begin{cases}
P \frac{1}{4} (M \pm 1)^2 (2 \mp M),\ \  \textnormal{~if} \ \ |M| \leq 1 \\
P \frac{1}{2} (M \pm |M|)/M,\ \  \textnormal{~~~~~~otherwise.}
\end{cases}
\end{align}
There is another type of Riemann solver named The Harten-Lax-van Leer (HLL) approximate Riemann solver which assumes that the lower and upper bounds on the characteristic speeds can be used in the solution of the Riemann problem involving the right and left states \citep{harten1983upstream,davis1988simplified,toro2013riemann}. Shortcomings of HLL scheme lead to a modified HLL scheme called HLLC scheme whereby the missing contact and shear waves in the Euler equations are restored \citep{toro2009hll}. Force scheme, Marquina scheme are also some examples of Riemann solver used in shock capturing schemes \citep{donat1996capturing,stecca2010upwind,dumbser2010force,donat1998flux,san2015evaluation}.
However, we haven't included these solvers into our study since the selected Riemann solvers may represent the lower and upper dissipation bounds within our ILES framework.

\subsection{6\textsuperscript{th} order Central Schemes and Relaxation Filtering ($CS+RF$)}
\label{CS+RF}

\begin{figure}[!ht]
\centering
\mbox{
\includegraphics[width=0.7\textwidth]{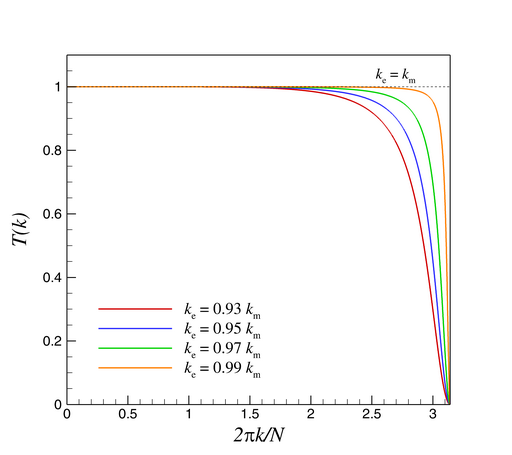}
}
\caption{Transfer function of Pad\'{e} filters (6\textsuperscript{th} order) for $CS+RF$ with relaxation filtering scale as a free parameter.}
\label{TF for RF}
\end{figure}

In this subsection, we consider a symmetric flux reconstruction approach using a purely central scheme (CS) combined with a classic discrete compact pad\'{e} filter \citep{lele1992compact} as a relaxation filter (RF). This can be considered as an explicit large eddy simulation technique for conservation laws \citep{mathew2003explicit}, where a $7$-point stencil non-dissipative scheme is used for face reconstruction of the conserved quantities \citep{hyman1992high}:
\begin{align}
q_{i+1/2} = a (q_{i+1}+q_{i}) + b (q_{i+2}+q_{i-1}) + c (q_{i+3}+q_{i-2})
\end{align}
where the stencil coefficients are given by:
\begin{align}
a = 37/60; \ \ b = -8/60; \ \ c = 1/60. \nonumber
\end{align}
Here, $q_i$ represents the inviscid spatial derivatives in $x$-direction of the Eq.~(\ref{ODE1}). The calculated fluxes from the relevant face quantities determined from the nodal values can be used in Eq.~(\ref{FVMEq}). In this approach, we assume that the explicit filtering removes the frequencies higher than a selected cut-off threshold through the use of a low-pass spatial filter (sixth-order pad\'{e} filter in our case). A low-pass filter is commonly used in explicit filtering approaches which can be considered as a free modeling parameter with a specific order of accuracy and a fixed filtering strength \citep{fauconnier2013performance}. The filtering operation is done at the end of every timestep to remove high frequency content from the solution which eventually prevents the oscillations \citep{lund2003use,bogey2006computation,bull2016explicit}. The motivation of using compact pad\'{e} filter is that it has been successfully implemented previously in many noteworthy literature \citep{san2015posteriori,stolz2001approximate,pruett2000priori,stolz1999approximate}. In our investigation, we use the following expression for compact pad\'{e} filter:
\begin{align}\label{pade}
\alpha \bar{f}_{j-1} + \bar{f}_{j} + \alpha \bar{f}_{j+1} = \sum_{m=0}^{3}{\frac{a_m}{2}(f_{j-m} + f_{j+m})}
\end{align}
where the discrete quantity $f_j$ yields the filtered value $\bar{f}_{j}$ and the filtering coefficients are:
\begin{equation}\label{coeff}
\left.\begin{aligned}
 a_0 &= \frac{11}{16} + \frac{5}{8} \alpha, \ a_1 = \frac{15}{32} + \frac{17}{16} \alpha, \\
a_2 &= -\frac{3}{16} + \frac{3}{8} \alpha, \ a_3 = \frac{1}{32} - \frac{1}{16} \alpha.
\end{aligned}
\right\}
\end{equation}
The free parameter, $\alpha$ controls filtering dissipation strength which can be expressed as:
\begin{align}
\alpha = -\frac{\cos{\left(\frac{k_e}{k_m}\pi\right)}}{2}
\end{align}
where $k_e$ is the effective wave number and $k_m $ is the grid cut-off wave number. The ratio between $k_e$ and $k_m $ can be called as relaxation filtering scale which should be in between the range of $0 < \frac{k_e}{k_m} < 1$ to yield $[-0.5,0.5]$ range for $\alpha$. A Fourier analysis can be carried out to analyze the behavior of the filter in the wave number space. Using the modified transfer function analysis, transfer function, $T(k)$ is determined in the Fourier domain as follows:
\begin{align}
\hat{\bar{f}} = T(k)\hat{f}
\end{align}
The discrete inverse Fourier transform of a function $f$, assumed to be periodic over the domain $[0,L]$ and defined only on a discrete set of $N$ grid points, can be written as:
\begin{align}
f_j = \sum_{m = - N/2}^{N/2 - 1}\hat{f}_m \ \text{exp}\left(\text{i}\frac{2\pi mj}{L}\right), \ j = 0,1,...,N-1
\end{align}
and the discrete forward Fourier transform is:
\begin{align}
\hat{f}_m = \frac{1}{N}\sum_{j = 0}^{N - 1}f_j \ \text{exp}\left(- \text{i} \frac{2\pi mj}{L}\right), \ m = - N/2, - N/2 + 1,...,N/2 - 1
\end{align}
where $\text{i} = \sqrt{-1}$ and the grid spacing $h = L/N$. We find the scaled wave number $\omega = 2\pi kh/L = 2\pi k/N$ over the domain $[0,\pi]$. With the definition of $k = \frac{2\pi}{L}m$, the following transfer function is obtained given by the Equation~(\ref{pade}):
\begin{align}
T(k) = \frac{a_0 + a_1\cos(hk) + a_2\cos(2hk) + a_3\cos(3hk)}{1 + 2\alpha\cos(hk)}
\end{align}
where the coefficients $a_i$ are defined in Eq.~(\ref{coeff}). The transfer functions of the Pad\'{e} filter for different values of relaxation filtering scale have been shown in Fig.~(\ref{TF for RF}). As shown in the figure, there is no dissipation at all when the effective wave number is equal to the maximum wave number. The maximum attenuation (when $T(k)$ becomes zero) happens for the Pad\'{e} filter at the largest or grid cut-off wave number. The explicit filtering models have the benefit over implicit ones that they allow the control of the numerical errors. We will show that how the $CS+RF$ technique performs for the classical numerical test problems in the later period of our study.

\subsection{Eddy Viscosity Models}

In eddy viscosity models, a turbulent eddy viscosity is used as a dissipation mechanism to account for the SGS contributions. For LES, the equations of motion are derived formally using a low-pass spatial filter and solving them for the filtered quantities. Although LES equations are determined from a low-pass filtering procedure, a filter is not specified in most functional model development. The filtered Euler equation solved in our computations with the SGS terms can be expressed as:
\begin{align}
 \frac{\partial \textbf{q}}{\partial t} + \frac{\partial \textbf{F}}{\partial x} + \frac{\partial \textbf{G}}{\partial y} + \frac{\partial \textbf{H}}{\partial z} = \textbf{K}^{sgs},
\end{align}
 where the turbulent stress contribution in energy equation $\textbf{K}_{sgs}$ is defined as:
\begin{align}
\textbf{K}^{sgs} = \frac{\partial \textbf{F}^{sgs}}{\partial x} + \frac{\partial \textbf{G}^{sgs}}{\partial y} + \frac{\partial \textbf{H}^{sgs}}{\partial z},
\end{align}

\begin{align}
\text{where} \ \textbf{F}^{sgs} = \begin{bmatrix}
           0 \\
           \tau_{xx} \\
           \tau_{xy} \\
           \tau_{xz} \\
           -\dot{q}_x
         \end{bmatrix}, \ \ \
\textbf{G}^{sgs} = \begin{bmatrix}
            0 \\
           \tau_{xy} \\
           \tau_{yy} \\
           \tau_{yz} \\
           -\dot{q}_y
         \end{bmatrix}, \ \ \
\textbf{H}^{sgs} = \begin{bmatrix}
           0 \\
           \tau_{xz} \\
           \tau_{yz} \\
           \tau_{zz} \\
           - \dot{q}_z
         \end{bmatrix}. \nonumber \ \ \
\end{align}
Here, $\textbf{F}^{sgs}$, $ \textbf{G}^{sgs}$, $\textbf{H}^{sgs}$ are the SGS fluxes. The turbulent SGS stress tensors $\tau_{ij}$ can be expressed as:

\begin{equation}\label{stress}
\left.\begin{aligned}
\tau_{xx} = \frac{2}{3} \nu_{e} \big(2\frac{\partial u}{\partial x} - \frac{\partial v}{\partial y} -\frac{\partial w}{\partial z}\big), \ \ \ \tau_{xy} = \nu_{e} \big(\frac{\partial u}{\partial y} + \frac{\partial v}{\partial x}\big), \\
\tau_{yy} = \frac{2}{3} \nu_{e} \big(2\frac{\partial v}{\partial y} - \frac{\partial u}{\partial x} -\frac{\partial w}{\partial z}\big), \ \ \ \tau_{xz} = \nu_{e} \big(\frac{\partial u}{\partial z} + \frac{\partial w}{\partial x}\big), \\
\tau_{zz} = \frac{2}{3} \nu_{e} \big(2\frac{\partial w}{\partial z} - \frac{\partial u}{\partial x} -\frac{\partial v}{\partial y}\big), \ \ \ \tau_{xy} = \nu_{e} \big(\frac{\partial v}{\partial z} + \frac{\partial w}{\partial y}\big).
\end{aligned}
\right\}
\end{equation}
In dimensionless for of the energy equation, the pressure-velocity and pressure-dilation subgrid terms are modeled together as a SGS heat flux term $\dot{\textbf{q}}$ based on the turbulent Prandtl number $Pr_t$ (i.e., $Pr_t$ is set $0.72$ unless otherwise stated in our study) which can be given by:
\begin{align}\label{heatflux}
\dot{q_i}= - \frac{\gamma \nu_e}{Pr_t} \frac{\partial \theta}{\partial x_i},
\end{align}
where dimensionless temperature field $\theta$ can be found by:
\begin{align}
\theta = \frac{P}{\rho (\gamma -1)}.
\end{align}
We use $7$-point stencil non-dissipative central scheme to compute flux derivatives at cell interfaces for SGS terms using the following expression (for $x$-direction):
\begin{align}
 \frac{\partial q_{i+1/2}}{\partial x} &= a\big(\frac{q_{i+1} - q_i}{\Delta x}\big) + b\big(\frac{q_{i+2} - q_{i-1}}{3\Delta x}\big) + c\big(\frac{q_{i+3} - q_{i-2}}{5\Delta x}\big),
\end{align}
where the coefficients are
\begin{align}
a &= \frac{245}{180}, \ b = \frac{-75}{180}, \ c = \frac{10}{180}. \nonumber
\end{align}
Here, $q$ accounts for the flow quantities such as $u$, $v$, $w$, and $\theta$ presented in Eq.~(\ref{stress}) and Eq.~(\ref{heatflux}). The expressions for $y$- and $z$-direction are similar to the expression for $x$-direction. Based on the approach to compute the artificial eddy viscosity $\nu_e$, various SGS models are proposed. We have investigated the Smagorinsky and localized dynamic eddy viscosity models in our study.

\subsubsection{Smagorinsky Model} \hfill\\

The Smagorinsky model is one of the most widely used eddy viscosity model over the decades which was originally proposed for simulating high $(Re)$ atmospheric flow \citep{smagorinsky1963general}. The model is based on the assumptions that large-scale carries the most of the energy whereas the smaller scales possess universality in their behavior. Smagorinsky assumed that the turbulence at the fine scale is strictly dissipative, homogeneous and the production equals the dissipation (the local equilibrium). Focusing only on the dissipating energy at a rate that is physically correct, the Smagorinsky model computes the eddy viscosity from the mixing length and the absolute value of the strain rate tensor of the resolved flow field in order to account for the appropriate dissipation at smaller scales. The Smagorinsky model gives the following expression for the SGS eddy-viscosity coefficient by setting the characteristic length to be the filter width $\Delta$:
\begin{align}
\nu_{sgs} &= C_S^2\Delta^2|S(\textbf{u})| \ \ \text{with} \ |S(\textbf{u})| = \sqrt{2S_{ij}S_{ij}}
\end{align}
where the strain rate $S_{ij} = \frac{1}{2}\big(\frac{\partial u_i}{\partial x_j}+\frac{\partial u_j}{\partial x_i} \big)$. In its explicit form, the absolute value of the strain rate tensor can be written as
\begin{align}
|S(\textbf{u})| &= \sqrt{2\Big( \frac{\partial u}{\partial x}\frac{\partial u}{\partial x} +\frac{\partial v}{\partial y}\frac{\partial v}{\partial y}+\frac{\partial w}{\partial z}\frac{\partial w}{\partial z} \Big) + \Big(\frac{\partial u}{\partial y} + \frac{\partial v}{\partial x} \Big)^2 +\Big(\frac{\partial u}{\partial z} + \frac{\partial w}{\partial x} \Big)^2 +\Big(\frac{\partial v}{\partial z} + \frac{\partial w}{\partial y} \Big)^2 }.
\end{align}
In this model, $C_S$ is the only non-dimensional constant that needs to be provided (we consider $C_s = 0.18$). In our study, we set constant grid spacing of $\Delta = h = (h_x h_y h_z)^{(1/3)}$ where $h$ is the cell size (i.e., a uniform equidistant mesh size in the present study).

\subsubsection{Dynamic Smagorinsky Model} \hfill\\

We here present the concept of the dynamic Smagorinsky model since the localized dynamic model we have used in our study is an extension of the dynamic Smagorinsky model. The Smagorinsky model gives good representation for the overall energy dissipation of the flow but cannot reproduce the flow locally. It also requires additional corrections for near-wall regions or for anisotropic flows. Also, modeling of various flow problems using Smagorinsky approach has revealed that the Smagorinsky constant is not single-valued \citep{smagorinsky1993some}. These shortcomings with Smagorinsky model lead to needs for developing models that determine the SGS eddy viscosity by utilizing both the grid-scale velocity gradients and the local grid-scale turbulence mechanism \citep{kajishima2017computational}. A new idea was introduced to determine $C_S$ dynamically from the grid-scale velocity field with the possibility of overcoming the limitations in the Smagorinsky model. Germano et al. \citep{germano1991dynamic} introduced the use of a test filter along with the original filter as grid filter to extract relatively small-scale structure from the grid-scale component. The dynamic model is further improved by \citep{lilly1992proposed} who proposed to calculate $C_S$ using the least square method. The approach taken by Lilly is to minimize the error in a least squares fashion and can usually avoid the division by zero by an ensemble averaging procedure with the homogeneous flow directions.

The test filter width is taken to be larger than that of the grid filter or the mesh size $h$. An optimized design for the test filters is necessary for constructing efficient and consistent dynamic model which will be discussed afterward. With the definition of a test filter, the dynamic computation of the $C_S$ parameter can be performed as
\begin{align}
C_S^2 \Delta^2 = \frac{\langle L_{ij}M_{ij}\rangle}{\langle M_{ij}M_{ij}\rangle}.
\end{align}
where the numerator and the denominator are averaged in general over the spatial homogeneous directions to prevent the numerical instabilities. Still, there might be numerical instabilities present due to a large fluctuation in the value of $C_S$. Therefore, several clipping ideas have also been introduced \citep{sagaut2006large}. There is another kind of dynamic model, referred as dynamic mixed model, which resolves the issue of negative eddy viscosity to represent the inverse cascade in dynamic models \citep{zang1993dynamic,salvetti1995priori}. Ghosal et al. \citep{ghosal1995dynamic} proposed a dynamic localization model to prevent the mathematical inconsistency of assuming $C_S$ as an invariant inside the filter where the Smagorinsky constant is computed as a function of position at each time-step \citep{piomelli1995large,chai2012dynamic,carati1995representation}. Following the definitions of the subgrid stress components in Eq.~(\ref{stress}), the definition of the $L_{ij}$ tensor in the setting of Euler equations can be written as
\begin{align}
L_{ij} = \bar{u}_i \bar{u}_j - \overline{(u_i u_j)}
\end{align}
where overbar refers to a test filtering operator to the resolved flow quantities. In explicit form this expression can be written as
\begin{align}
L_{11} &= \bar{u} \bar{u} - \overline{(u u)} \\
L_{22} &= \bar{v} \bar{v} - \overline{(v v)} \\
L_{33} &= \bar{w} \bar{w} - \overline{(w w)} \\
L_{12} &= \bar{u} \bar{v} - \overline{(u v)} \\
L_{13} &= \bar{u} \bar{w} - \overline{(u w)} \\
L_{23} &= \bar{v} \bar{w} - \overline{(v w)}
\end{align}
The definition of $M_{ij}$ tensor is given by
\begin{align}
M_{ij} = \kappa^2 |S(\bar{\textbf{u}})| \big(\frac{\partial \bar{u}_i}{\partial x_j} + \frac{\partial \bar{u}_j}{\partial x_i} -\frac{2}{3}\frac{\partial \bar{u}_k}{\partial x_k}\delta_{ij} \big) - \overline{|S({\textbf{u}})| \big(\frac{\partial u_i}{\partial x_j} + \frac{\partial u_j}{\partial x_i} -\frac{2}{3}\frac{\partial u_k}{\partial x_k}\delta_{ij} \big)},
\end{align}
which can be further rewritten for each component as
\begin{align}\label{Mstress}
M_{11} &= \frac{2}{3} \left(\kappa^2 |S(\bar{\textbf{u}})| \big(2\frac{\partial \bar{u}}{\partial x} - \frac{\partial \bar{v}}{\partial y} -\frac{\partial \bar{w}}{\partial z}\big) - |\overline{S(\textbf{u})| \big(2\frac{\partial u}{\partial x} - \frac{\partial v}{\partial y} -\frac{\partial w}{\partial z}} \big) \right), \\
M_{22} &= \frac{2}{3} \left(\kappa^2 |S(\bar{\textbf{u}})| \big(2\frac{\partial \bar{v}}{\partial y} - \frac{\partial \bar{u}}{\partial x} -\frac{\partial \bar{w}}{\partial z}\big) - |\overline{S(\textbf{u})| \big(2\frac{\partial v}{\partial y} - \frac{\partial u}{\partial x} -\frac{\partial w}{\partial z}} \big) \right), \\
M_{33} &= \frac{2}{3} \left(\kappa^2 |S(\bar{\textbf{u}})| \big(2\frac{\partial \bar{w}}{\partial z} - \frac{\partial \bar{u}}{\partial x} -\frac{\partial \bar{v}}{\partial y}\big) - |\overline{S(\textbf{u})| \big(2\frac{\partial w}{\partial z} - \frac{\partial u}{\partial x} -\frac{\partial v}{\partial y}} \big) \right), \\
M_{12} &= \kappa^2 |S(\bar{\textbf{u}})| \big(\frac{\partial \bar{u}}{\partial y} +\frac{\partial \bar{v}}{\partial x}\big) - |\overline{S(\textbf{u})| \big(\frac{\partial u}{\partial y} +\frac{\partial v}{\partial x}} \big),  \\
M_{13} &= \kappa^2 |S(\bar{\textbf{u}})| \big(\frac{\partial \bar{u}}{\partial z} +\frac{\partial \bar{w}}{\partial x}\big) - |\overline{S(\textbf{u})| \big(\frac{\partial u}{\partial z} +\frac{\partial w}{\partial x}} \big),  \\
M_{23} &= \kappa^2 |S(\bar{\textbf{u}})| \big(\frac{\partial \bar{v}}{\partial z} +\frac{\partial \bar{w}}{\partial y}\big) - |\overline{S(\textbf{u})| \big(\frac{\partial v}{\partial z} +\frac{\partial w}{\partial y}} \big).
\end{align}
The ratio $\kappa = \bar{\Delta}/\Delta > 1$ which is the only input parameter in the dynamic Smagorinsky model. Here, $\Delta$ and $\bar{\Delta}$ are the characteristic lengths of the grid filter and the test filter, respectively. In the following section we will develop a consistent filtering operator.

\subsubsection{Localized Dynamic Eddy-viscosity Model} \hfill\\
\label{localized}

As we have mentioned above, the issues associated with the dynamic model which lead to instability and the mathematical inconsistency in pulling $C_S$ outside of the test filter \citep{kajishima2017computational} make it necessary to formulate more efficient dynamic models like dynamic mixed model or dynamic localization model. Furthermore, a global averaging operator for homogeneous directions defined in the Lilly's model becomes impractical in parallel computing since each processing element has only local field information (e.g., see \cite{brehm2017consistent} for a recent discussion for the parallelization of stencil-based computing). Therefore, we propose here a dynamic modeling approach using the ensemble average over the local neighboring grid points in a $7$-point stencil scheme to calculate $C_S^2 \Delta^2$:
\begin{align}
(C_S^2 \Delta^2)_{i+1/2, j+1/2, k+1/2} = \frac{\sum\limits_{p={i-2}}^{i+3}\sum\limits_{q={j-2}}^{j+3}\sum\limits_{r={k-2}}^{k+3}(L_{mn} M_{mn})_{p,q,r}}{\sum\limits_{p={i-2}}^{i+3}\sum\limits_{q={j-2}}^{j+3}\sum\limits_{r={k-2}}^{k+3}(M_{mn} M_{mn})_{p,q,r}},
\end{align}
where $p$, $q$, and $r$ refer to local grid index in each spatial direction. Here, we also use the Einstein summation notation for tensor indices (i.e., $m = 1, 2, 3$ and $n = 1, 2, 3$).
Then the eddy viscosity is calculated using $\nu_{e} = C_S^2 \Delta^2|S(\textbf{u})|$ which can be used in Eq.~(\ref{stress}) to calculate the subgrid stress terms. Since the coefficient is calculated locally, we refer to this approach as localized dynamic Smagorinsky (LDS) model. This approach shows significant improvement in the estimation of higher resolution flow physics for both compressible and incompressible flows which will be illustrated in Section~\ref{sec:4} based on a-posteriori analysis of the numerical test cases.

\subsubsection{Optimized Filter Design for the Dynamic Model} \hfill\\

\begin{figure}[!ht]
\centering
\mbox{
\subfigure[Regular Gaussian filters (see Eq.~(\ref{optG}) and Eq.~(\ref{coeff1}))]{\includegraphics[width=0.48\textwidth]{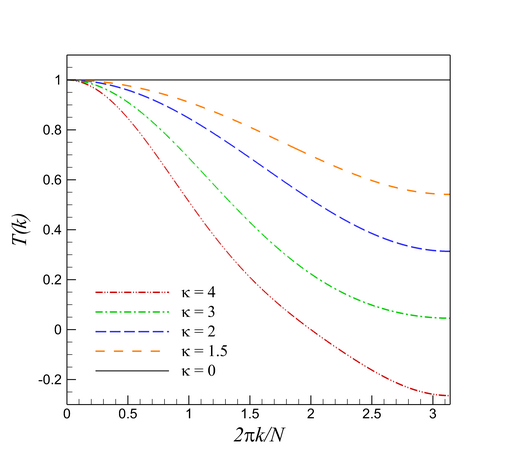}}
\subfigure[Optimized Gaussian filters (see Eq.~(\ref{optG}) and Eq.~(\ref{coeff2}))]{\includegraphics[width=0.48\textwidth]{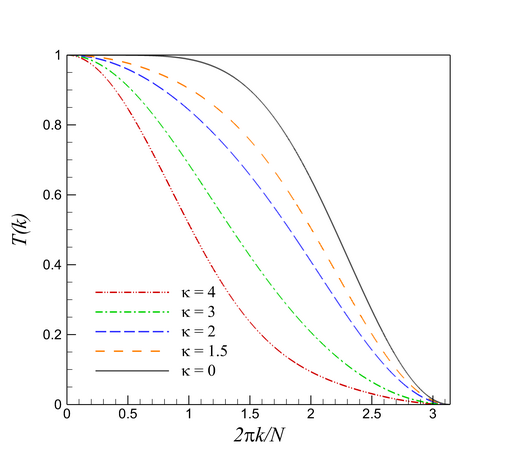}}
}\\
\caption{Transfer functions of optimized low-pass spatial filters designed for dynamic eddy viscosity models.}
\label{fig:optimized filter}
\end{figure}

Low-pass filters play a central role in LES and have been widely used in most of the dynamic eddy viscosity, approximate deconvolution, and mixed model approaches. Since most of the state-of-the-art closure models require specification of a low-pass filter, a consistent implementation of these filters has become increasingly important for accurate LES computations in geophysical and engineering applications. Although an ideal sharp cut-off filtering approach is utilized in pseudo-spectral methods, a smoothing operator based approach or discrete filtering implementations have been considered extensively in physical space based finite difference/volume/element
formulations \citep{de2002sharp,vasilyev1998general,sagaut1999discrete,najjar1996study}. Discrete filters are mostly used in dynamic models as test filter, and in explicit filtering models as relaxation filter. The dynamic model utilizes the test filter to approximate the Smagorinsky constant dynamically from the resolved flow field and requires specification of the filter ratio parameter $\kappa$ between the grid scale and test filter scale. However, many discrete filters designed to eliminate high frequency content of the flows consider a filtering strength parameter with lack of consistent definition of the parameter $\kappa$, which has been set $\kappa$ = 2 by most of the LES practitioners \citep{lund1997use} and also, some filters do not attenuate fully within the computational domain \citep{san2015posteriori,san2014dynamic,san2016analysis}. Again, the pad\'{e} filter is found to show complete attenuation at the grid cut-off scale \citep{maulik2018explicit}. However, it can be observed in Fig.~(\ref{TF for RF}) of Section~\ref{CS+RF}, the transfer function for pad\'{e} filter shows less dissipative characteristics even though the transfer functions for all the filter ratio are attenuating completely. Adding less dissipation can produce an oscillatory or non-smooth solution of the flow field. To resolve these issues associated with regular discrete filters, we develop an optimized form of regular Gaussian discrete filter and utilize it in our proposed localized dynamic model. Gaussian-type filters are frequently used in well established digital image processing community \citep{oppenheim1999discrete,jahne1997digital}. The expression for a one-dimensional filter can be expressed in the following discrete form \citep{kajishima2017computational}:
\begin{align}\label{gauss}
\bar{f}_j = f_j + \gamma_2 f_j^{(2)} + \gamma_4 f_j^{(4)} + \gamma_6 f_j^{(6)} + ...,
\end{align}
where for the Gaussian filter,
\begin{align}
\gamma_2 = \frac{\bar{\Delta}^2}{24}, \ \gamma_4 = \frac{\bar{\Delta}^4}{1152}, \ \gamma_6 = \frac{\bar{\Delta}^6}{82944}, \ ... \nonumber
\end{align}
Using the expansion of the derivatives, we can control the order of accuracy of the filters. The test filter length $\bar{\Delta}$ can also be used as a control parameter. To formulate the regular Gaussian filter using $7$-point stencil central scheme for high-order derivatives:
\begin{align}
f_j^{(6)} &= \frac{1(f_{j-3} + f_{j+3}) - 6 (f_{j-2} + f_{j+2}) + 15 (f_{j-1}  + f_{j+1}) - 20f_j}{\Delta^6} + O(\Delta^2), \\
f_j^{(4)} &= \frac{-1(f_{j-3} + f_{j+3}) + 12 (f_{j-2} + f_{j+2}) - 39 (f_{j-1}  + f_{j+1}) +  56f_j}{6\Delta^4} + O(\Delta^4), \\
f_j^{(2)} &= \frac{2(f_{j-3} + f_{j+3}) - 27 (f_{j-2} + f_{j+2}) + 270 (f_{j-1}  + f_{j+1}) -  490f_j}{180\Delta^2} + O(\Delta^6).
\end{align}
Here, $\Delta$ is the grid size which is considered as the characteristic length for grid filter. If we use the above equations for high-order derivatives in Eq.~(\ref{gauss}), we get the following approximation up to the third term assuming the test filter ratio, $\kappa = \bar{\Delta} / \Delta$:
\begin{align}\label{optG}
\bar{f}_j = a_0 f_j + a_1(f_{j-1}  + f_{j+1}) + a_2(f_{j-2} + f_{j+2}) + a_3(f_{j-3} + f_{j+3}),
\end{align}
where the coefficients are
\begin{equation}\label{coeff1}
\left.\begin{aligned}
&a_0 = 1 - \frac{490}{180}\left(\frac{\kappa^2}{24}\right) + \frac{56}{6}\left(\frac{\kappa^4}{1152}\right) - 20 \left(\frac{\kappa^6}{82944}\right), \\
&a_1 = \frac{270}{180}\left(\frac{\kappa^2}{24}\right) - \frac{39}{6}\left(\frac{\kappa^4}{1152}\right) + 15 \left(\frac{\kappa^6}{82944}\right), \\
&a_2 = -\frac{27}{180}\left(\frac{\kappa^2}{24}\right) + 2\left(\frac{\kappa^4}{1152}\right) - 6 \left(\frac{\kappa^6}{82944}\right), \\
&a_3 = \frac{2}{180}\left(\frac{\kappa^2}{24}\right) - \frac{1}{6}\left(\frac{\kappa^4}{1152}\right) + \left(\frac{\kappa^6}{82944}\right).
\end{aligned}
\right\}
\end{equation}
Since the discrete filters are developed in physical space, we transform the filter equation into wave space by a Fourier analysis. This transformation yields the transfer function using the modified transfer function analysis that correlates the Fourier coefficients of the filtered variable , $\bar{f}_j$ and unfiltered variable, $f_j$. In Fig.~(\ref{fig:optimized filter}), the transfer functions of the regular Gaussian filter show a good dissipative property, but does not get fully attenuated for some values of $\kappa$. To develop an optimized Gaussian filter, we reduce or optimize the accuracy of higher order derivatives and consider up to the fourth term in Eq.~(\ref{gauss}). This gives a free modeling parameter $\alpha$. The new equations for derivatives can be expressed as:
\begin{align}
f_j^{(4)} &=  \frac{4 + \alpha}{15\Delta^4}(f_{j-3} + f_{j+3}) - \frac{3+2\alpha}{5\Delta^4}(f_{j-2} + f_{j+2}) + \frac{\alpha}{\Delta^4}(f_{j-1}  + f_{j+1}) + \frac{2-4\alpha}{3\Delta^4}f_j + O(\Delta^2), \\
f_j^{(2)} &= \frac{-4 + 3\alpha}{45\Delta^2}(f_{j-3} + f_{j+3}) + \frac{9-8\alpha}{20\Delta^2}(f_{j-2} + f_{j+2}) + \frac{\alpha}{\Delta^2}(f_{j-1}  + f_{j+1}) - \frac{13+24\alpha}{18\Delta^2}f_j + O(\Delta^4).
\end{align}
We use the derivatives above in Eq.~(\ref{gauss}) to get the filter equation similar to Eq.~(\ref{optG}) with new coefficients given by
\begin{equation}\label{coeff2}
\left.\begin{aligned}
&a_0 = 1-\left(\frac{13+24\alpha}{18}\right)\left(\frac{\kappa^2}{24}\right) + \left(\frac{2-4\alpha}{3}\right)\left(\frac{\kappa^4}{1152}\right),\\
&a_1 = \alpha\left(\frac{\kappa^2}{24}\right) + \alpha\left(\frac{\kappa^4}{1152}\right), \\
&a_2 = \left(\frac{9-8\alpha}{20}\right)\left(\frac{\kappa^2}{24}\right) - \left(\frac{3+2\alpha}{5}\right)\left(\frac{\kappa^4}{1152}\right), \\
&a_3 = \left(\frac{-4 + 3\alpha}{45}\right)\left(\frac{\kappa^2}{24}\right) + \left(\frac{4 + \alpha}{15}\right)\left(\frac{\kappa^4}{1152}\right).
\end{aligned}
\right\}
\end{equation}
We obtain the transfer function of the filter equation using a Fourier transformation. To achieve full attenuation, we set the condition for the transfer function to be $0$ at position $\pi$ in wavenumber space, and get the following expression for $\alpha$ for full attenuation condition:
\begin{align}
\alpha = \frac{1080 + 16\kappa^2 - \kappa^4}{4\kappa^2(48 + \kappa^2)}.
\end{align}
The transfer functions of the optimized Gaussian filter is shown in Fig.~(\ref{fig:optimized filter}) along with the transfer functions for regular Gauss filter to illustrate the improvement achieved through the optimization. It can be clearly observed that the optimized Gaussian filter fully attenuates for all the values of $\kappa$ at high wavenumber component. Also, the filter is adding dissipation in a consistent manner that with the increase of the filter ratio, $\kappa$ value, the optimized Gaussian filter is adding more dissipation. We emphasize that the optimized Gaussian filter we presented here can be utilized as a tool to develop more efficient numerical schemes like the proposed localized dynamic model in this paper.

\section{Numerical Test Problems and Results}
\label{sec:4}

The following two numerical test cases have been considered to illustrate the potential of the localized dynamic model proposed in Section~(\ref{localized}) through the comparative analysis between different numerical methods. All the test simulation results are computed in coarse resolutions and compared with the high-resolution WENO-Roe ILES simulation for same test conditions. We denote the high-resolution reference simulation results as HRRS in this paper. Since the physical $Re$ tends to infinity for Euler turbulence, obtaining a reference DNS data is not possible using the available computational resources until now. We, therefore, consider the high-resolution data obtained by the widely used WENO-Roe ILES solver capture most of the physical scales. In this section, we first define the initial conditions of the test cases which is followed by the quantification of the performance of different numerical approaches outlined above in the form of field variables, kinetic energy spectra, and total energy plots.

\subsection{$3$D Taylor-Green Vortex Problem}
\label{TGVs}

\begin{figure}[!ht]
\centering
\mbox{
\subfigure[$t=5$]{\includegraphics[width=0.48\textwidth]{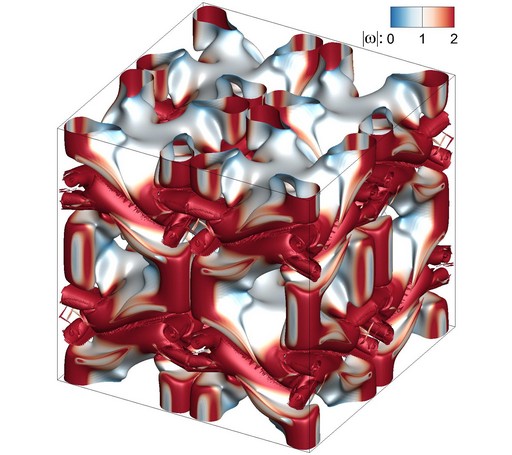}}
\subfigure[$t=10$]{\includegraphics[width=0.48\textwidth]{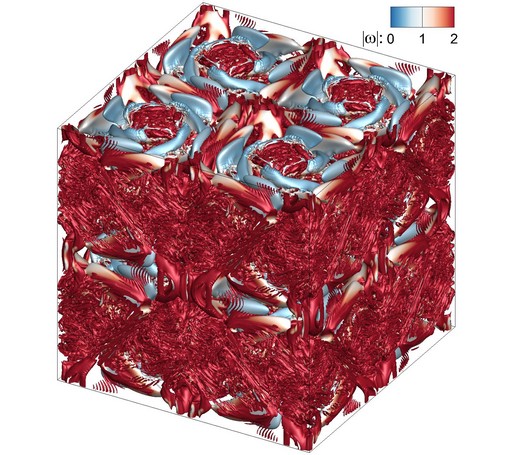}}
}\\
\mbox{
\subfigure[$t=15$]{\includegraphics[width=0.48\textwidth]{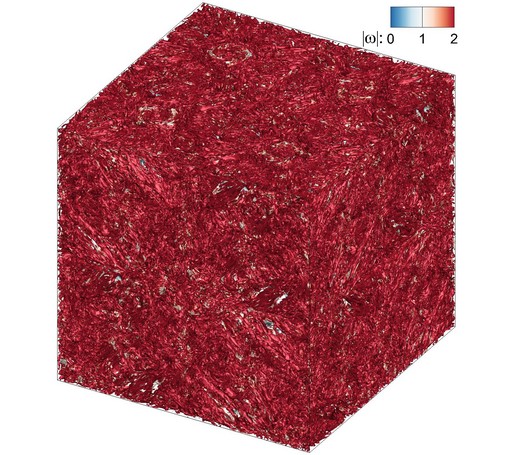}}
\subfigure[$t=20$]{\includegraphics[width=0.48\textwidth]{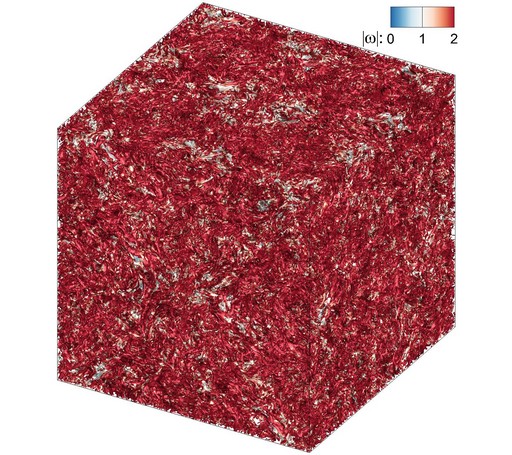}}
}
\caption{Iso-surfaces for zero $Q$ criterion for the classical 3D TGV decaying problem obtained by the WENO-Roe ILES simulation using $512^3$ grid resolution. Evolution of the z-component of the vorticity is shown, and the colors indicate absolute vorticity level.}
\label{fig:tgvwenoroe}
\end{figure}

As a first step in the comparative analysis of the models, a detailed study is made for a simple case of canonical turbulence case given by the TGV in a nearly incompressible setting. TGV is a well-defined flow that has been used as a prototype for the enhancement of vorticity by vortex stretching, and investigating the dynamics of transition to the turbulence through the nonlinear transfer of kinetic energy among eddies from low wave numbers (large-scales) to high wave numbers (small-scales) \citep{ae1937mechanism,shu2005numerical,frisch1995turbulence,brachet1991direct,brachet1992numerical}. The single-mode initial condition for TGV involves smooth triple-periodic boundary condition and consists of a first-degree trigonometric polynomial in all three directions. The uniformly spaced computational domain used in all the numerical tests involving TGV is a cubic box with an edge length of $2\pi$. The flow is driven by the following velocity field conditions:
\begin{equation}\label{TVGIC1}
  \left.\begin{aligned}
     u(x,y,z,t=0) &= \frac{2}{\sqrt{3}}(\phi + \frac{2\pi}{3})\sin(x)\cos(y)\cos(z),  \\
     v(x,y,z,t=0) &= \frac{2}{\sqrt{3}}(\phi - \frac{2\pi}{3})\cos(x)\sin(y)\cos(z),  \\
     w(x,y,z,t=0) &= \frac{2}{\sqrt{3}}\sin(\phi)\cos(x)\cos(y)\sin(z).
\end{aligned}
\right\}
\end{equation}
By setting $\phi = 0$ in Eq.~(\ref{TVGIC1}), the initial flow configuration is found which has 2D streamlines. But for all $t\geq0$, the flow is three-dimensional. The pressure as a solution of the Poisson equation for the above-given velocity can be expressed as:
\begin{align}
\rho(x,y,z,t=0) &= 1.0 , \\
 P(x,y,z,t=0) &= P_0 + \frac{\rho}{16}[( \cos(2x) + \cos(2y) ) (\cos(2z) + 2) -2].
\end{align}
The value for $P_0$ is chosen to the limit of reference Mach number ($M_0$) of $0.08$ in our study (i.e., $P_0 = \rho_0 /( \gamma M_0^2)$ for the dimensionless constant term). The z-component vorticity iso-surface evolution for our reference high fidelity simulation obtained by using the WENO-Roe solver (HRRS) on $512^3$ grid resolution is shown in Fig.~(\ref{fig:tgvwenoroe}). It can be clearly observed that the scheme captures the behavior of the flow field accurately by $t = 10$ since from a qualitative point of view, we can observe a considerable scale separation. It is also seen that as the flow evolves according to the nonlinear Euler equations, the flow starts rotating about the vertical $z$-axis at $t=10$. And eventually, at time $t=20$, the radius of the vortex core tends to zero and forms a flow singularity. This implies the flow has evolved into turbulent by the time $t=10$.
The first criterion we use to evaluate the numerical schemes is the time evolution of the total kinetic energy of the solution field, which can be obtained by spatially averaging the instantaneous kinetic energy at all points,
\begin{align}\label{TKE}
  E(t) = \frac{1}{V} \int \int \int E(x,y,z,t) dx dy dz,
\end{align}
where $V$ is the volume of the physical domain and $E(x,y,z,t)$ is the instantaneous kinetic energy per unit mass at a particular point in the solution field (i.e., $E=1/2(u^2 + v^2 + w^2)$). Considering the cell centered values for the velocity components, Eq.~(\ref{TKE}) can be expressed as following discrete form:
\begin{align}\label{TKE2}
  E(t) = \frac{1}{N_x N_y N_z} \sum_{i=1}^{N_x}\sum_{j=1}^{N_y}\sum_{k=1}^{N_z} \frac{1}{2} \left( u^2_{i,j,k}(t) + v^2_{i,j,k}(t) + w^2_{i,j,k}(t) \right).
\end{align}
Numerical simulations are limited to times to resolve all of the scales with sufficient accuracy since the nonlinear interactions keep generating smaller scales successively. Still, the total kinetic energy can be a very useful tool to measure the numerical dissipation inherent in a scheme by observing the rate of decrease in the total kinetic energy from its initial value $E(0)=1/8$.
The second measure of comparison can be obtained from averaged kinetic energy spectral scaling which may be calculated using the following definition in wave number space \citep{kida1990energy}:
\begin{align}\label{Spec1}
  \hat{E}(\mathbf{k},t) &= \frac{1}{2} \lvert \hat{\mathbf{u}} (\mathbf{k},t)\rvert^2, \\ \nonumber
  &= \frac{1}{2} \left( \lvert \hat{u} (\mathbf{k},t)\rvert^2 + \lvert \hat{v} (\mathbf{k},t)\rvert^2 + \lvert \hat{w} (\mathbf{k},t)\rvert^2 \right)
\end{align}
where the velocity components $\hat{u} (\mathbf{k},t), \hat{v} (\mathbf{k},t)$ and $\hat{w} (\mathbf{k},t)$ can be computed using a fast Fourier transform algorithm \citep{press1992numerical}. An angle averaging of the spectra is then carried out by:
\begin{align}\label{Spec3}
  E(k,t) = \sum_{k-\frac{1}{2} \leq |\textbf{\'{k}}| < k+\frac{1}{2}} \hat{E} (\textbf{\'{k}},t), \\
\end{align}
where
\begin{align}
 k = |\textbf{k}| = \sqrt{k_x^2 + k_y^2 + k_z^2}. \nonumber
\end{align}
The final criterion we consider is the time evolution of vortices by the $Q$ criterion as an a-posteriori analysis. The $Q$ criterion represents the local balance between the shear strain rate and the vorticity magnitude where vortices are defined as areas showing vorticity magnitude dominates the shear strain rate \citep{kajishima2017computational,kolavr2007vortex}. When the flow becomes turbulent, there can be seen a considerable increase in smaller size vortical structures with the time evolution of $Q$ criterion and also, relative uniformity of the distribution can be seen for TGV due to its isotropic property. Fig.~(\ref{fig:tgvwenoroe}) shows the $Q$ criterion for the high fidelity reference simulation where it is apparent that the flow is started to become turbulent at $t=10$ with the increase in small-scales. The $Q$ criterion can be expressed as:
\begin{align}\label{QCrita}
  Q = \frac{1}{4} \left( \lvert \lvert \mathbf{\Omega}\rvert \rvert^2 - \lvert \lvert \mathbf{S}\rvert \rvert^2 \right),
\end{align}
where $\lvert \lvert \mathbf{\Omega}\rvert \rvert = \sqrt{2 \Omega_{ij}\Omega_{ij}}$ and $\lvert \lvert \mathbf{S}\rvert \rvert = \sqrt{2 S_{ij}S_{ij}}$ are the rate-of-rotation and rate-of-strain tensors, respectively. Using the indicial notation, $Q$ criterion can be rewritten as
\begin{align}\label{QCrit}
  Q = -\frac{1}{2}\frac{\partial u_i}{\partial x_j} \frac{\partial u_j}{\partial x_i}.
\end{align}

Fig.~(\ref{fig:TGV, ILES,SG,DS:energy}) shows a comparison of kinetic energy evolution plots between different ILES and eddy viscosity schemes for various grid resolutions at $t = 10$. For ILES cases, it can be easily seen that Rusanov solvers to be more dissipative than the Roe and AUSM solvers, which is expected based on the previous literature \citep{san2015evaluation,maulik2017resolution}. For coarser resolutions, the Roe and AUSM solvers do not capture the physics well. But the result improves with the increase of grid resolutions. All the eddy viscosity approach results show a good agreement with reference and also capture a wide range of scales. The $128$ resolution results provide a good estimation of the high-resolution reference data for all the approaches considered. If we look closely, it can be seen that the localized dynamic model result shows quite a similar result as the reference result at $t =20$ whereas the Smagorinsky and the localized dynamic model with relaxation filtering deviates a little. These can be because of the issues associated with the Smagorinsky model considering the model coefficient constant, and the relaxation filtering adds unwanted dissipation in the model which reduces the performance of the model. On the other hand, Fig.~(\ref{fig:TGV, relaxation, energy}) shows the kinetic energy evolution for $CS+RF$ scheme which displays a dependence of the dissipative characteristic of this approach on the relaxation filtering scale discussed in Section~\ref{CS+RF}. Similar to other schemes, the estimation of the flow physics improves with the increment of grid resolutions in $CS+RF$ scheme too. The $128$ resolution results seem to capture all the scales very well for $CS+RF$ scheme. But the spectra plot for $k_e = 0.99 k_m$ filtering scale in Fig.~(\ref{fig:TGV, ILES,SG,DS:spectram}) shows artificial accumulation of energy even in $128$ resolution simulations which is not desirable. The lower values of filtering scale produce a better result because more numerical dissipation is added. Adding more dissipation in the numerical algorithm resolves the issue with accumulating excess energy pile up, but if the system becomes too dissipative, it can not capture the smaller scales in higher wave numbers. From the spectra figures of $CS+RF$ scheme with lower filtering scale, it can be observed that for $k_e = 0.93 k_m$, the curve gets dissipated earlier than the higher filtering scale results. Then again from the energy spectra plots for ILES and eddy viscosity schemes in Fig.~(\ref{fig:TGV, ILES,SG,DS:spectram}) reveal the ILES schemes are able to capture more scales than $CS+RF$ schemes. But since they are over-dissipative, they do not capture the energy containing eddies. Also, it is again seen that the Rusanov solver is more dissipative than the AUSM and Roe solvers. The eddy viscosity models provide a better approximation than the ILES and $CS+RF$ schemes in a sense that it captures the energy containing eddies because of not being too dissipative like the ILES schemes and also it captures a greater range of scales. It can be observed from the figures that the proposed localized dynamic model provides overall a better estimation with the reference result than the other eddy viscosity models. One can observe the scaling behavior in spectra plots that corresponds to classical Kolmogorov $k^{-5/3}$ scaling law \citep{kolmogorov1991local} which is expected from 3D isotropic homogeneous turbulence.

\begin{figure}[!ht]
\centering
\mbox{
\subfigure[$k_e = 0.93 k_m$]{\includegraphics[width=0.48\textwidth]{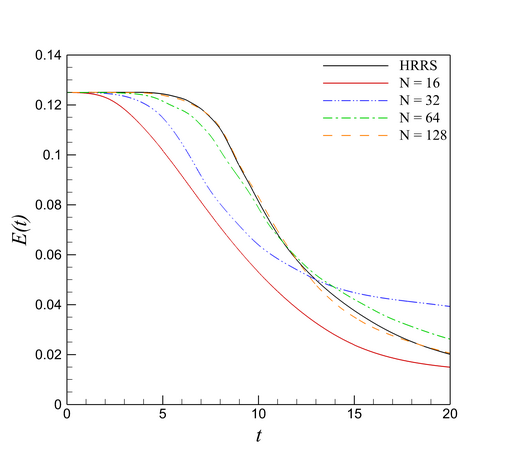}}
\subfigure[$k_e = 0.95 k_m$]{\includegraphics[width=0.48\textwidth]{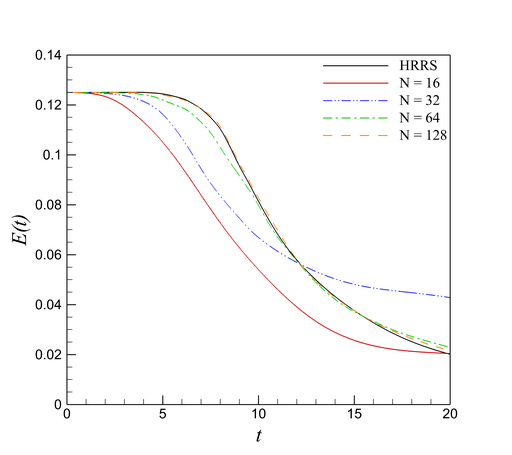}}
}\\
\mbox{
\subfigure[$k_e = 0.97 k_m$]{\includegraphics[width=0.48\textwidth]{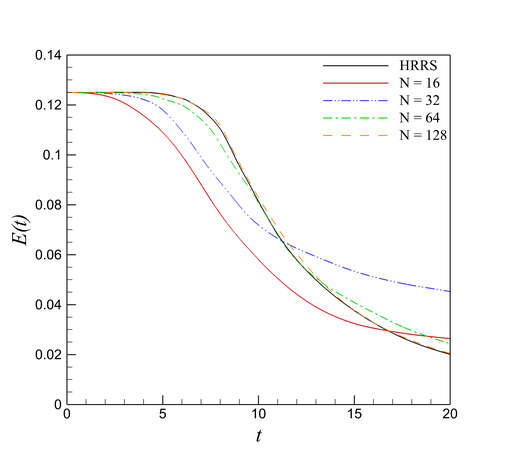}}
\subfigure[$k_e = 0.99 k_m$]{\includegraphics[width=0.48\textwidth]{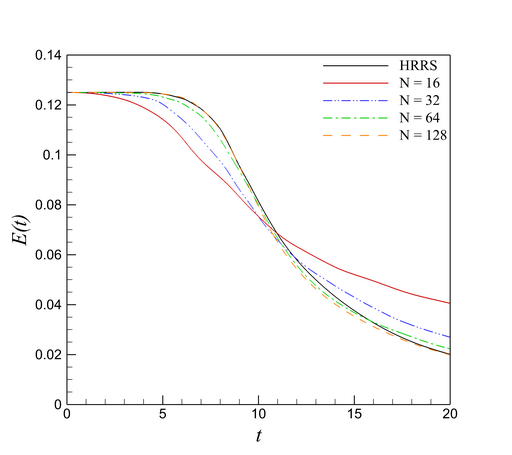}}
}
\caption{Time evolution of total kinetic energy of the 3D TGV problem on different grid resolutions for $CS+RF$ scheme. N: Grid resolution in each direction, HRRS: High Resolution Reference Simulation.}
\label{fig:TGV, relaxation, energy}
\end{figure}
\newpage

\begin{figure}[!ht]
\centering
\mbox{
\subfigure[$k_e = 0.93 k_m$]{\includegraphics[width=0.48\textwidth]{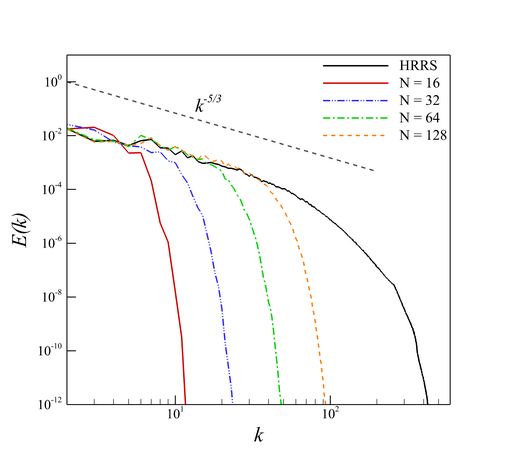}}
\subfigure[$k_e = 0.95 k_m$]{\includegraphics[width=0.48\textwidth]{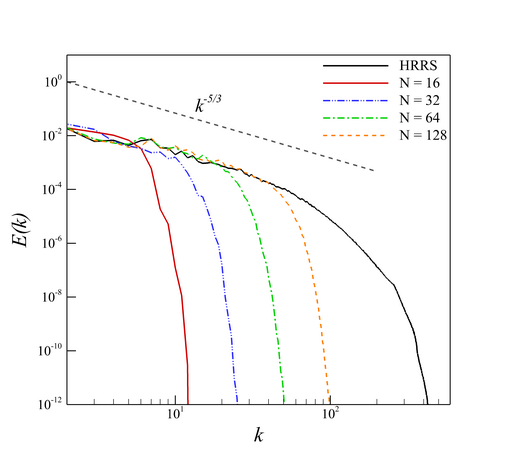}}
}\\
\mbox{
\subfigure[$k_e = 0.97 k_m$]{\includegraphics[width=0.48\textwidth]{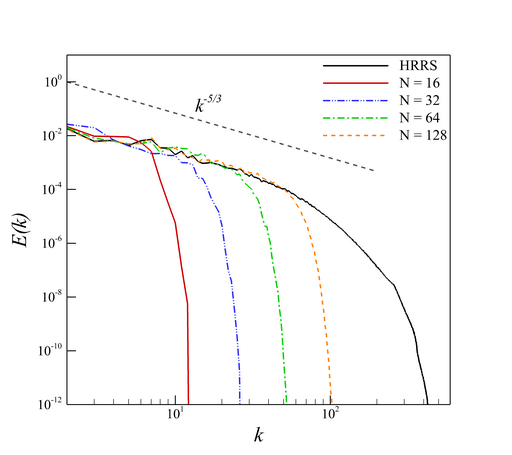}}
\subfigure[$k_e = 0.99 k_m$]{\includegraphics[width=0.48\textwidth]{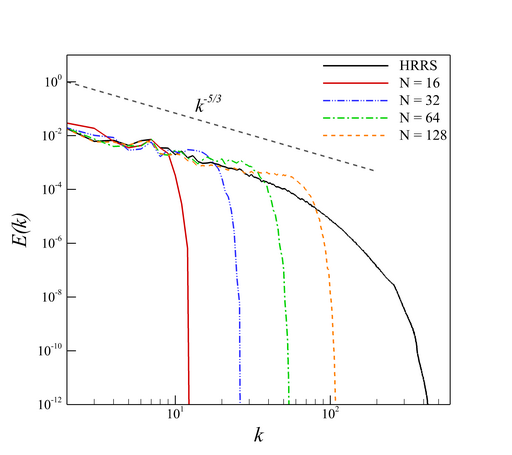}}
}
\caption{Angle-averaged kinetic energy spectra of the 3D TGV problem on different grid resolutions at $t = 10$ for $CS+RF$ scheme. The spherical-averaged energy spectra in the inertial range flattens towards the $k^{-5/3}$ scaling that corresponds to classical Kolmogorov theory. N: Grid resolution in each direction, HRRS: High Resolution Reference Simulation.}
\label{fig:TGV, relaxation, spectra}
\end{figure}

\begin{figure}[!ht]
\centering
\mbox{
\subfigure[ILES-Rusanov]{\includegraphics[width=0.48\textwidth]{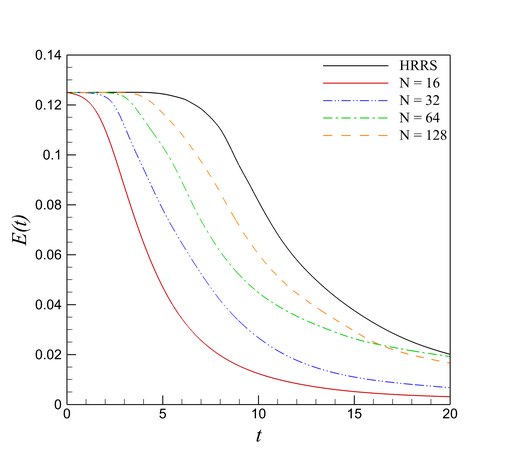}}
\subfigure[ILES-Roe]{\includegraphics[width=0.48\textwidth]{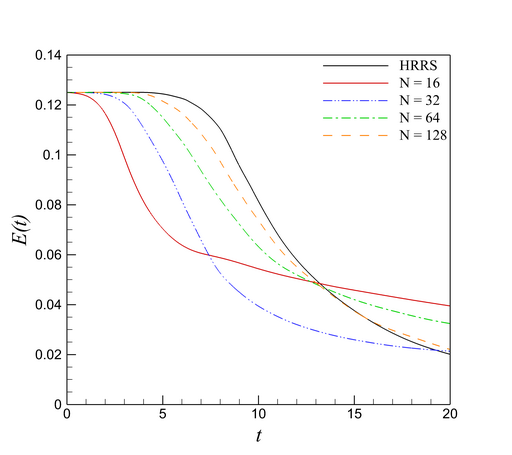}}
}\\
\mbox{
\subfigure[ILES-AUSM]{\includegraphics[width=0.48\textwidth]{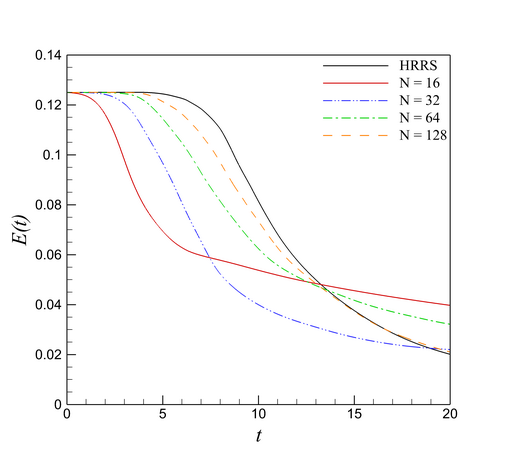}}
\subfigure[Smagorinsky]{\includegraphics[width=0.48\textwidth]{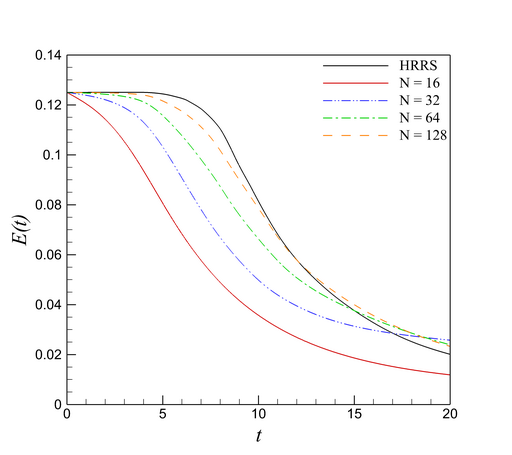}}
}\\
\mbox{
\subfigure[Localized Dynamic Smagorinsky]{\includegraphics[width=0.48\textwidth]{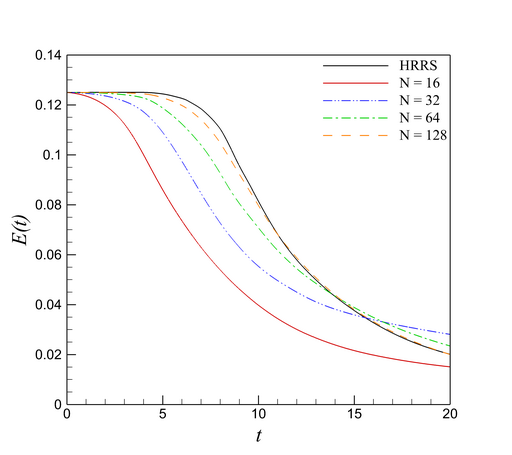}}
\subfigure[Localized Dynamic Smagorinsky with RF]{\includegraphics[width=0.48\textwidth]{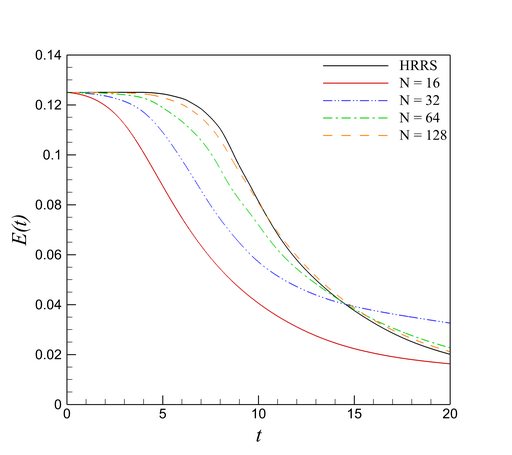}}
}\\
\caption{Time evolution of total kinetic energy of the 3D TGV problem on different grid resolutions for ILES and Eddy viscosity schemes. N: Grid resolution in each direction, HRRS: High Resolution Reference Simulation.}
\label{fig:TGV, ILES,SG,DS:energy}
\end{figure}

\begin{figure}[!ht]
\centering
\mbox{
\subfigure[ILES-Rusanov]{\includegraphics[width=0.48\textwidth]{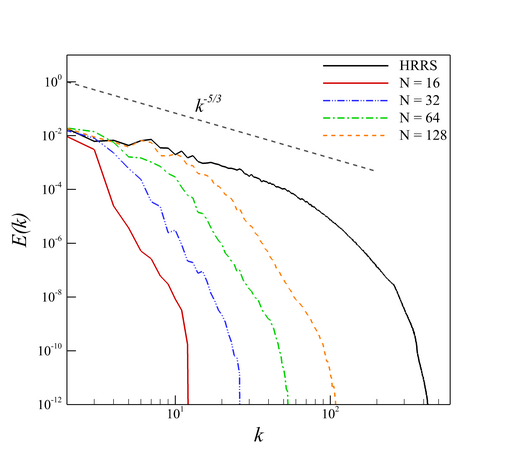}}
\subfigure[ILES-Roe]{\includegraphics[width=0.48\textwidth]{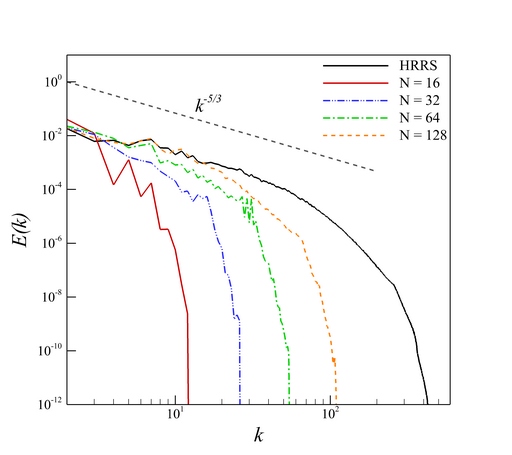}}
}\\
\mbox{
\subfigure[ILES-AUSM]{\includegraphics[width=0.48\textwidth]{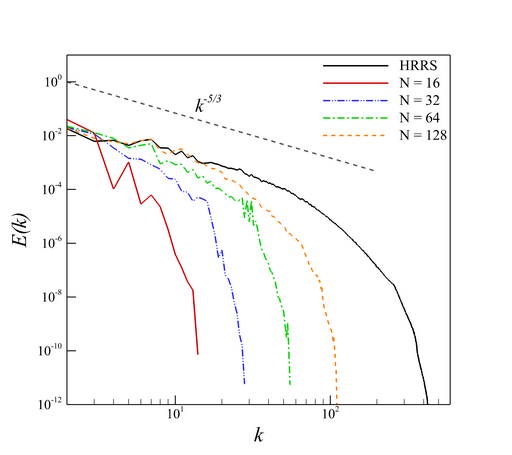}}
\subfigure[Smagorinsky]{\includegraphics[width=0.48\textwidth]{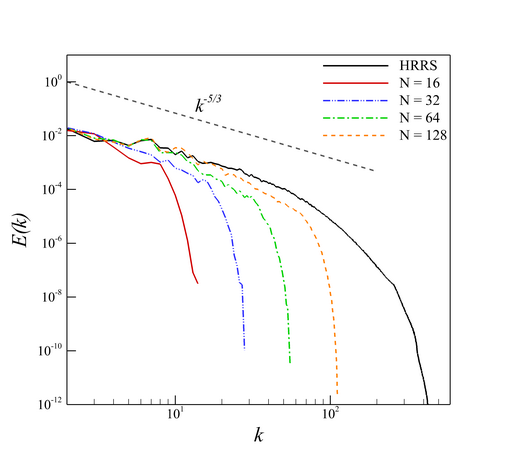}}
}\\
\mbox{
\subfigure[Localized Dynamic Smagorinsky]{\includegraphics[width=0.48\textwidth]{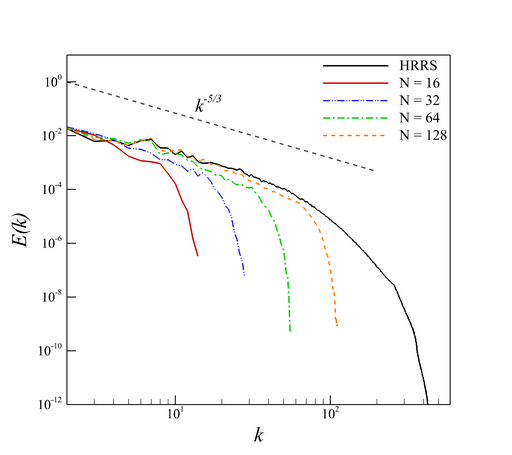}}
\subfigure[Localized Dynamic Smagorinsky with RF]{\includegraphics[width=0.48\textwidth]{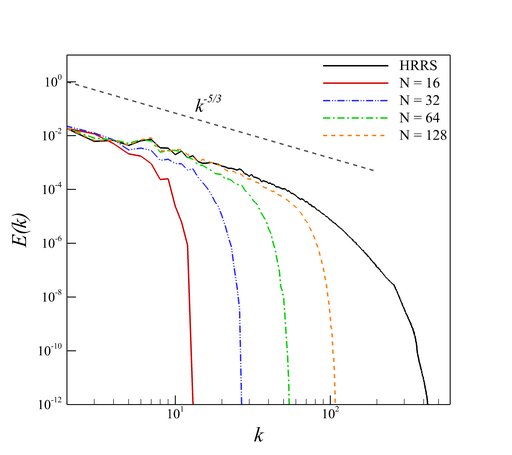}}
}\\
\caption{Angle-averaged kinetic energy spectra of the 3D TGV problem on different grid resolutions at $t = 10$ for ILES and Eddy viscosity schemes. The spherical-averaged energy spectra in the inertial range flatten towards the $k^{-5/3}$ scaling that corresponds to classical Kolmogorov theory. N: Grid resolution in each direction, HRRS: High Resolution Reference Simulation.}
\label{fig:TGV, ILES,SG,DS:spectram}
\end{figure}

Fig.~(\ref{fig:TGV, Q}) shows the $Q$ criterion for TGV flows at time $t = 10$ by applying different numerical schemes on it as an a-posteriori analysis. We choose to represent the spectra plots at $t=10$ since there is more physics to visualize at that time. We consider ILES-Roe scheme as a representative of ILES schemes since it performs better than most of the other ILES schemes. It again shows the ILES schemes are not capturing the energy containing eddies since its too dissipative which validates the result from the statistical analysis. It can also be observed from the $CS+RF$ scheme plots that there are not much scales present for $k_e = 0.93 k_m$ and there are too many smaller scales present for $k_e = 0.99 k_m$. If we compare the $Q$ criterion plot of $k_e = 0.99 k_m$ with reference high-resolution figure at time $t = 10$, it can be clearly observed the scheme is showing excess energy containing eddies. These results from $CS+RF$ scheme also support the results from the spectra and energy plots that the $CS+RF$ scheme with filtering scale $k_e = 0.99 k_m$ provides an over-estimation of the true flow physics. Among the eddy viscosity models, the localized dynamic model is showing more scales similar to the high-resolution simulation result. All of these results lead to the conclusion that overall the localized dynamic model provides a better estimation of the flow physics than all the other schemes of interest in this study. If we observe the comparison spectra plots on $128^3$ resolutions in Fig.~(\ref{tgv comparison, 128 resolution}), it supports all the statements we put forth above that the ILES schemes are too dissipative, the $CS+RF$ scheme with lower filtering scale does not capture most of the smaller scales whereas higher filtering scale results are accumulating unrealistic energy and the localized dynamic model is representing the true physics better than the other eddy viscosity and numerical models.

\begin{figure}[!ht]
\centering
\mbox{
\subfigure[ILES-Roe]{\includegraphics[width=0.48\textwidth]{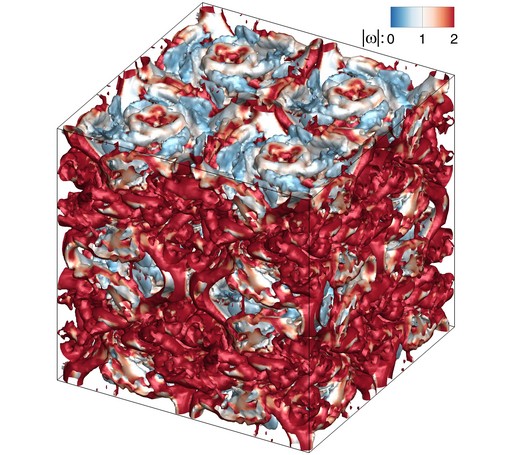}}
\subfigure[Smagorinsky]{\includegraphics[width=0.48\textwidth]{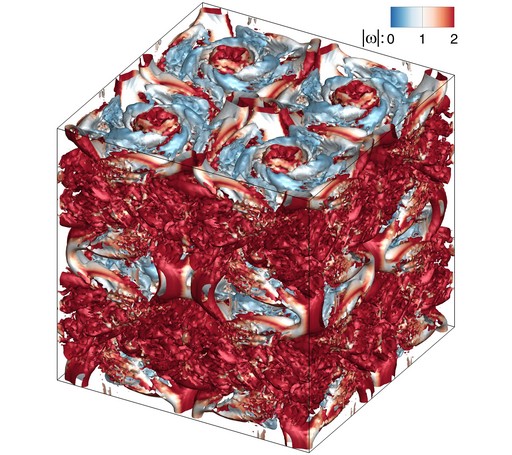}}
}\\
\mbox{
\subfigure[Localized Dynamic Smagorinsky]{\includegraphics[width=0.48\textwidth]{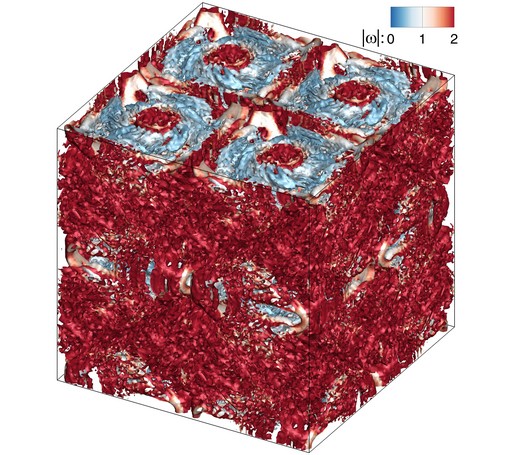}}
\subfigure[Localized Dynamic Smagorinsky with RF]{\includegraphics[width=0.48\textwidth]{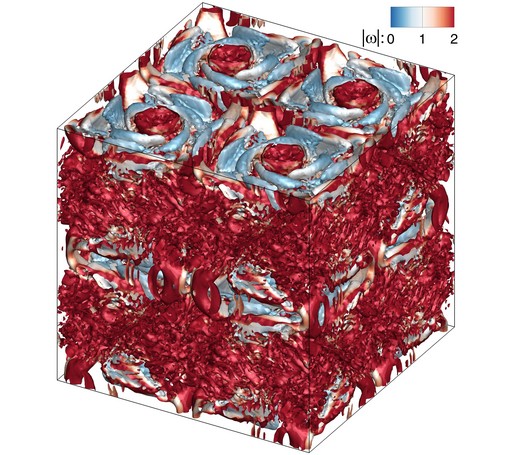}}
}\\
\mbox{
\subfigure[$CS+RF~(k_e = 0.93 k_m)$]{\includegraphics[width=0.48\textwidth]{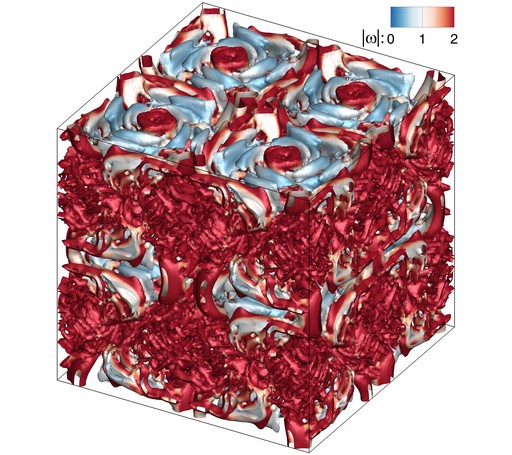}}
\subfigure[$CS+RF~(k_e = 0.99 k_m)$]{\includegraphics[width=0.48\textwidth]{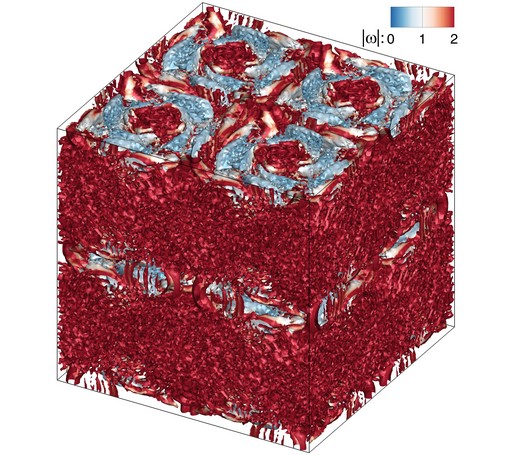}}
}\\
\caption{Iso-surfaces for zero $Q$ criterion for the 3D TGV decaying problem on the $128^3$ grid resolution at $t = 10$ from different schemes. Evolution of the z-component of the vorticity is shown and the colors indicate absolute vorticity level.}
\label{fig:TGV, Q}
\end{figure}

\begin{figure}[!ht]
\centering
\mbox{
\includegraphics[width=0.48\textwidth]{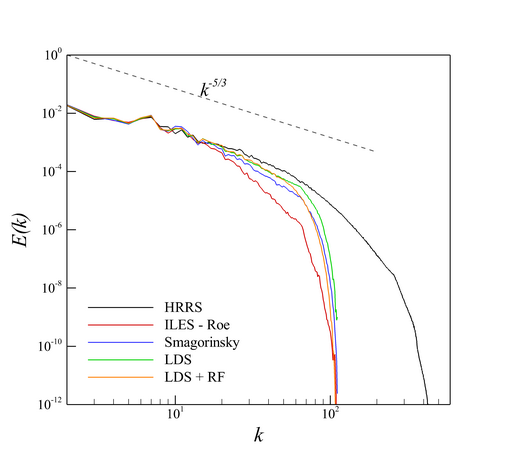}
\includegraphics[width=0.48\textwidth]{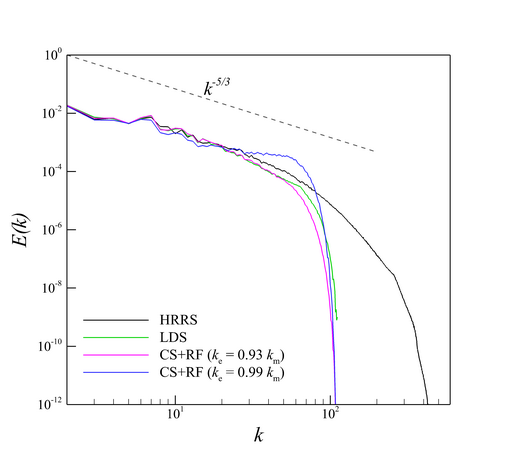}
}
\caption{Comparison of the angle-averaged kinetic energy spectra for the 3D TGV decaying problem on the $128^3$ grid resolution at $t = 10$ computed by different schemes. Left: Comparison of ILES and eddy viscosity model with the LDS model. Right: Comparison of $CS+RF$ models with \textcolor{blue}{the} LDS model. HRRS: High Resolution Reference Simulation, LDS: Localized Dynamic Smagorinsky and LDS+RF: Localized Dynamic Smagorinsky with Relaxation Filtering.}
\label{tgv comparison, 128 resolution}
\end{figure}

\subsection{Shear Layer Turbulence: Kelvin-Helmholtz Instability (KHI)}

\begin{figure}[!ht]
\centering
\mbox{
\subfigure[$t=1$]{\includegraphics[width=0.48\textwidth]{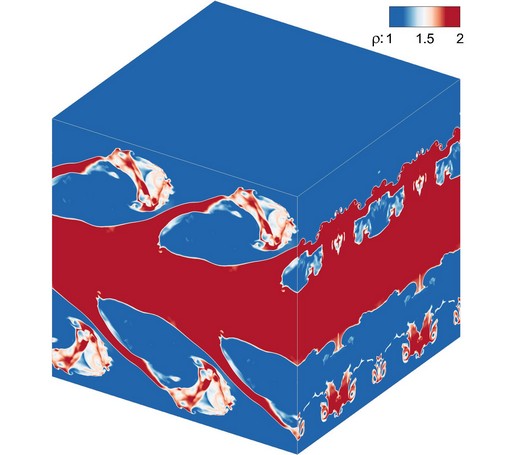}}
\subfigure[$t=2$]{\includegraphics[width=0.48\textwidth]{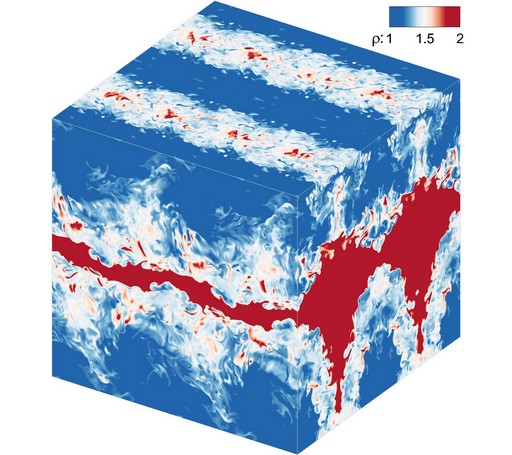}}
}\\
\mbox{
\subfigure[$t=4$]{\includegraphics[width=0.48\textwidth]{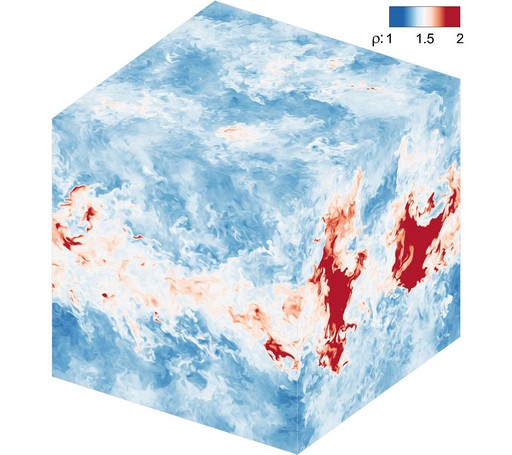}}
\subfigure[$t=5$]{\includegraphics[width=0.48\textwidth]{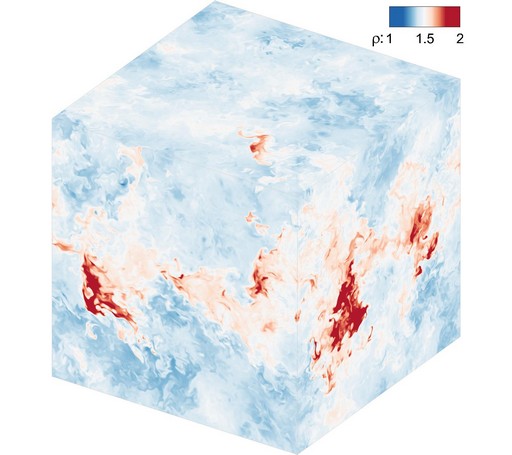}}
}
\caption{Time evolution of density field for the 3D KHI problem with $u_{KHI} = 1.0$ obtained by the WENO-Roe ILES simulation using $512^3$ grid resolution. Colors indicate absolute density level.}
\label{fig:khidns}
\end{figure}

Since highly compressible flows are governed by Eulerian hyperbolic conservation laws, we consider the classical KHI problem to assess the performance of aforementioned schemes for compressible turbulent flows. The KHI phenomenon is ubiquitous in nature and can be observed through both the experimental observations and numerical simulations. KHI can be referred to as a two-layer system in which a lower density layer floats over a heavier layer, and the layers flow at different velocities. It can occur in a continuous field where velocity shear is present or at the interface between two fluids in different densities or velocities \citep{thomson1871xlvi,hwang2012first}. The instability is triggered through the growing discontinuous wave (sharp density or discontinuous velocity) at the interface which induces vorticity and eventually transitions into a turbulent field through nonlinear interactions or mixing. The evolution of this instability can be compared across different numerical schemes through the simulation results to give a qualitative and quantitative measurement on the performance of that model \citep{maulik2017resolution,san2015evaluation}. In our study, we use the averaged kinetic energy spectra and density contour to assess the models statistically and visually. We use a triple periodic square domain with the size of $L$ in each side for our three-dimensional simulations. The $ \mathbb{R}^3$ simulation domain for the test cases is set $(x,y,z) \in [-0.5,0.5]\times[-0.5,0.5]\times[-0.5,0.5]$. The initial conditions are specified as follows:
\begin{align}\label{KHI}
 \rho(x,y,z) &=
\begin{cases}
 1.0,\ \  \textnormal{if} \ \ |y| \geq 0.25 \\
 2.0,\ \  \textnormal{if} \ \ |y| < 0.25  \\
\end{cases} \\
P(x,y,z) &= 2.5, \\
u(x,y,z) &=
\begin{cases}
  u_{KHI},\ \  \textnormal{if} \ \ |y| \geq 0.25 \\
 -u_{KHI},\ \  \textnormal{if}  \ \ |y| < 0.25  \\
\end{cases} \\
v(x,y,z) &= \lambda \sin(2\pi nx/L), \\
w(x,y,z) &= \lambda \sin(2\pi nz/L).
\end{align}
Here, the vertical component of the velocity is perturbed using a single-mode sine wave $(n =2, \ L=1)$ and the amplitude of the perturbation is set at $\lambda = 0.01$. Here, $u_{KHI}$ is the shearing velocity magnitude which is used to increase or decrease the compressibility effect in the flow. For the lower value of $u_{KHI}$, the flow is less compressible due to less shear velocity magnitude at the interface. Fig.~(\ref{fig:khidns}) shows the evolution of the density field of high-resolution $(512^3)$ WENO-Roe ILES scheme for a shearing velocity magnitude of $1.0$ i.e., $u_{KHI} = 1.0$. The transition to turbulence can be clearly noticed once the initial instability develops. At $t = 5$, almost full mixing is observed for this case. We consider the results obtained from this simulation as reference data to quantify the coarser resolution simulations of other numerical schemes. The expression to calculate the average kinetic energy spectrum is formulated similarly as the TGV case (Section~(\ref{TGVs})).

Fig.~(\ref{fig:KHI, u = $0.1$, ILES,SG,DS:spectrum}) to Fig.~(\ref{fig:KHI, u = $1.0$, ILES,SG,DS:spectrum}) show the angle-averaged kinetic energy spectra for ILES and eddy viscosity models at different resolutions for various shearing velocity magnitudes. We consider presenting the results at $t = 5$ where the flow becomes fully developed turbulent, and a larger amount of mixing happens. We include the expected $k^{-5/3}$ scaling given by Kolmogorov in the spectra plots. All the spectra plots obey the Kolmogorov's $5/3^{rd}$ scaling since a flat inertial range is observed in each plot. In Fig.~(\ref{fig:KHI, u = $0.1$, ILES,SG,DS:spectrum}), it can be seen that the Rusanov solver is more dissipative than the Roe and AUSM schemes which is expected. Also, the inertial subrange is not developed for Rusanov solver. All the ILES schemes are more dissipative than the eddy viscosity schemes. Another point should be noticed for $u_{KHI} = 0.1$ plots that means in lower compressibility simulations, both the Smagorinsky and localized dynamic model show a narrow spike in the spectrum at high wave number which does not appear later for higher shear velocity simulations. Also, the noises smoothen up for the localized dynamic model with relaxation filter. In Fig.~(\ref{fig:KHI, u = $0.25$, ILES,SG,DS:spectra}), we see the improvement in the eddy viscosity model results from the $u_{KHI} = 0.1$ case while the ILES schemes still remain dissipative. For Fig.~(\ref{fig:KHI, u = $0.5$, ILES,SG,DS:spectrum}), more range of scales is captured by the schemes where both the localized dynamic and the localized dynamic model with relaxation filtering show a good estimation of the reference result. For $u_{KHI} = 1.0$ in Fig.~(\ref{fig:KHI, u = $1.0$, ILES,SG,DS:spectrum}), the localized dynamic model as well as other eddy viscosity models capture a wide inertial subrange. It can also be observed that the coarser $16$ resolution results do not behave well for almost any of the test runs. However, the estimations improve with the increase of grid resolutions for all the schemes, and the $128$ resolution runs give the most stable and accurate results for both the ILES and eddy viscosity schemes. For all the values of $u_{KHI}$, ILES schemes produce over-dissipative results and again, capture less amount of scales than the eddy viscosity schemes. Based on the figures, it can be concluded that the eddy viscosity models produce better approximation statistically with the reference data than ILES schemes.

From Fig.~(\ref{fig:KHI, relaxation, spectra, u = $0.1$}) to Fig.~(\ref{fig:KHI, relaxation, spectra, u = $1.0$}), it can be observed that for higher relaxation filtering scale values, results are less dissipative but also less stable. Overall, the $CS+RF$ schemes provide a good agreement with the reference simulation data and capture the inertial range very well with the increase of compressibility. But it is also observed that adding less amount of dissipation leads to the nonphysical Gibbs oscillations. For this reason, most of the simulations for $k_e = 0.99 k_m$ show nonphysical behavior, and do not converge. In Fig.~(\ref{fig:KHI, relaxation, spectra, u = $0.25$}) and Fig.~(\ref{fig:KHI, relaxation, spectra, u = $0.5$}), the results for $CS+RF$ scheme with $k_e = 0.99 k_m$ filtering scale show an excess energy accumulation over the reference spectrum which leads to a set of full non-converging results in Fig.~(\ref{fig:KHI, relaxation, spectra, u = $1.0$}) for $u_{KHI} = 1.0$. Also, we find the results improve with the increase of grid resolutions. $CS+RF$ scheme with $k_e = 0.93 k_m$ shows the better approximation for all the values of $u_{KHI}$ than the other $CS+RF$ simulations with different relaxation filtering scales.

\begin{figure}[!ht]
\centering
\mbox{
\subfigure[ILES-Rusanov]{\includegraphics[width=0.48\textwidth]{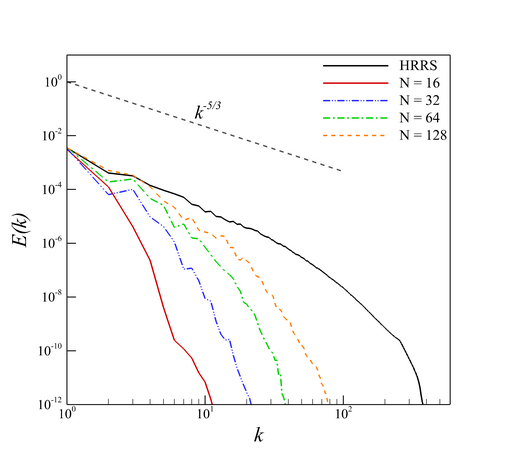}}
\subfigure[ILES-Roe]{\includegraphics[width=0.48\textwidth]{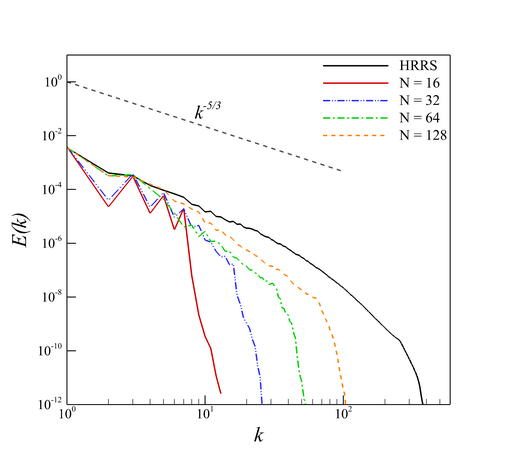}}
}\\
\mbox{
\subfigure[ILES-AUSM]{\includegraphics[width=0.48\textwidth]{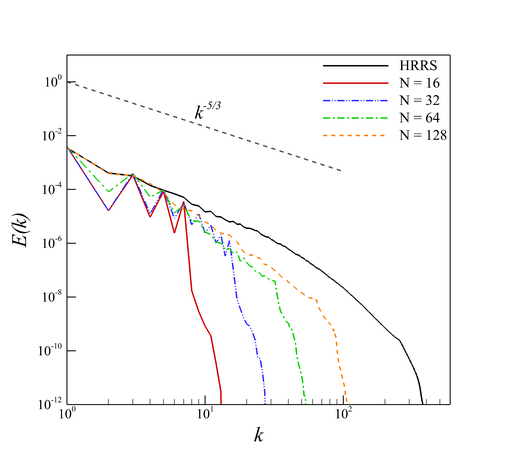}}
\subfigure[Smagorinsky]{\includegraphics[width=0.48\textwidth]{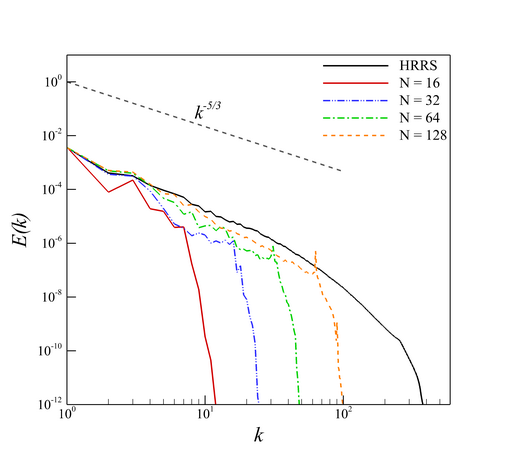}}
}\\
\mbox{
\subfigure[Localized Dynamic Smagorinsky]{\includegraphics[width=0.48\textwidth]{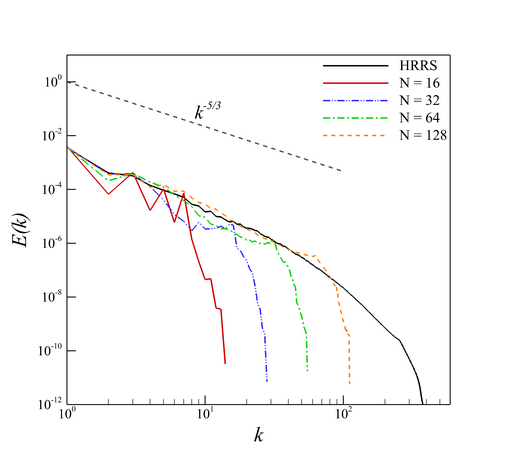}}
\subfigure[Localized Dynamic Smagorinsky with RF]{\includegraphics[width=0.48\textwidth]{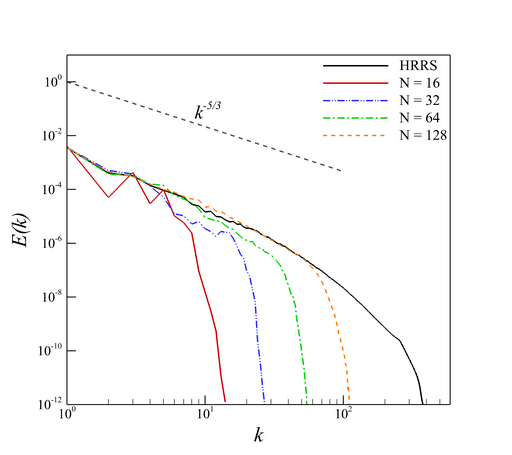}}
}\\
\caption{Angle-averaged kinetic energy spectra of the 3D KHI problem with $u_{KHI} = 0.1$ on different grid resolutions at $t = 5$ for ILES and Eddy viscosity schemes. The spherical-averaged energy spectra in the inertial range flatten towards the $k^{-5/3}$ scaling that corresponds to classical Kolmogorov theory. N: Grid resolution in each direction, HRRS: High Resolution Reference Simulation.}
\label{fig:KHI, u = $0.1$, ILES,SG,DS:spectrum}
\end{figure}

\begin{figure}[!ht]
\centering
\mbox{
\subfigure[ILES-Rusanov]{\includegraphics[width=0.48\textwidth]{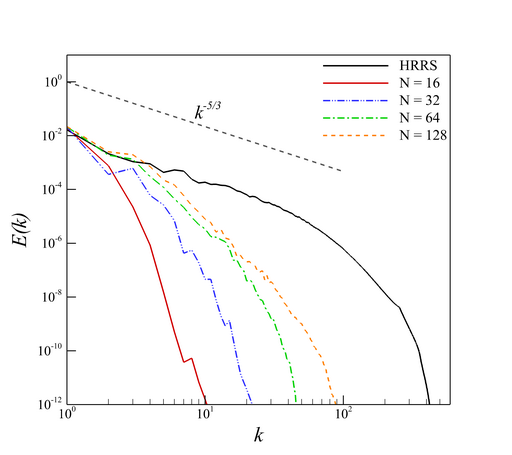}}
\subfigure[ILES-Roe]{\includegraphics[width=0.48\textwidth]{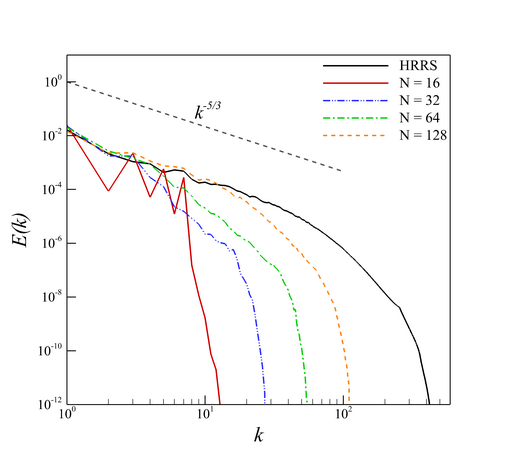}}
}\\
\mbox{
\subfigure[ILES-AUSM]{\includegraphics[width=0.48\textwidth]{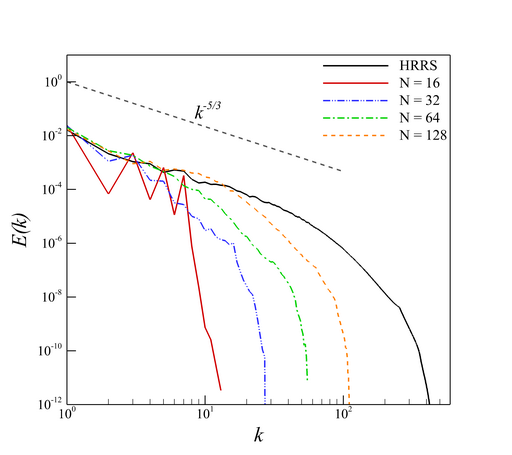}}
\subfigure[Smagorinsky]{\includegraphics[width=0.48\textwidth]{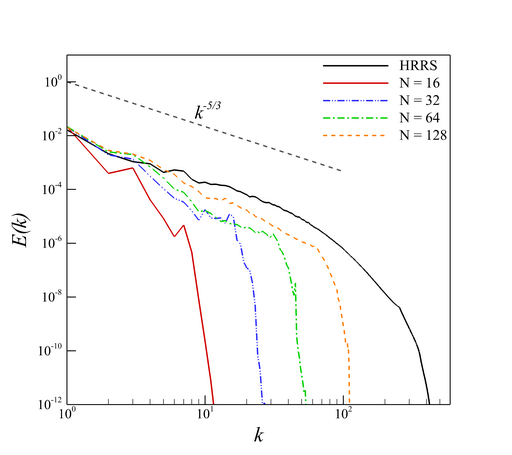}}
}\\
\mbox{
\subfigure[Localized Dynamic Smagorinsky]{\includegraphics[width=0.48\textwidth]{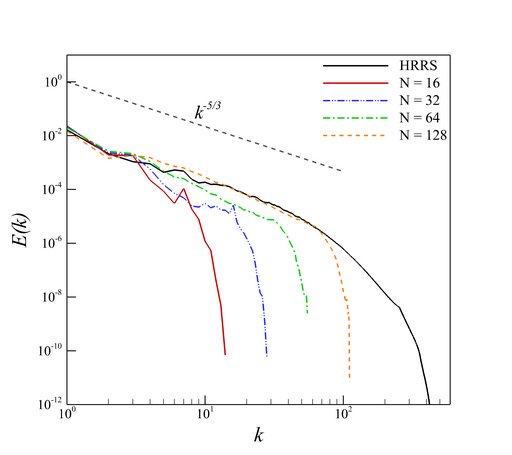}}
\subfigure[Localized Dynamic Smagorinsky with RF]{\includegraphics[width=0.48\textwidth]{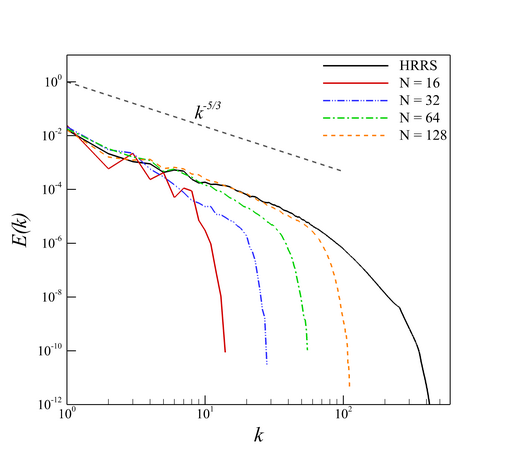}}
}\\
\caption{Angle-averaged kinetic energy spectra of the 3D KHI problem with $u_{KHI} = 0.25$ on different grid resolutions at $t = 5$ for ILES and Eddy viscosity schemes. The spherical-averaged energy spectra in the inertial range flatten towards the $k^{-5/3}$ scaling that corresponds to classical Kolmogorov theory. N: Grid resolution in each direction, HRRS: High Resolution Reference Simulation.}
\label{fig:KHI, u = $0.25$, ILES,SG,DS:spectra}
\end{figure}

\begin{figure}[!ht]
\centering
\mbox{
\subfigure[ILES-Rusanov]{\includegraphics[width=0.48\textwidth]{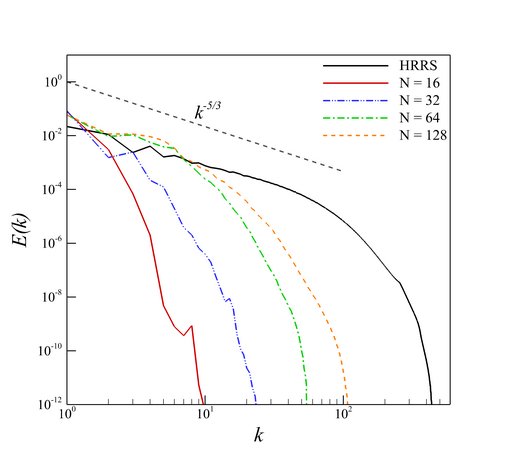}}
\subfigure[ILES-Roe]{\includegraphics[width=0.48\textwidth]{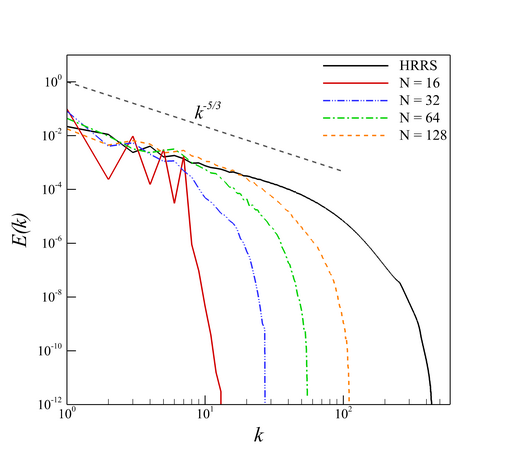}}
}\\
\mbox{
\subfigure[ILES-AUSM]{\includegraphics[width=0.48\textwidth]{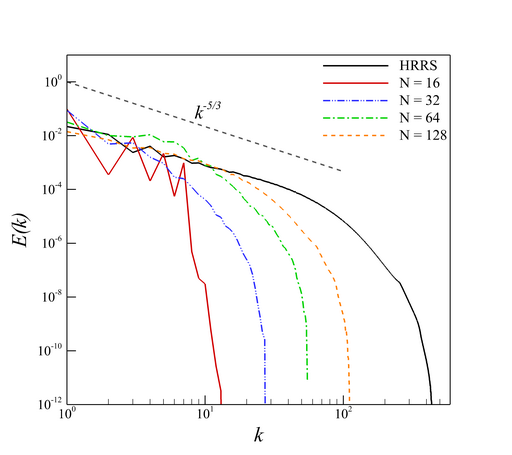}}
\subfigure[Smagorinsky]{\includegraphics[width=0.48\textwidth]{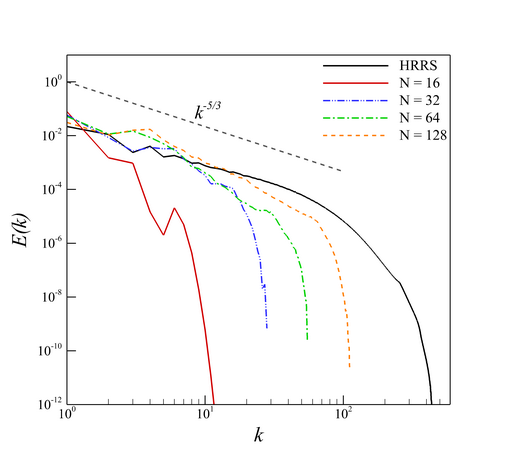}}
}\\
\mbox{
\subfigure[Localized Dynamic Smagorinsky]{\includegraphics[width=0.48\textwidth]{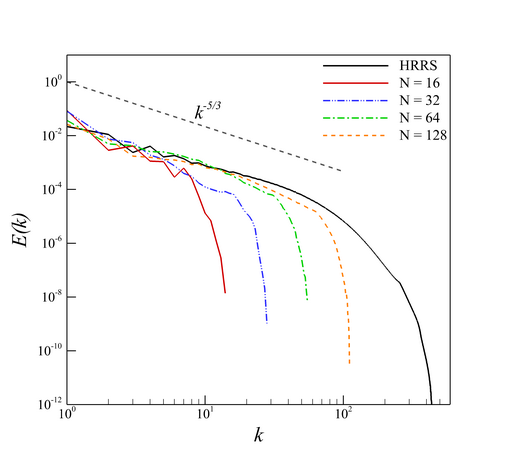}}
\subfigure[Localized Dynamic Smagorinsky with RF]{\includegraphics[width=0.48\textwidth]{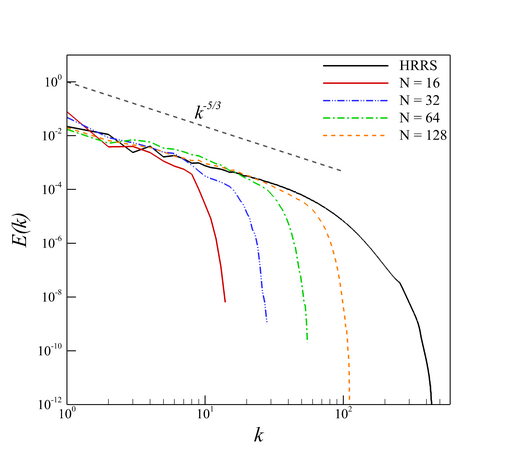}}
}\\
\caption{Angle-averaged kinetic energy spectra of the 3D KHI problem with $u_{KHI} = 0.5$ on different grid resolutions at $t = 5$ for ILES and Eddy viscosity schemes. The spherical-averaged energy spectra in the inertial range flatten towards the $k^{-5/3}$ scaling that corresponds to classical Kolmogorov theory. N: Grid resolution in each direction, HRRS: High Resolution Reference Simulation.}
\label{fig:KHI, u = $0.5$, ILES,SG,DS:spectrum}
\end{figure}

\begin{figure}[!ht]
\centering
\mbox{
\subfigure[ILES-Rusanov]{\includegraphics[width=0.48\textwidth]{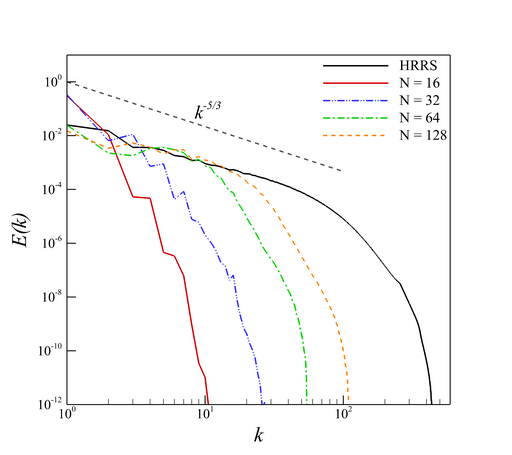}}
\subfigure[ILES-Roe]{\includegraphics[width=0.48\textwidth]{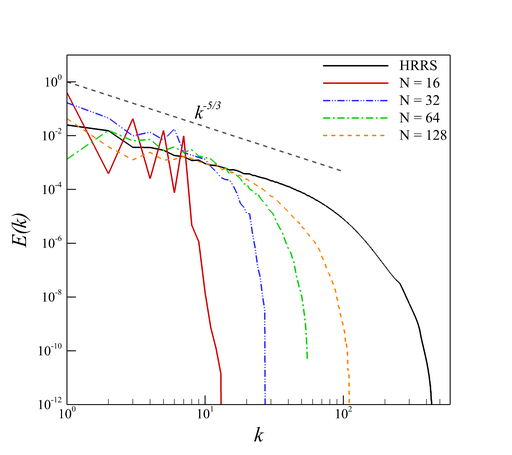}}
}\\
\mbox{
\subfigure[ILES-AUSM]{\includegraphics[width=0.48\textwidth]{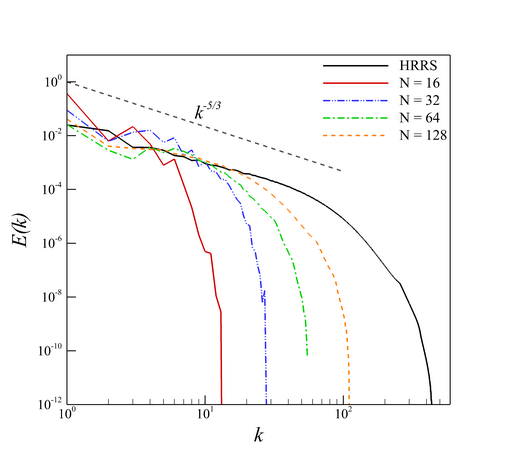}}
\subfigure[Smagorinsky]{\includegraphics[width=0.48\textwidth]{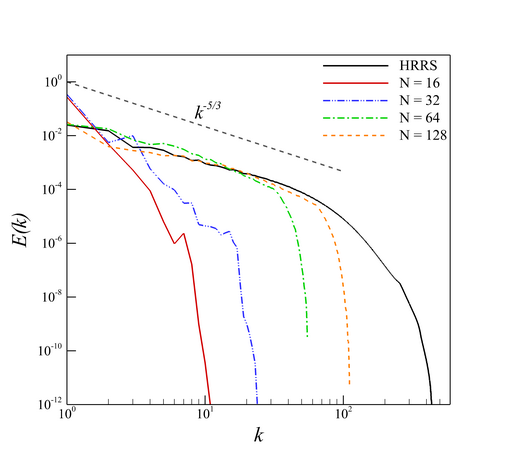}}
}\\
\mbox{
\subfigure[Localized Dynamic Smagorinsky]{\includegraphics[width=0.48\textwidth]{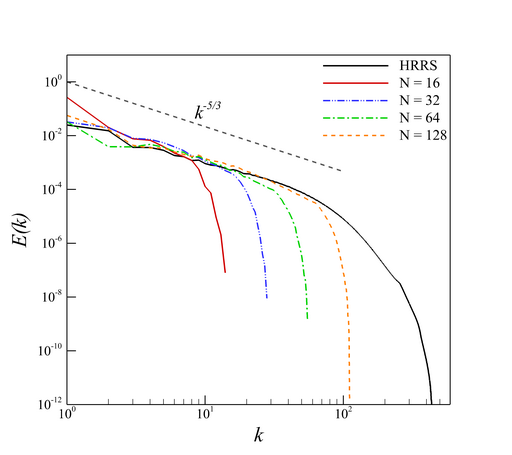}}
\subfigure[Localized Dynamic Smagorinsky with RF]{\includegraphics[width=0.48\textwidth]{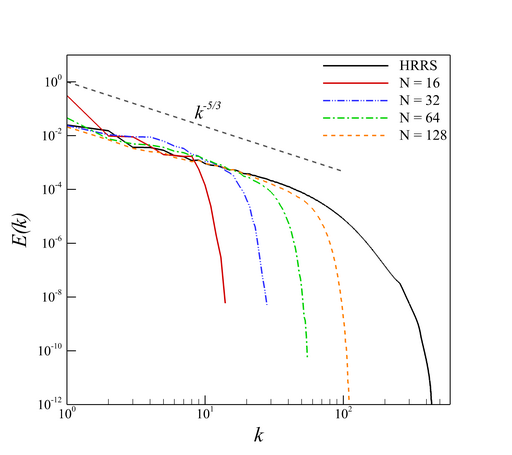}}
}\\
\caption{Angle-averaged kinetic energy spectra of the 3D KHI problem with $u_{KHI} = 1.0$ on different grid resolutions at $t = 5$ for ILES and Eddy viscosity schemes. The spherical-averaged energy spectra in the inertial range flatten towards the $k^{-5/3}$ scaling that corresponds to classical Kolmogorov theory. N: Grid resolution in each direction, HRRS: High Resolution Reference Simulation.}
\label{fig:KHI, u = $1.0$, ILES,SG,DS:spectrum}
\end{figure}

\begin{figure}[!ht]
\centering
\mbox{
\subfigure[$k_e = 0.93 k_m$]{\includegraphics[width=0.48\textwidth]{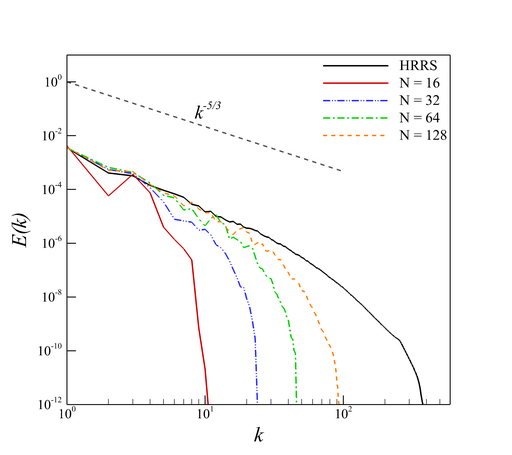}}
\subfigure[$k_e = 0.95 k_m$]{\includegraphics[width=0.48\textwidth]{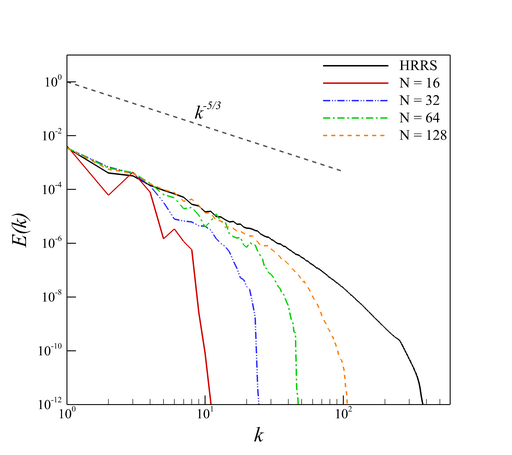}}
}\\
\mbox{
\subfigure[$k_e = 0.97 k_m$]{\includegraphics[width=0.48\textwidth]{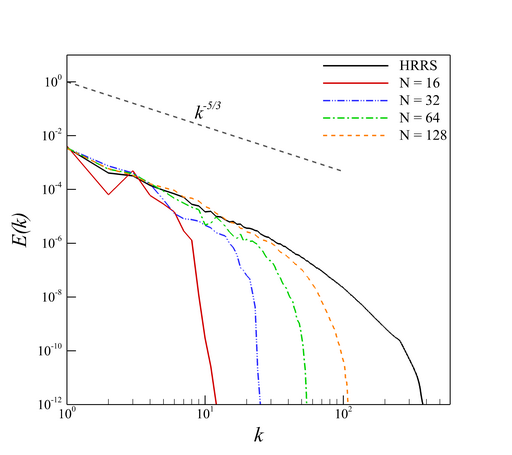}}
\subfigure[$k_e = 0.99 k_m$]{\includegraphics[width=0.48\textwidth]{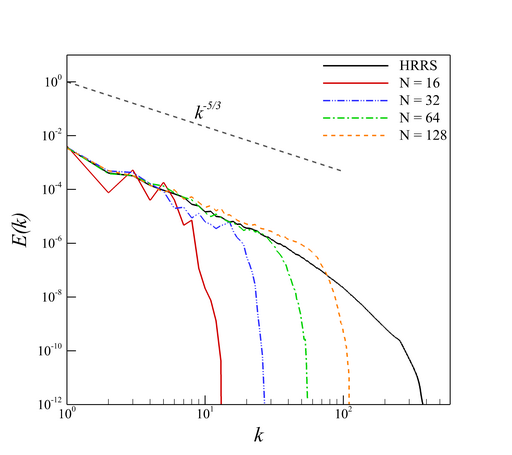}}
}
\caption{Angle-averaged kinetic energy spectra of the 3D KHI problem with $u_{KHI} = 0.1$ on different grid resolutions at $t = 5$ for $CS+RF$ scheme. The spherical-averaged energy spectra in the inertial range flatten towards the $k^{-5/3}$ scaling that corresponds to classical Kolmogorov theory. N: Grid resolution in each direction, HRRS: High Resolution Reference Simulation.}
\label{fig:KHI, relaxation, spectra, u = $0.1$}
\end{figure}

\begin{figure}[!ht]
\centering
\mbox{
\subfigure[$k_e = 0.93 k_m$]{\includegraphics[width=0.48\textwidth]{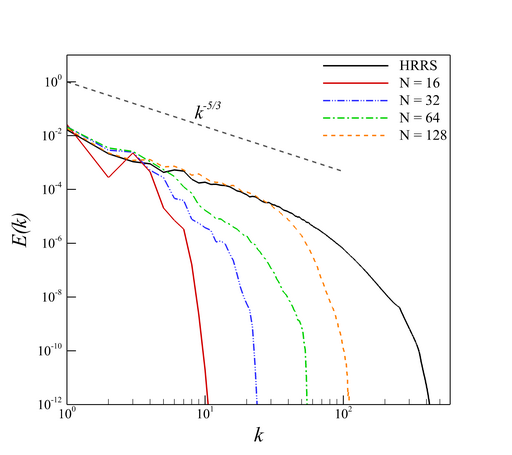}}
\subfigure[$k_e = 0.95 k_m$]{\includegraphics[width=0.48\textwidth]{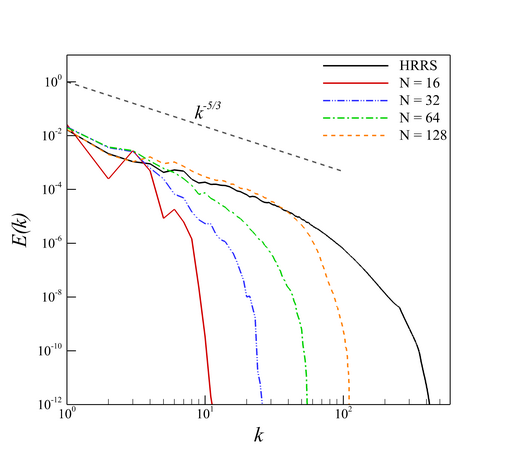}}
}\\
\mbox{
\subfigure[$k_e = 0.97 k_m$]{\includegraphics[width=0.48\textwidth]{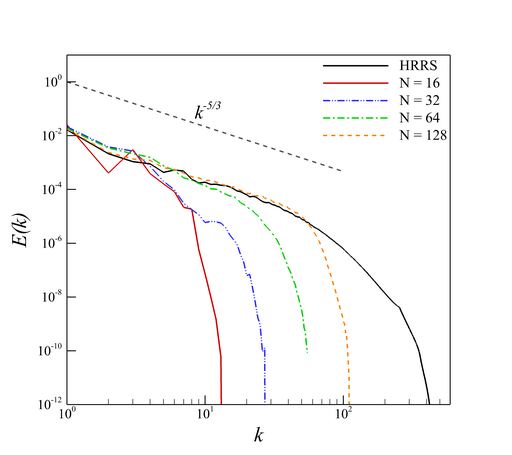}}
\subfigure[$k_e = 0.99 k_m$]{\includegraphics[width=0.48\textwidth]{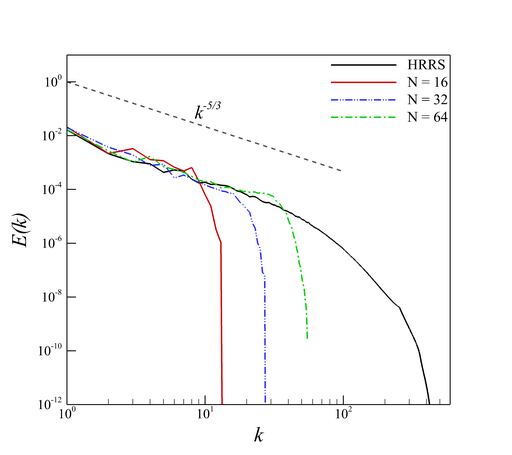}}
}
\caption{Angle-averaged kinetic energy spectra of the 3D KHI problem with $u_{KHI} = 0.25$ on different grid resolutions at $t = 5$ for $CS+RF$ scheme. The spherical-averaged energy spectra in the inertial range flatten towards the $k^{-5/3}$ scaling that corresponds to classical Kolmogorov theory. N: Grid resolution in each direction, HRRS: High Resolution Reference Simulation.}
\label{fig:KHI, relaxation, spectra, u = $0.25$}
\end{figure}

\begin{figure}[!ht]
\centering
\mbox{
\subfigure[$k_e = 0.93 k_m$]{\includegraphics[width=0.48\textwidth]{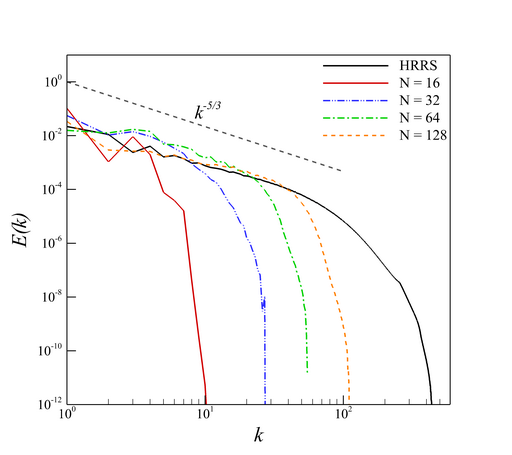}}
\subfigure[$k_e = 0.95 k_m$]{\includegraphics[width=0.48\textwidth]{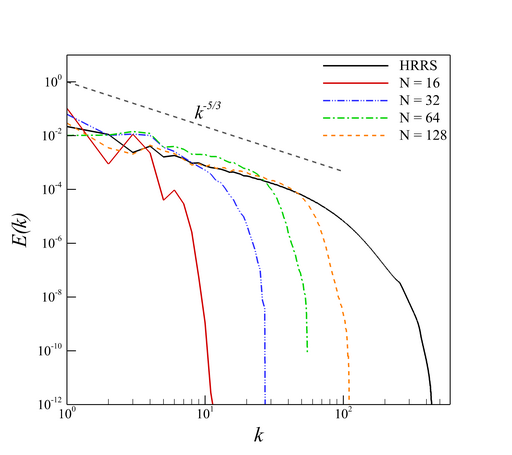}}
}\\
\mbox{
\subfigure[$k_e = 0.97 k_m$]{\includegraphics[width=0.48\textwidth]{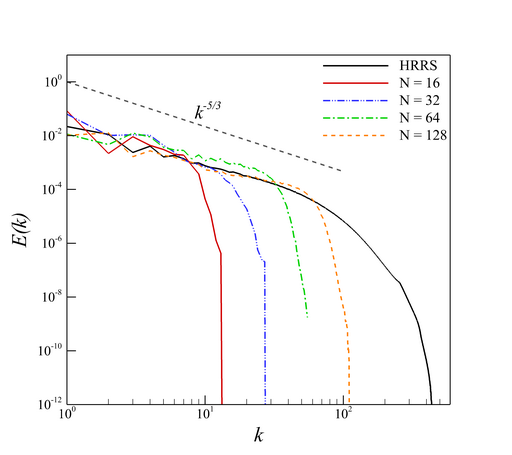}}
\subfigure[$k_e = 0.99 k_m$]{\includegraphics[width=0.48\textwidth]{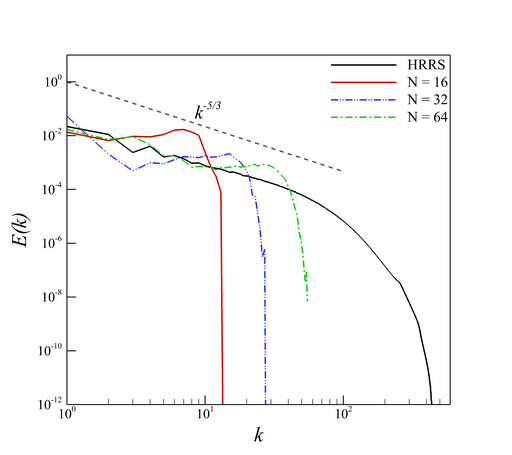}}
}
\caption{Angle-averaged kinetic energy spectra of the 3D KHI problem with $u_{KHI} = 0.5$ on different grid resolutions at $t = 5$ for $CS+RF$ scheme. The spherical-averaged energy spectra in the inertial range flatten towards the $k^{-5/3}$ scaling that corresponds to classical Kolmogorov theory. N: Grid resolution in each direction, HRRS: High Resolution Reference Simulation.}
\label{fig:KHI, relaxation, spectra, u = $0.5$}
\end{figure}

\begin{figure}[!ht]
\centering
\mbox{
\subfigure[$k_e = 0.93 k_m$]{\includegraphics[width=0.48\textwidth]{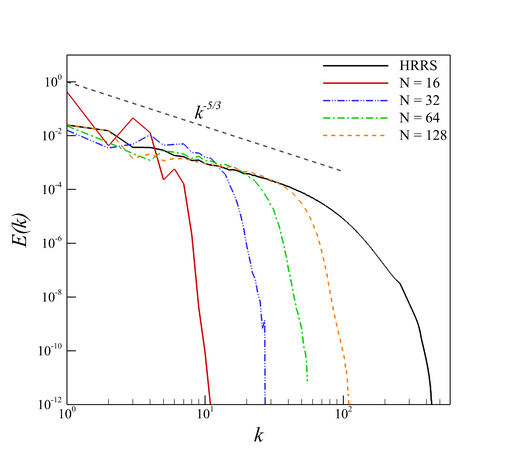}}
\subfigure[$k_e = 0.95 k_m$]{\includegraphics[width=0.48\textwidth]{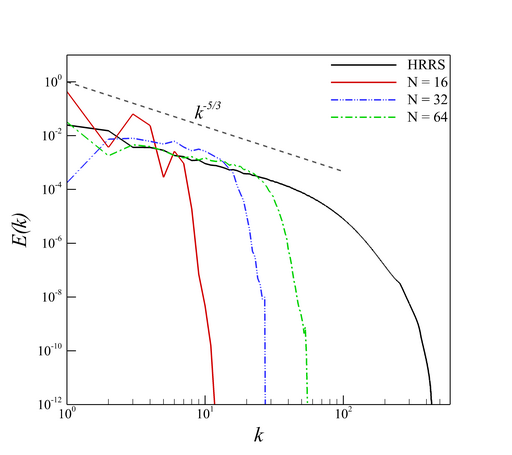}}
}\\
\mbox{
\subfigure[$k_e = 0.97 k_m$]{\includegraphics[width=0.48\textwidth]{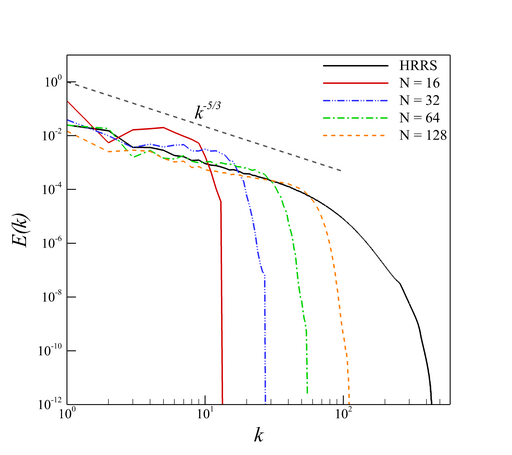}}
}
\caption{Angle-averaged kinetic energy spectra of the 3D KHI problem with $u_{KHI} = 1.0$ on different grid resolutions at $t = 5$ for $CS+RF$ scheme. The spherical-averaged energy spectra in the inertial range flatten towards the $k^{-5/3}$ scaling that corresponds to classical Kolmogorov theory. N: Grid resolution in each direction, HRRS: High Resolution Reference Simulation.}
\label{fig:KHI, relaxation, spectra, u = $1.0$}
\end{figure}

For further quantification of the conclusions we have reached so far, we present the time evolution of the density field for various values of $u_{KHI}$ at two different instances (i.e., $t=1$ and $t=5$) in Fig.~(\ref{fig:KHI, u = $0.1$, 128 resolution}) to Fig.~(\ref{fig:t = $5$,KHI, u = $1$, 128 resolution}). We select to visualize the density field variable because the density is not a constant for compressible flows. It can be observed that the transition to turbulence starts around $t=1$. With the increase of time for a finite value of $u_{KHI}$, the density gets dissipated which represents the mixing characteristic of compressible flow. And with the increase of compressibility through $u_{KHI}$, the dissipation gets prominent that means more mixing of density happens in a short time. In Fig.~(\ref{fig:KHI, u = $0.1$, 128 resolution}), we see the presence of high-density fluid layer in between the lower density fluid layers and the interface is quite smooth at $u_{KHI}=0.1$. At $t = 5$, very small amount of vertical mixing is seen for the same less compressible flow in Fig.~(\ref{fig: t = $5$, KHI, u = $0.1$, 128 resolution,}) but the formation of the wave at the interface can be clearly identified. Even though the wave structure is different for all the schemes, still there are similarities in wave structure for ILES-Roe, Smagorinsky and $CS+RF$ scheme with $k_e = 0.93 k_m$ which is because these three schemes are most dissipative at $u_{KHI} =0.1$. At $u_{KHI} =0.25$ and $u_{KHI}=0.5$, the wave starts forming early in time, and more mixing can be seen at $t = 0.5$ for all the schemes. In Fig.~(\ref{fig:KHI, u = $0.5$, 128 resolution}), the less dissipative $CS+RF$ scheme shows the full mixing of the densities due to excess accumulation of energy whereas ILES-Roe scheme can be clearly visualized as over-dissipative.

Fig.~(\ref{fig:t = $5$,KHI, u = $1$, 128 resolution}) shows that full mixing has happened for all of the schemes at $t =5$ except the localized dynamic model which is still showing the presence of the central belt of high density. The reason behind it can be the localized dynamic model is capturing the physics well even for highly compressible flow. It can also be seen the dissipation of the acoustic waves and density around the central belt of the localized dynamic model. Since the reference high-resolution simulation is computed using over-dissipative WENO-Roe ILES scheme, it is expected to get a more dissipative result after a finite time. It is impressive from the localized dynamic model showing better performance at coarse resolution than the higher resolution ILES simulation which indicates the strength of the model than other models under consideration. The comparison figures of the kinetic energy spectra in Fig.~(\ref{fig:KHI, comparison plots1}) to Fig.~(\ref{fig:KHI, comparison plots4}) summarize the discussions above. As we stated before, the ILES spectra on the left are too dissipative for all the values of $u_{KHI}$. Even though the $u_{KHI} = 0.1$ simulations show some sort of ill prediction in the results, most of the LDS simulations perform better than the results produced by other eddy viscosity models. The right-hand side plots verify that the $CS+RF$ scheme blows up if the added dissipation is not sufficient enough and capture less amount of scales than the LDS model. $CS+RF~(k_e = 0.93 k_m)$ shows over-dissipative result whereas $CS+RF~(k_e = 0.99 k_m)$ results blow up after $u_{KHI} = 0.1$ due to accumulating excess energy pile up at higher wavenumber and eventually, end up making the results unstable.

\begin{figure}[!ht]
\centering
\mbox{
\subfigure[ILES-Roe]{\includegraphics[width=0.48\textwidth]{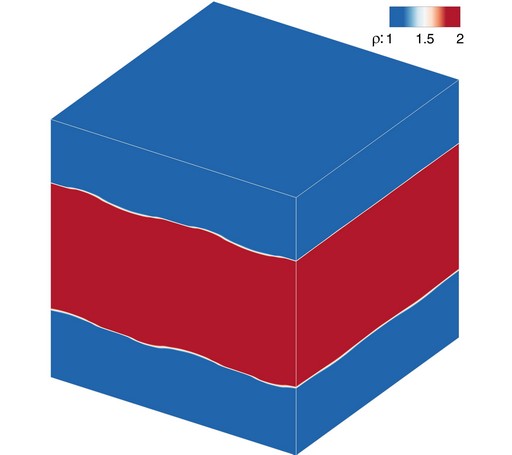}}
\subfigure[Smagorinsky]{\includegraphics[width=0.48\textwidth]{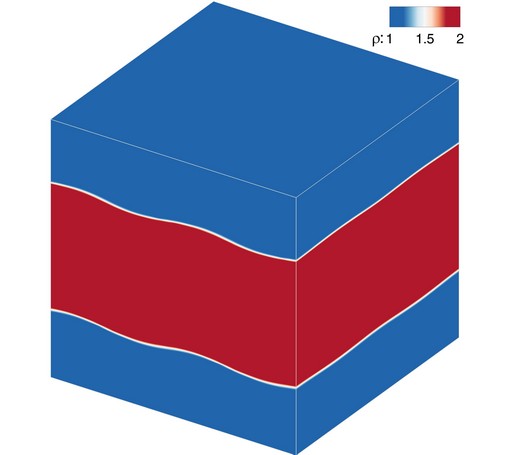}}
}\\
\mbox{
\subfigure[Localized Dynamic Smagorinsky]{\includegraphics[width=0.48\textwidth]{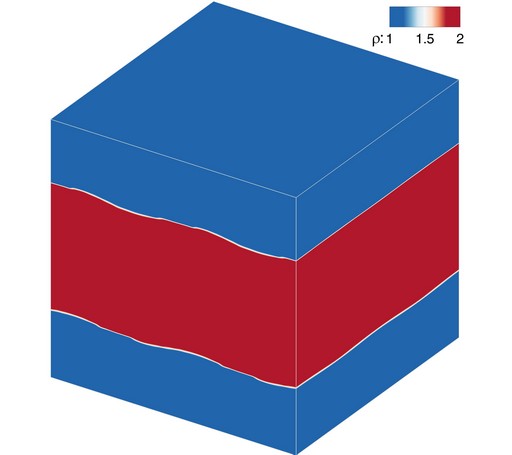}}
\subfigure[Localized Dynamic Smagorinsky with RF]{\includegraphics[width=0.48\textwidth]{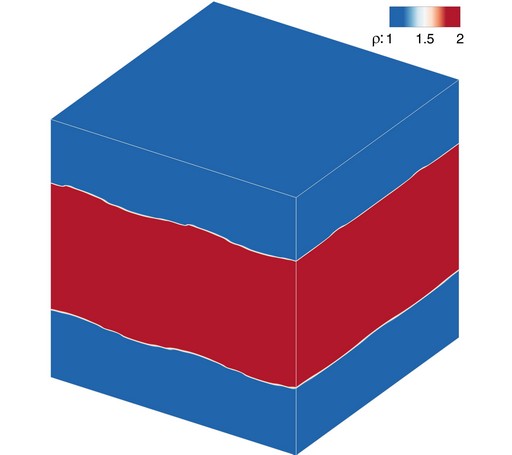}}
}\\
\mbox{
\subfigure[$CS+RF~(k_e = 0.93 k_m)$]{\includegraphics[width=0.48\textwidth]{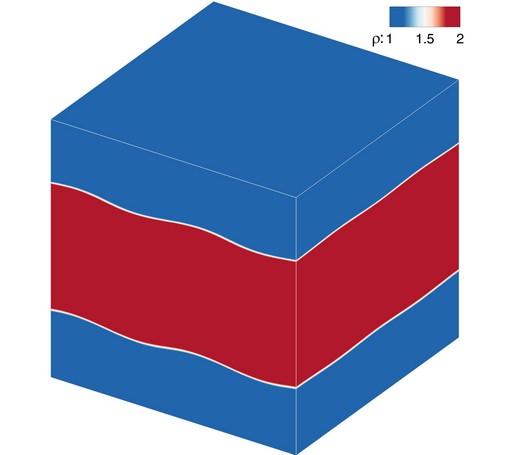}}
\subfigure[$CS+RF~(k_e = 0.99 k_m)$]{\includegraphics[width=0.48\textwidth]{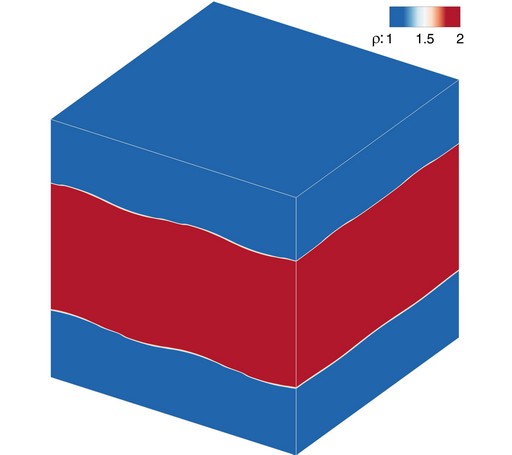}}
}
\caption{Time evolution of density field for the 3D KHI problem with $u_{KHI} = 0.1$ on the $128^3$ grid resolution at $t = 1$ from different schemes. Colors indicate absolute density level.}
\label{fig:KHI, u = $0.1$, 128 resolution}
\end{figure}

\begin{figure}[!ht]
\centering
\mbox{
\subfigure[ILES-Roe]{\includegraphics[width=0.48\textwidth]{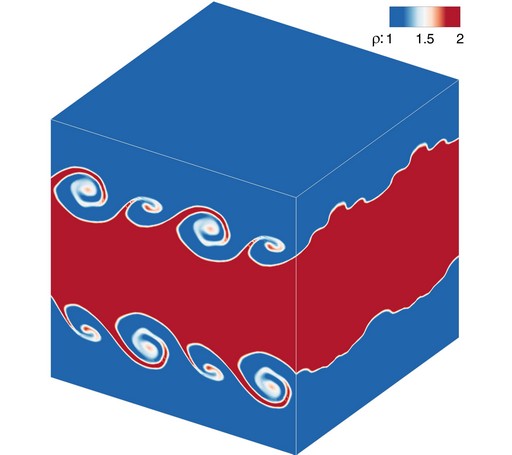}}
\subfigure[Smagorinsky]{\includegraphics[width=0.48\textwidth]{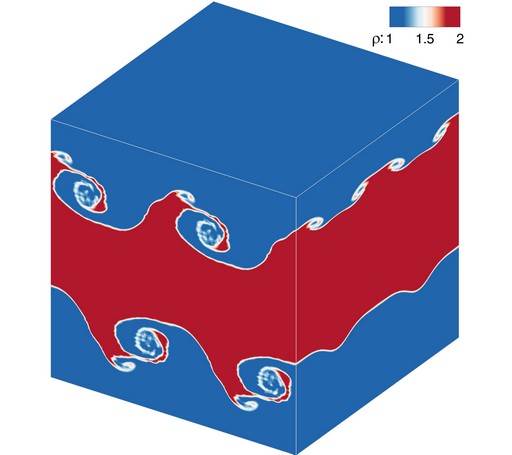}}
}\\
\mbox{
\subfigure[Localized Dynamic Smagorinsky]{\includegraphics[width=0.48\textwidth]{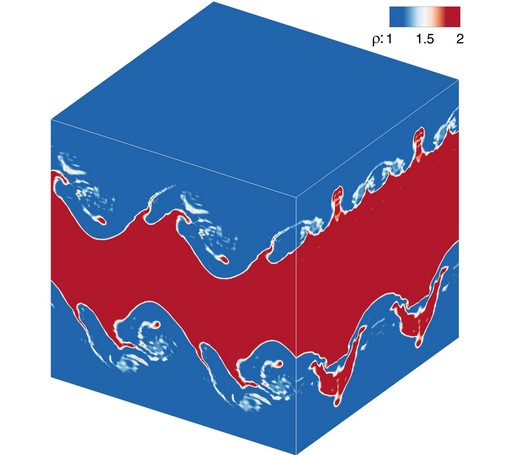}}
\subfigure[Localized Dynamic Smagorinsky with RF]{\includegraphics[width=0.48\textwidth]{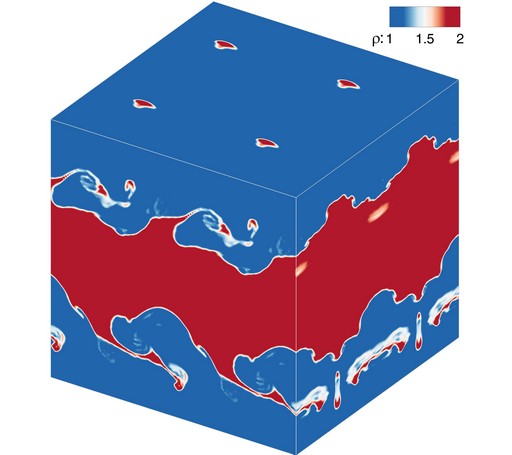}}
}\\
\mbox{
\subfigure[$CS+RF~(k_e = 0.93 k_m)$]{\includegraphics[width=0.48\textwidth]{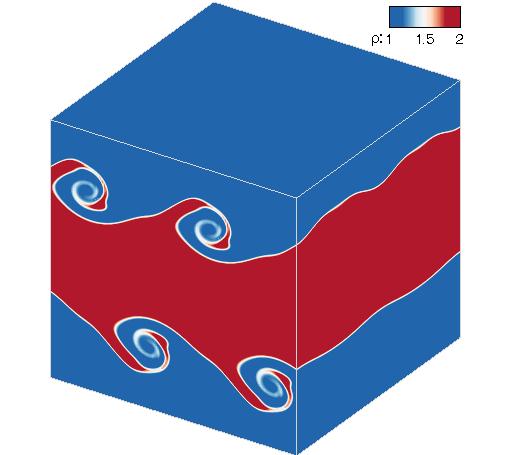}}
\subfigure[$CS+RF~(k_e = 0.99 k_m)$]{\includegraphics[width=0.48\textwidth]{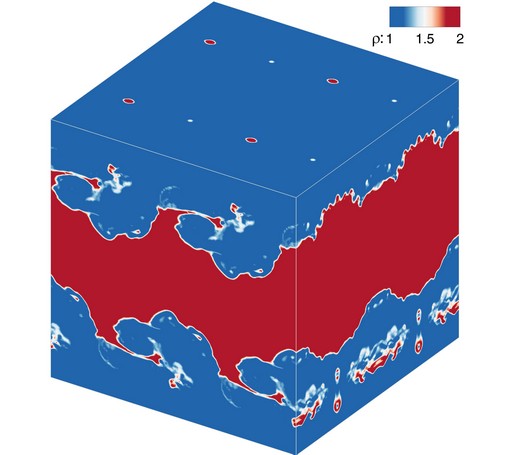}}
}
\caption{Time evolution of density field for the 3D KHI problem with $u_{KHI} = 0.1$ on the $128^3$ grid resolution at $t = 5$ from different schemes. Colors indicate absolute density level.}
\label{fig: t = $5$, KHI, u = $0.1$, 128 resolution,}
\end{figure}

\begin{figure}[!ht]
\centering
\mbox{
\subfigure[ILES-Roe]{\includegraphics[width=0.48\textwidth]{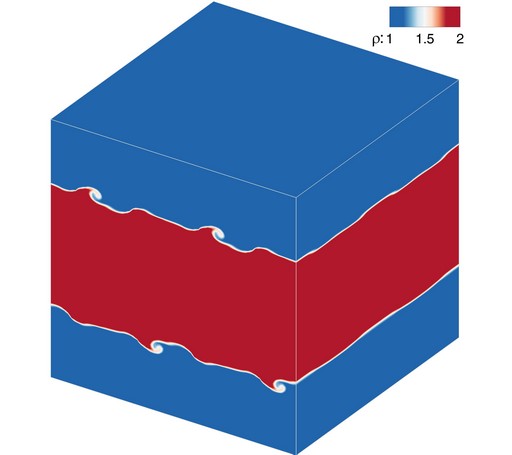}}
\subfigure[Smagorinsky]{\includegraphics[width=0.48\textwidth]{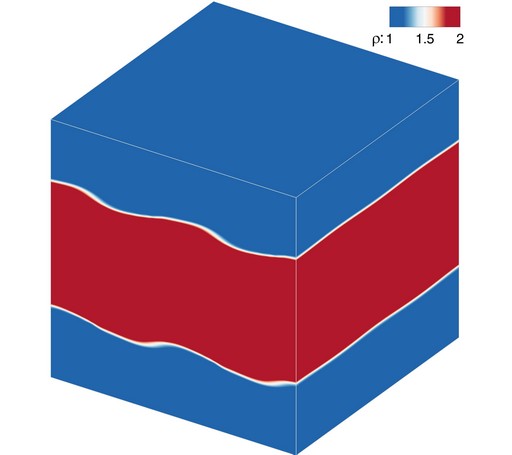}}
}\\
\mbox{
\subfigure[Localized Dynamic Smagorinsky]{\includegraphics[width=0.48\textwidth]{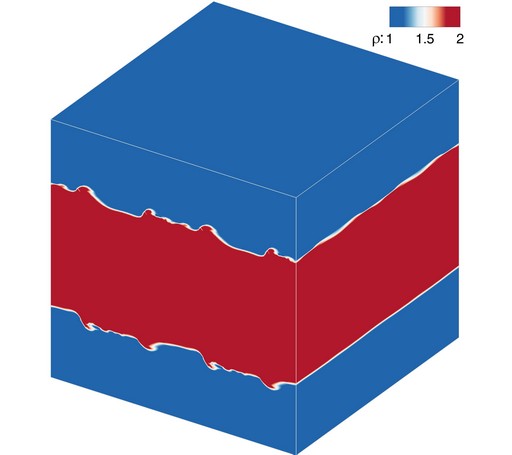}}
\subfigure[Localized Dynamic Smagorinsky with RF]{\includegraphics[width=0.48\textwidth]{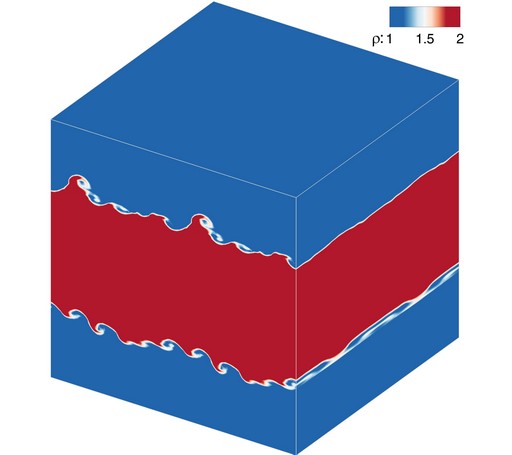}}
}\\
\mbox{
\subfigure[$CS+RF~(k_e = 0.93 k_m)$]{\includegraphics[width=0.48\textwidth]{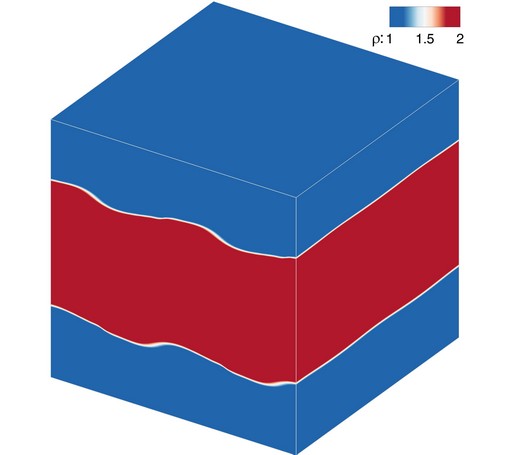}}
\subfigure[$CS+RF~(k_e = 0.99 k_m)$]{\includegraphics[width=0.48\textwidth]{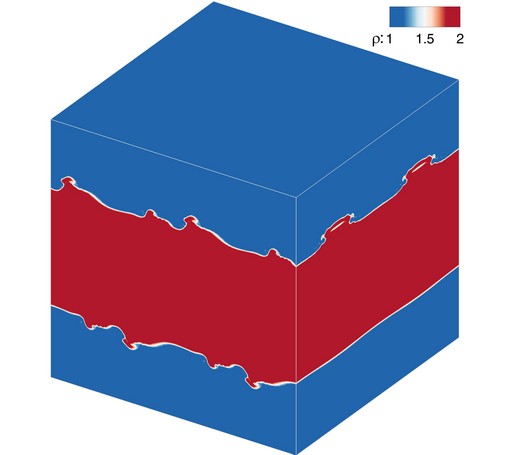}}
}
\caption{Time evolution of density field for the 3D KHI problem with $u_{KHI} = 0.25$ on the $128^3$ grid resolution at $t = 1$ from different schemes. Colors indicate absolute density level.}
\label{fig:KHI, u = $0.25$, 128 resolution}
\end{figure}

\begin{figure}[!ht]
\centering
\mbox{
\subfigure[ILES-Roe]{\includegraphics[width=0.48\textwidth]{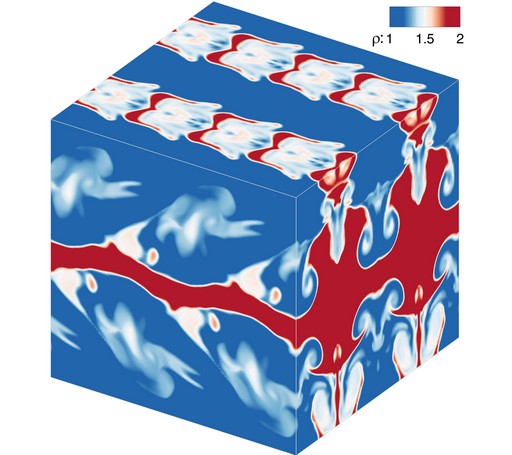}}
\subfigure[Smagorinsky]{\includegraphics[width=0.48\textwidth]{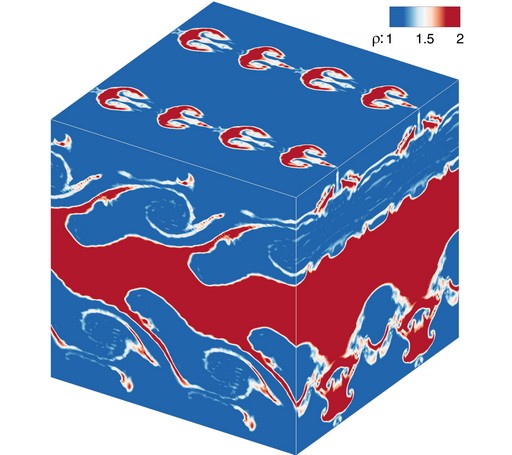}}
}\\
\mbox{
\subfigure[Localized Dynamic Smagorinsky]{\includegraphics[width=0.48\textwidth]{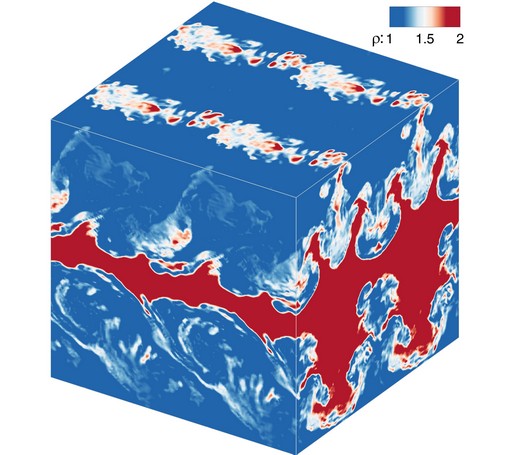}}
\subfigure[Localized Dynamic Smagorinsky with RF]{\includegraphics[width=0.48\textwidth]{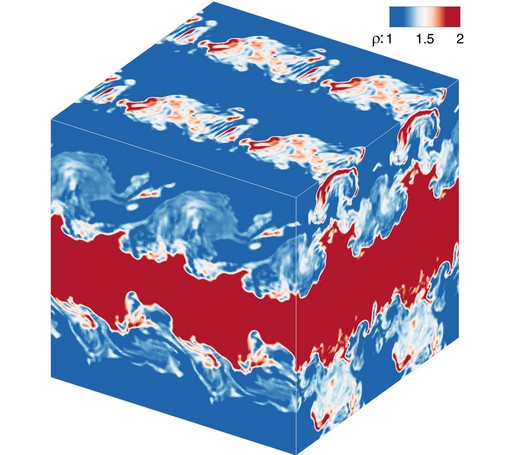}}
}\\
\mbox{
\subfigure[$CS+RF~(k_e = 0.93 k_m)$]{\includegraphics[width=0.48\textwidth]{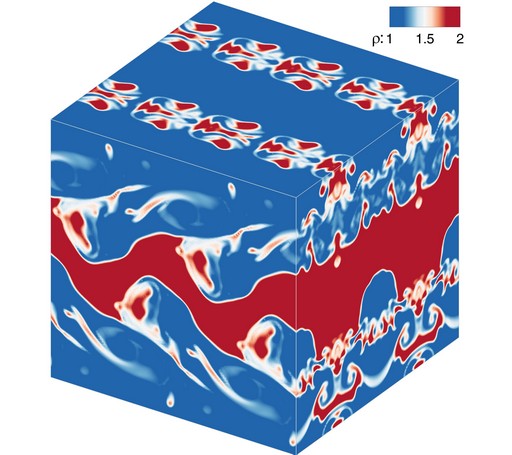}}
\subfigure[$CS+RF~(k_e = 0.97 k_m)$]{\includegraphics[width=0.48\textwidth]{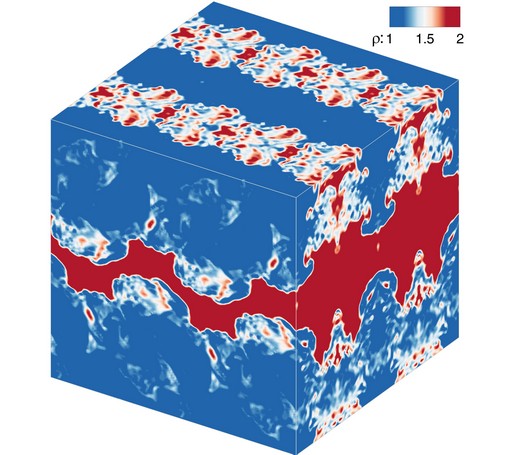}}
}
\caption{Time evolution of density field for the 3D KHI problem with $u_{KHI} = 0.25$ on the $128^3$ grid resolution at $t = 5$ from different schemes. Colors indicate absolute density level.}
\label{fig: t = $5$, KHI, u = $0.25$, 128 resolution}
\end{figure}

\begin{figure}[!ht]
\centering
\mbox{
\subfigure[ILES-Roe]{\includegraphics[width=0.48\textwidth]{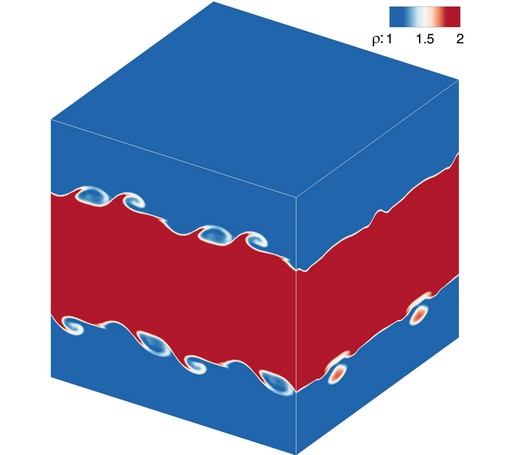}}
\subfigure[Smagorinsky]{\includegraphics[width=0.48\textwidth]{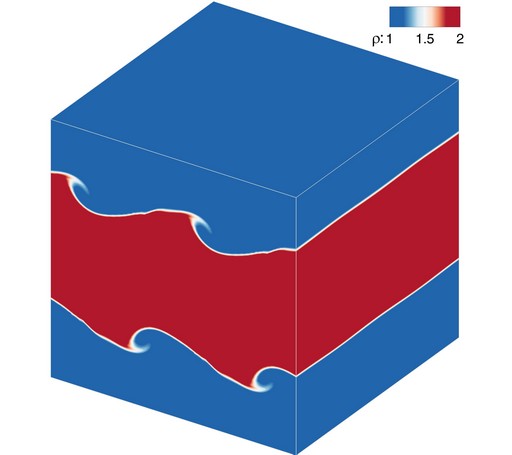}}
}\\
\mbox{
\subfigure[Localized Dynamic Smagorinsky]{\includegraphics[width=0.48\textwidth]{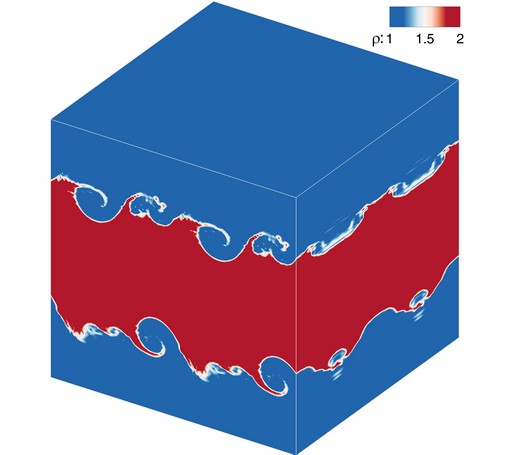}}
\subfigure[Localized Dynamic Smagorinsky with RF]{\includegraphics[width=0.48\textwidth]{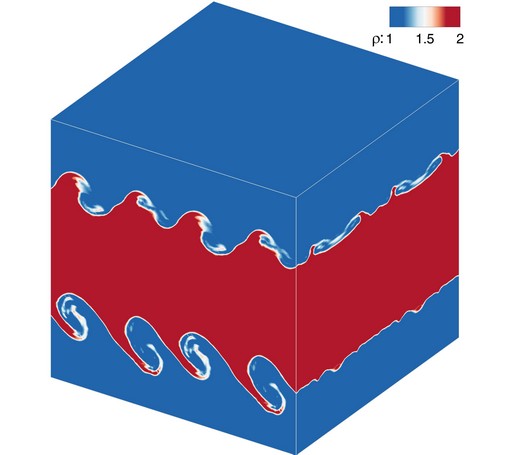}}
}\\
\mbox{
\subfigure[$CS+RF~(k_e = 0.93 k_m)$]{\includegraphics[width=0.48\textwidth]{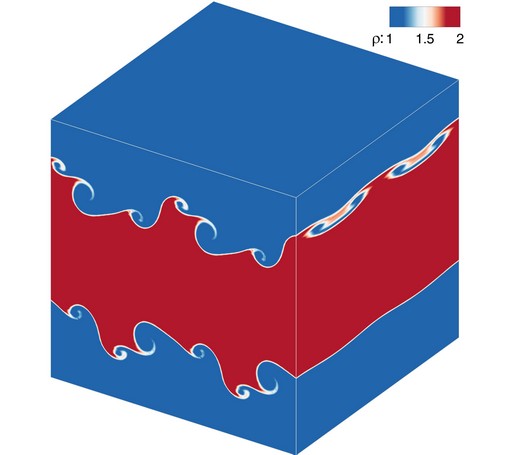}}
\subfigure[$CS+RF~(k_e = 0.97 k_m)$]{\includegraphics[width=0.48\textwidth]{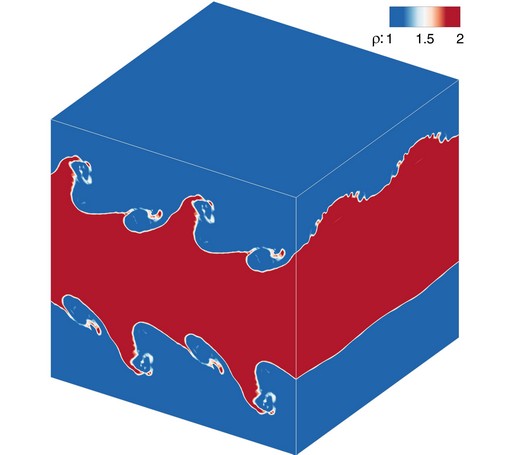}}
}
\caption{Time evolution of density field for the 3D KHI problem with $u_{KHI} = 0.5$ on the $128^3$ grid resolution at $t = 1$ from different schemes. Colors indicate absolute density level.}
\label{fig:KHI, u = $0.5$, 128 resolution}
\end{figure}

\begin{figure}[!ht]
\centering
\mbox{
\subfigure[ILES-Roe]{\includegraphics[width=0.48\textwidth]{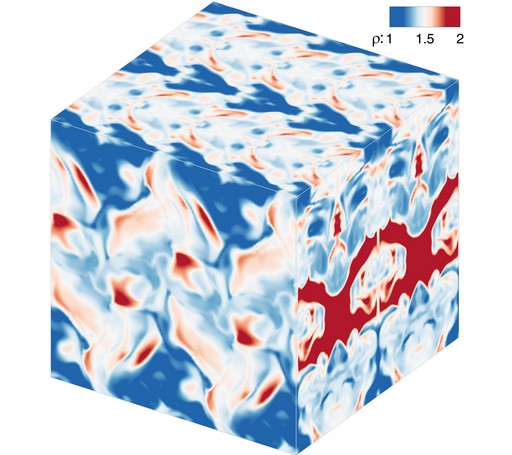}}
\subfigure[Smagorinsky]{\includegraphics[width=0.48\textwidth]{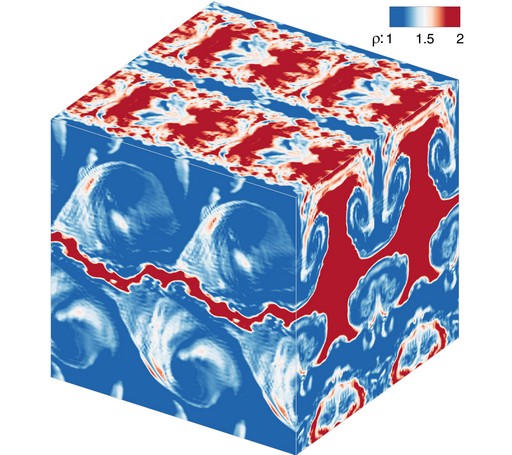}}
}\\
\mbox{
\subfigure[Localized Dynamic Smagorinsky]{\includegraphics[width=0.48\textwidth]{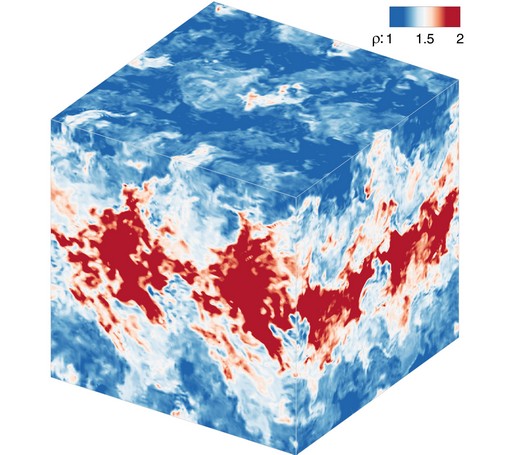}}
\subfigure[Localized Dynamic Smagorinsky with RF]{\includegraphics[width=0.48\textwidth]{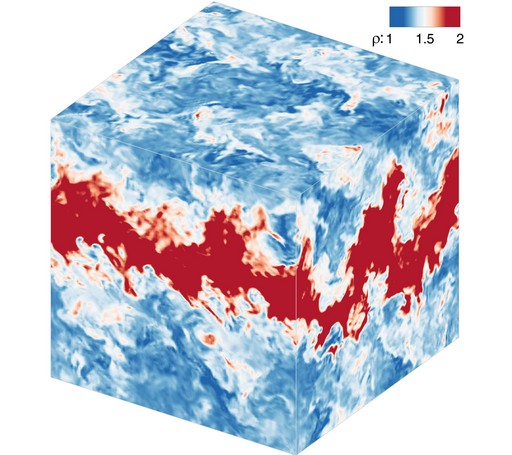}}
}\\
\mbox{
\subfigure[$CS+RF~(k_e = 0.93 k_m)$]{\includegraphics[width=0.48\textwidth]{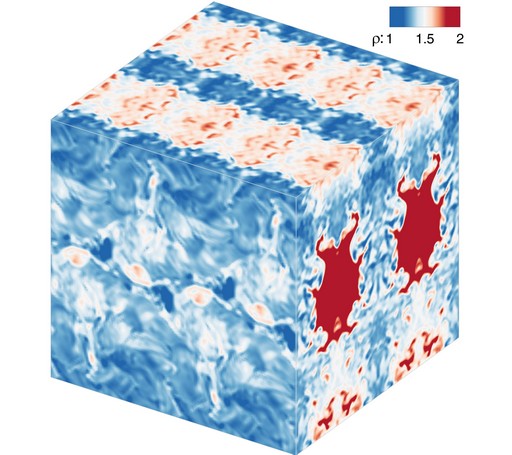}}
\subfigure[$CS+RF~(k_e = 0.97 k_m)$]{\includegraphics[width=0.48\textwidth]{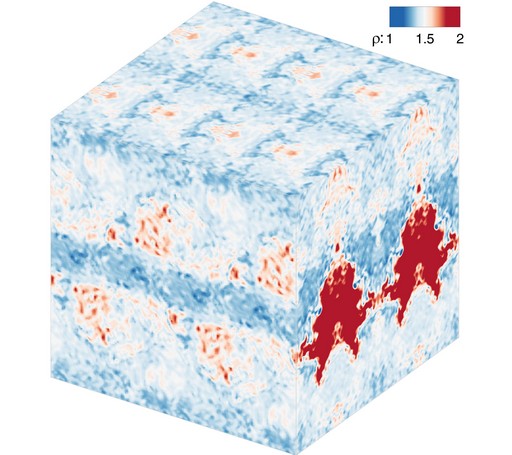}}
}
\caption{Time evolution of density field for the 3D KHI problem with $u_{KHI} = 0.5$ on the $128^3$ grid resolution at $t = 5$ from different schemes. Colors indicate absolute density level.}
\label{fig: t = $5$,KHI, u = $0.5$, 128 resolution,}
\end{figure}

\begin{figure}[!ht]
\centering
\mbox{
\subfigure[ILES-Roe]{\includegraphics[width=0.48\textwidth]{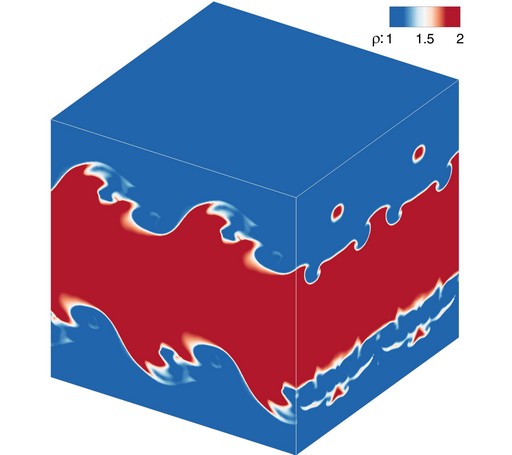}}
\subfigure[Smagorinsky]{\includegraphics[width=0.48\textwidth]{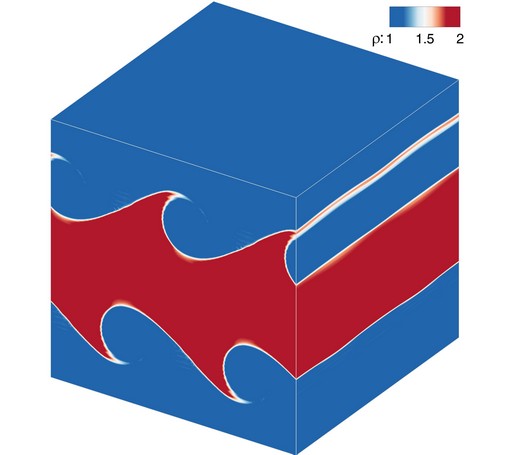}}
}\\
\mbox{
\subfigure[Localized Dynamic Smagorinsky]{\includegraphics[width=0.48\textwidth]{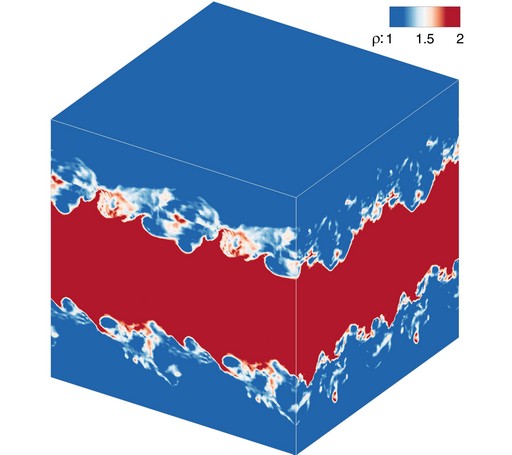}}
\subfigure[Localized Dynamic Smagorinsky with RF]{\includegraphics[width=0.48\textwidth]{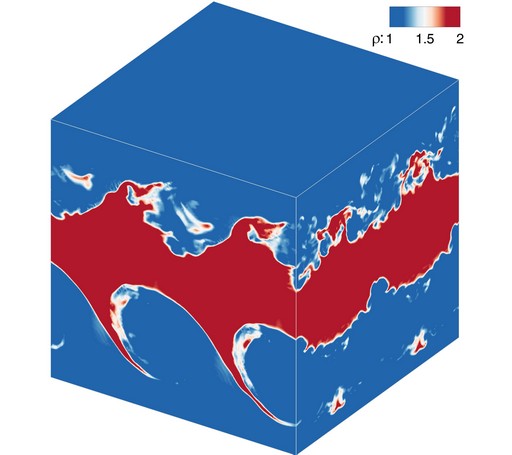}}
}\\
\mbox{
\subfigure[$CS+RF~(k_e = 0.93 k_m)$]{\includegraphics[width=0.48\textwidth]{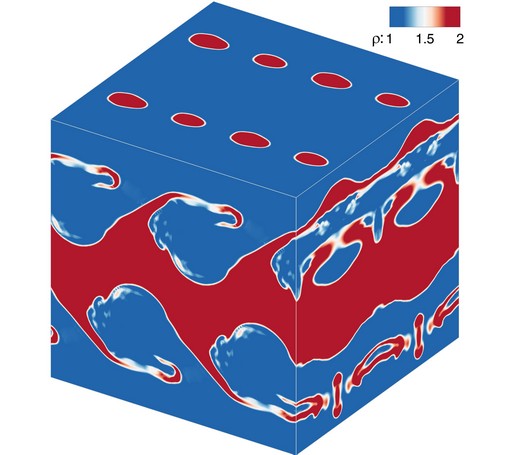}}
\subfigure[$CS+RF~(k_e = 0.97 k_m)$]{\includegraphics[width=0.48\textwidth]{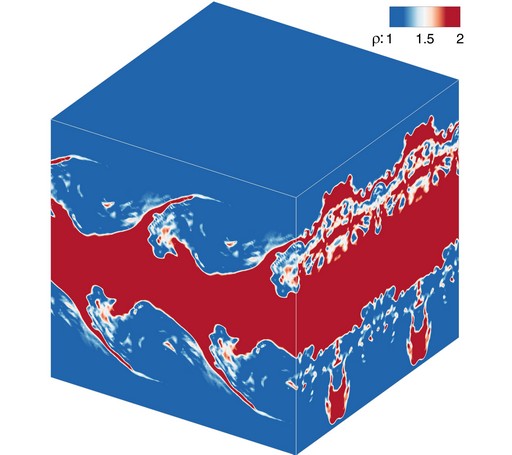}}
}
\caption{Time evolution of density field for the 3D KHI problem with $u_{KHI} = 1.0$ on the $128^3$ grid resolution at $t = 1$ from different schemes. Colors indicate absolute density level.}
\label{fig:KHI, u = $1$, 128 resolution}
\end{figure}

\begin{figure}[!ht]
\centering
\mbox{
\subfigure[ILES-Roe]{\includegraphics[width=0.48\textwidth]{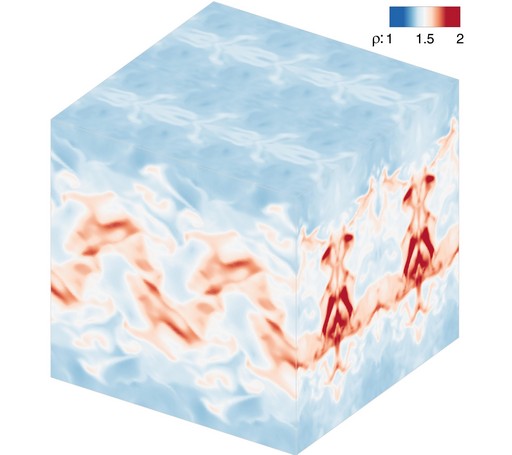}}
\subfigure[Smagorinsky]{\includegraphics[width=0.48\textwidth]{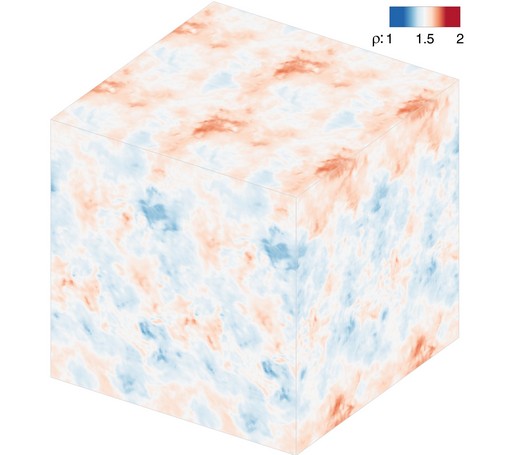}}
}\\
\mbox{
\subfigure[Localized Dynamic Smagorinsky]{\includegraphics[width=0.48\textwidth]{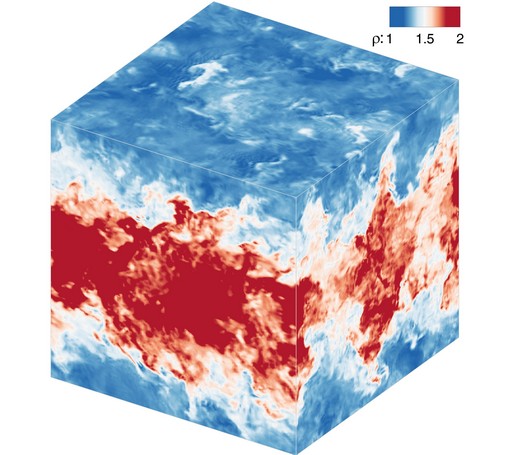}}
\subfigure[Localized Dynamic Smagorinsky with RF]{\includegraphics[width=0.48\textwidth]{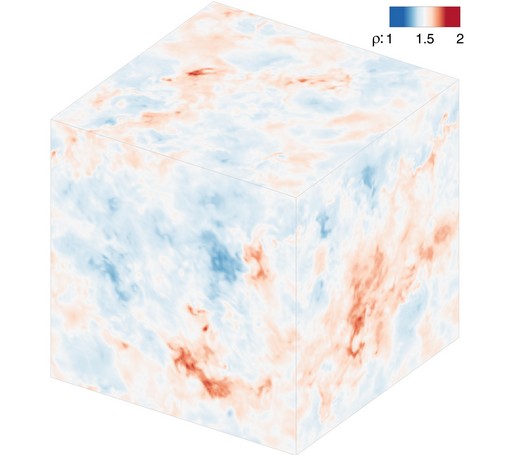}}
}\\
\mbox{
\subfigure[$CS+RF~(k_e = 0.93 k_m)$]{\includegraphics[width=0.48\textwidth]{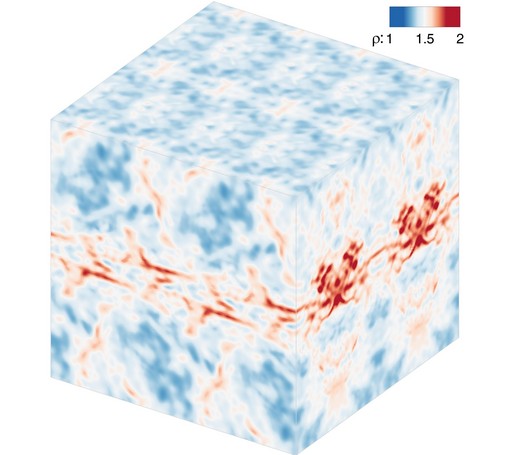}}
\subfigure[$CS+RF~(k_e = 0.97 k_m)$]{\includegraphics[width=0.48\textwidth]{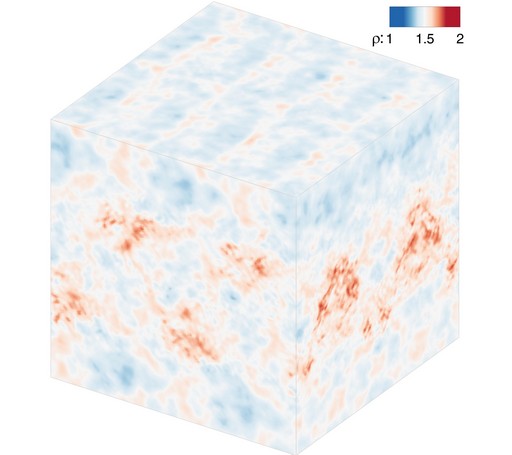}}
}
\caption{Time evolution of density field for the 3D KHI problem with $u_{KHI} = 1.0$ on the $128^3$ grid resolution at $t = 5$ from different schemes. Colors indicate absolute density level.}
\label{fig:t = $5$,KHI, u = $1$, 128 resolution}
\end{figure}

\begin{figure}[!ht]
\centering
\mbox{
{\includegraphics[width=0.48\textwidth]{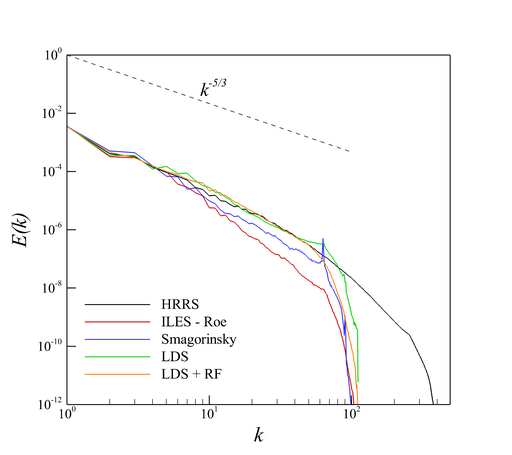}}
{\includegraphics[width=0.48\textwidth]{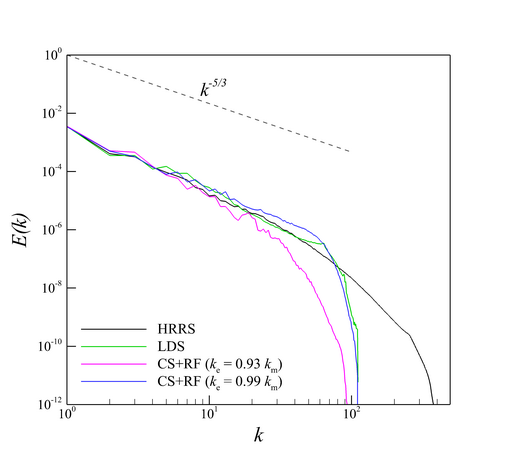}}
}\\
\caption{Comparison of the angle-averaged kinetic energy spectra for the 3D KHI problem on the $128^3$ grid resolution at $t = 5$ and $u_{KHI}=0.1$ obtained by different schemes. Left: Comparison of ILES and eddy viscosity model with the LDS model. Right: Comparison of CS+RF models with LDS model. HRRS: High Resolution Reference Simulation, LDS: Localized Dynamic Smagorinsky and LDS+RF: Localized Dynamic Smagorinsky with Relaxation Filtering.}
\label{fig:KHI, comparison plots1}
\end{figure}
\begin{figure}[!ht]
\centering
\mbox{
{\includegraphics[width=0.48\textwidth]{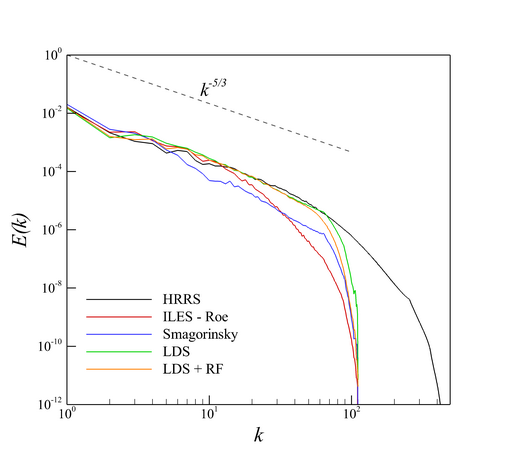}}
{\includegraphics[width=0.48\textwidth]{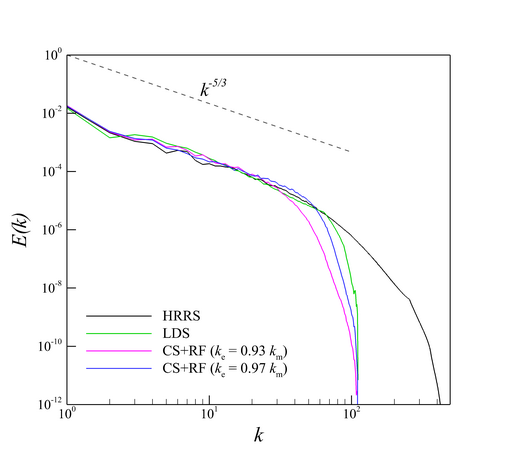}}
}\\
\caption{Comparison of the angle-averaged kinetic energy spectra for the 3D KHI problem on the $128^3$ grid resolution at $t = 5$ and $u_{KHI}=0.25$ obtained by different schemes. Left: Comparison of ILES and eddy viscosity model with the LDS model. Right: Comparison of $CS+RF$ models with LDS model. HRRS: High Resolution Reference Simulation, LDS: Localized Dynamic Smagorinsky and LDS+RF: Localized Dynamic Smagorinsky with Relaxation Filtering.}
\label{fig:KHI, comparison plots2}
\end{figure}
\begin{figure}[!ht]
\centering
\mbox{
{\includegraphics[width=0.48\textwidth]{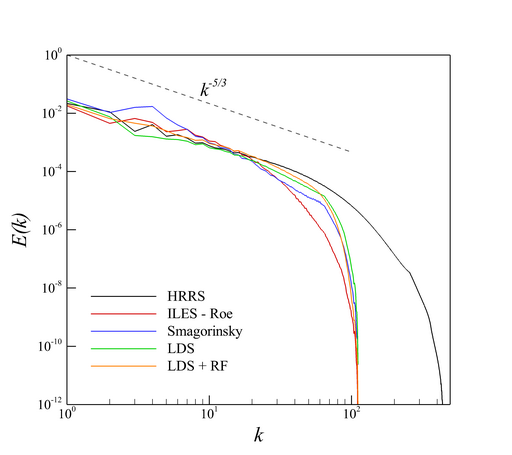}}
{\includegraphics[width=0.48\textwidth]{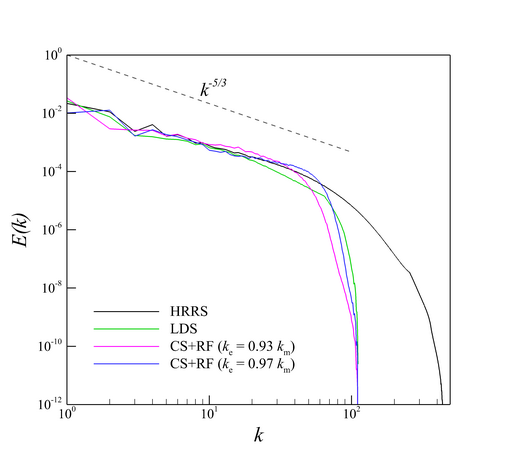}}
}
\caption{Comparison of the angle-averaged kinetic energy spectra for the 3D KHI problem on the $128^3$ grid resolution at $t = 5$ and $u_{KHI}=0.5$ obtained by different schemes. Left: Comparison of ILES and eddy viscosity model with the LDS model. Right: Comparison of $CS+RF$ models with LDS model. HRRS: High Resolution Reference Simulation, LDS: Localized Dynamic Smagorinsky and LDS+RF: Localized Dynamic Smagorinsky with Relaxation Filtering.}
\label{fig:KHI, comparison plots3}
\end{figure}

\begin{figure}[!ht]
\centering
\mbox{
{\includegraphics[width=0.48\textwidth]{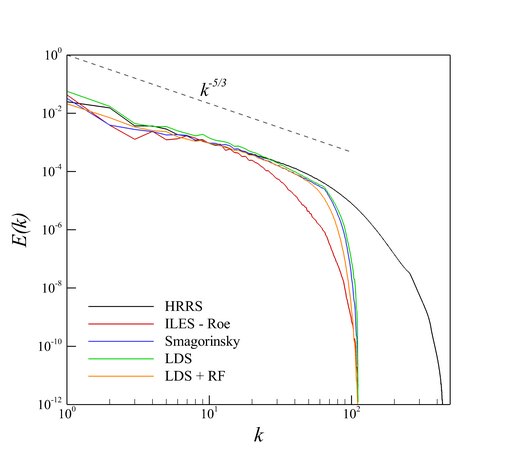}}
{\includegraphics[width=0.48\textwidth]{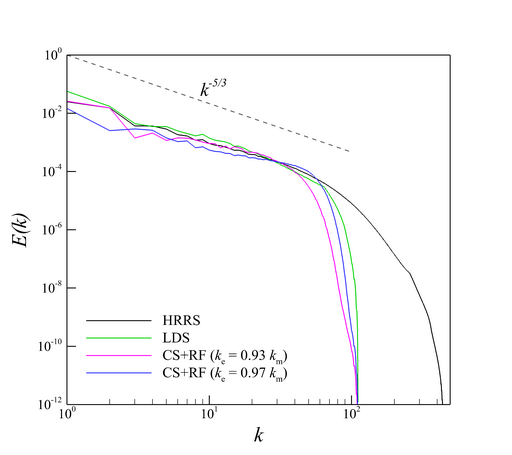}}
}
\caption{Comparison of the angle-averaged kinetic energy spectra for the 3D KHI problem on the $128^3$ grid resolution at $t = 5$ and $u_{KHI}=1.0$ obtained by different schemes. Left: Comparison of ILES and eddy viscosity model with the LDS model. Right: Comparison of $CS+RF$ models with LDS model. HRRS: High Resolution Reference Simulation, LDS: Localized Dynamic Smagorinsky and LDS+RF: Localized Dynamic Smagorinsky with Relaxation Filtering.}
\label{fig:KHI, comparison plots4}
\end{figure}

To provide a comparison between the traditional dynamic eddy viscosity model and proposed localized dynamic eddy viscosity model, we present the Smagorinsky constant and relative eddy viscosity against time in Fig.~(\ref{fig:relative viscosity, tgv}) to Fig.~(\ref{fig:relative viscosity, khi4}) based on a-posteriori analysis of the above numerical test cases. We compute the Smagorinsky constat, $C_s$ homogeneously in all the directions, i.e., we compute $C_s$ in a global sense for the traditional dynamic Smagorinsky model whereas for the localized dynamic Smagorinsky model, we rely on the information of the neighboring stencils to compute $C_s$. In Fig.~(\ref{fig:relative viscosity, tgv}), we observe that the $C_s$ values asymptotically reach a level between 0.1 and 0.14 for this relatively isotropic flow case. It is expected not to have a substantial difference between traditional dynamic Smagorinsky model and localized dynamic Smagorinsky model for TGV flow because of its isotropic nature. The eddy viscosity plots show that less amount of eddy viscosity is added with the increase of the resolution which indicates the credibility of the high-resolution results.

When the global averaging procedure is applied in all directions for the KHI problem the global dynamic model yields unstable results for high-resolution due to the stratification in the normal direction of the flow field as we can see in Fig.~(\ref{fig:relative viscosity, khi1}) to Fig.~(\ref{fig:relative viscosity, khi4}). Also, as we increase the level of stratification through $u_{KHI}$ parameter, the model becomes unstable earlier for much coarser resolution. One might develop another averaging procedure for only $x-$ and $z-$ directions to stabilize the standard dynamic model in the KHI case. On the other hand, the localized dynamic Smagorinsky model produces stable results even for highly stratified case because of calculating nodal values using the local information of the neighboring nodes. Another benefit of the localized dynamic model over the traditional dynamic model is the MPI parallelization of the scheme since the data transfer for the localized model is required only for the boundary point calculation of the local domain. Therefore, in our approach, we don't require to know all the values of all data points far away from the local point. We have preferred such a local stencil averaging operation primarily due to the underlying firth-order WENO scheme. Therefore, any particular point in the present processing element has already connected those neighboring points. Therefore, it will be computational more tractable than the classical dynamic modeling approach (i.e., in terms of both communication and arithmetic costs). Since communication cost is dominating the arithmetics, we believe that our 7-point averaging stencil arrangement is able to capture the localized eddy viscosity contribution in a computationally effective way.
\begin{figure}[!ht]
\centering
\mbox{
\subfigure[Dynamic Smagorinsky (Global)]{\includegraphics[width=0.48\textwidth]{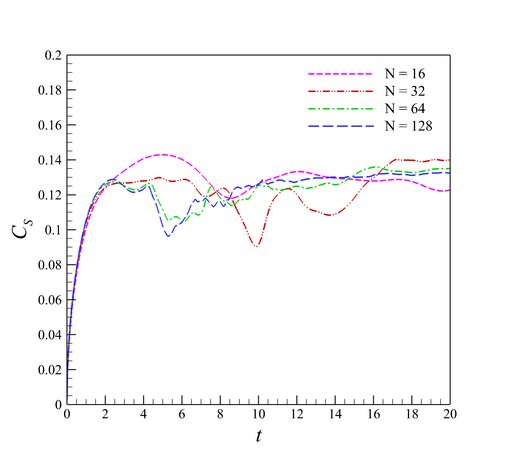}}
\subfigure[Localized Dynamic Smagorinsky]{\includegraphics[width=0.48\textwidth]{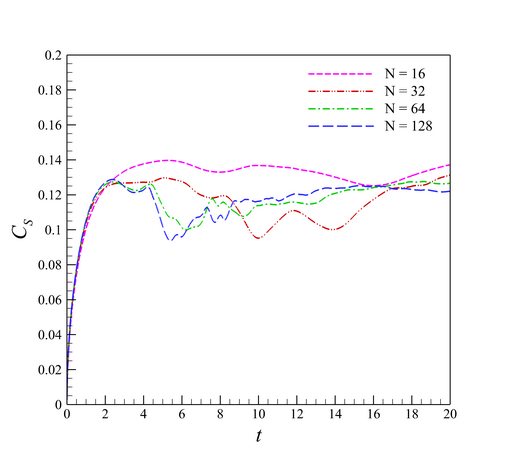}}
}\\
\mbox{
\subfigure[Dynamic Smagorinsky (Global)]{\includegraphics[width=0.48\textwidth]{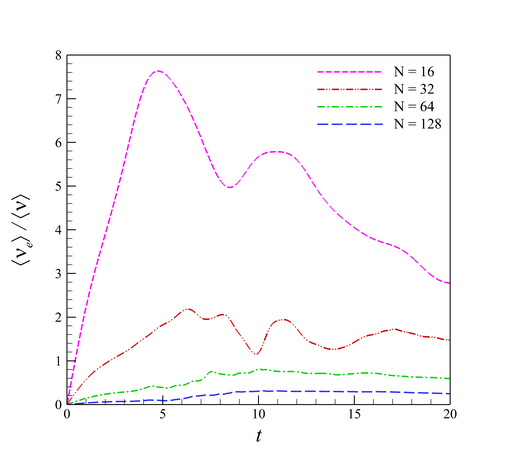}}
\subfigure[Localized Dynamic Smagorinsky]{\includegraphics[width=0.48\textwidth]{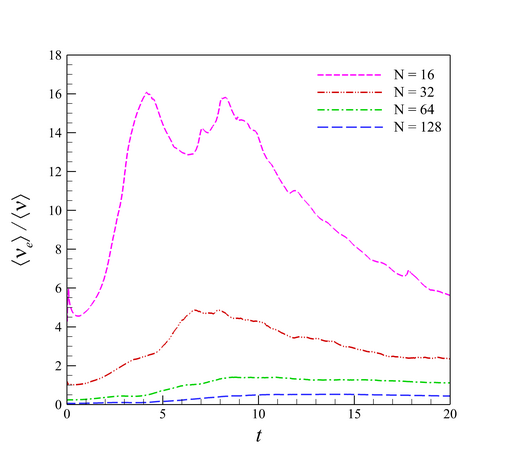}}
}
\caption{Evolution of Smagorinsky constant, $C_s$ and eddy viscosity behavior for the TGV problem with varying grid resolution, N.}
\label{fig:relative viscosity, tgv}
\end{figure}

\begin{figure}[!ht]
\centering
\mbox{
\subfigure[Dynamic Smagorinsky (Global)]{\includegraphics[width=0.48\textwidth]{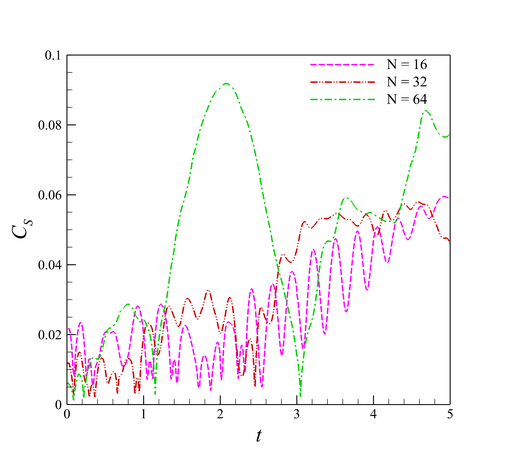}}
\subfigure[Localized Dynamic Smagorinsky]{\includegraphics[width=0.48\textwidth]{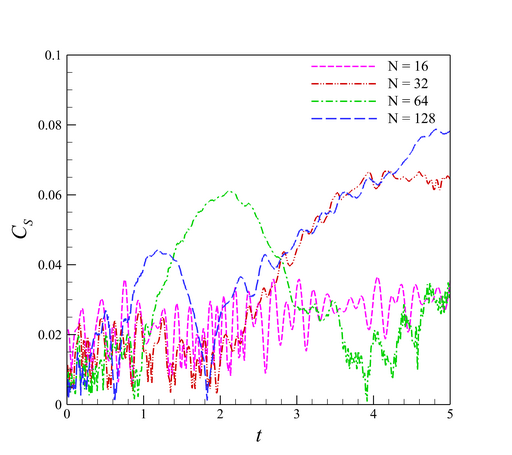}}
}\\
\mbox{
\subfigure[Dynamic Smagorinsky (Global)]{\includegraphics[width=0.48\textwidth]{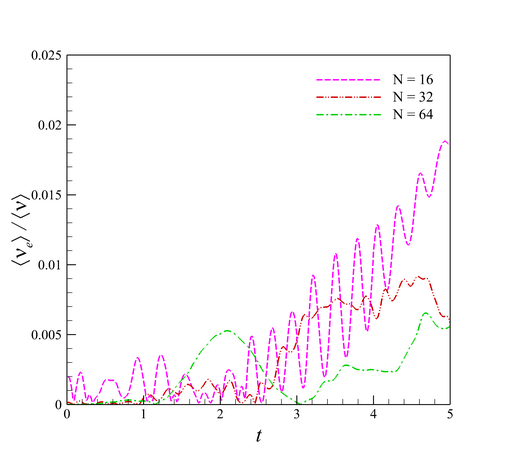}}
\subfigure[Localized Dynamic Smagorinsky]{\includegraphics[width=0.48\textwidth]{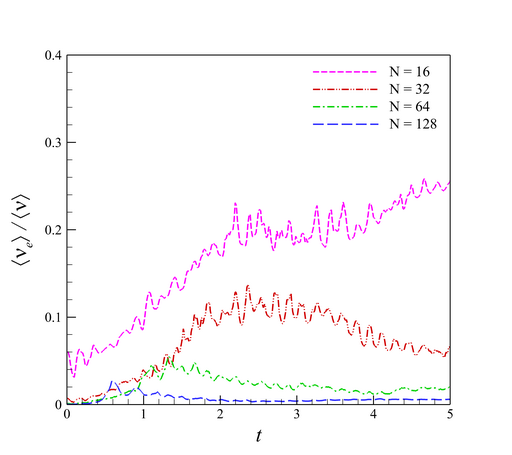}}
}
\caption{Evolution of Smagorinsky constant, $C_s$ and eddy viscosity behavior for the KHI problem ($u_{KHI}= 0.1$) with varying grid resolution, N.}
\label{fig:relative viscosity, khi1}
\end{figure}

\begin{figure}[!ht]
\centering
\mbox{
\subfigure[Dynamic Smagorinsky (Global)]{\includegraphics[width=0.48\textwidth]{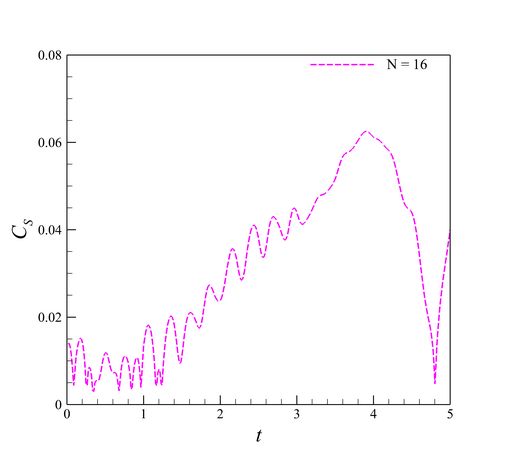}}
\subfigure[Localized Dynamic Smagorinsky]{\includegraphics[width=0.48\textwidth]{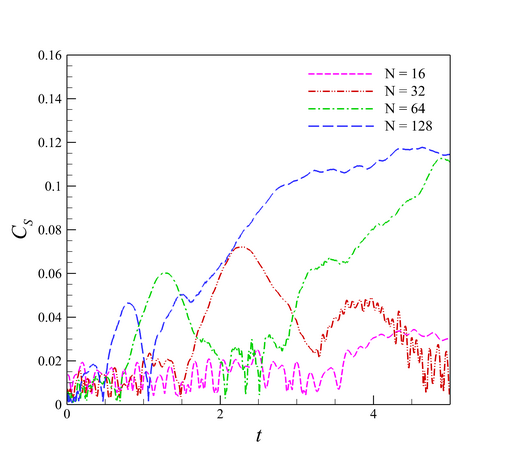}}
}\\
\mbox{
\subfigure[Dynamic Smagorinsky (Global)]{\includegraphics[width=0.48\textwidth]{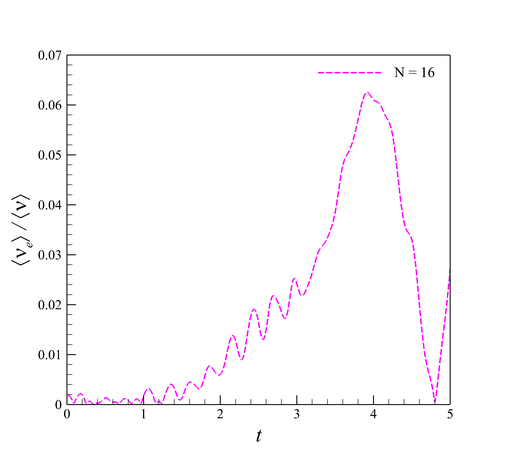}}
\subfigure[Localized Dynamic Smagorinsky]{\includegraphics[width=0.48\textwidth]{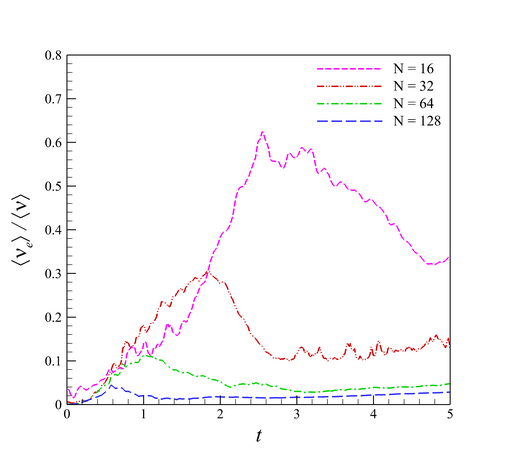}}
}
\caption{Evolution of Smagorinsky constant, $C_s$ and eddy viscosity behavior for the KHI problem ($u_{KHI}= 0.25$) with varying grid resolution, N.}
\label{fig:relative viscosity, khi2}
\end{figure}

\begin{figure}[!ht]
\centering
\mbox{
\subfigure[Dynamic Smagorinsky (Global)]{\includegraphics[width=0.48\textwidth]{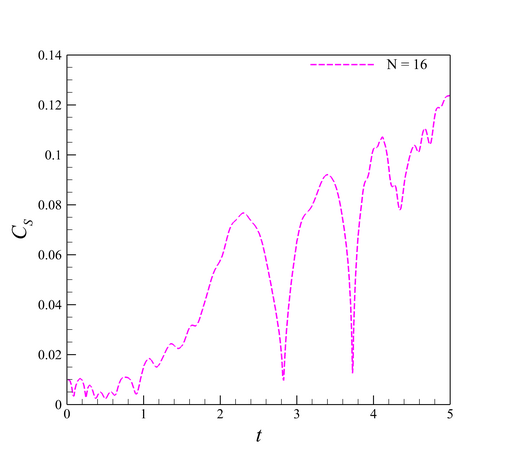}}
\subfigure[Localized Dynamic Smagorinsky]{\includegraphics[width=0.48\textwidth]{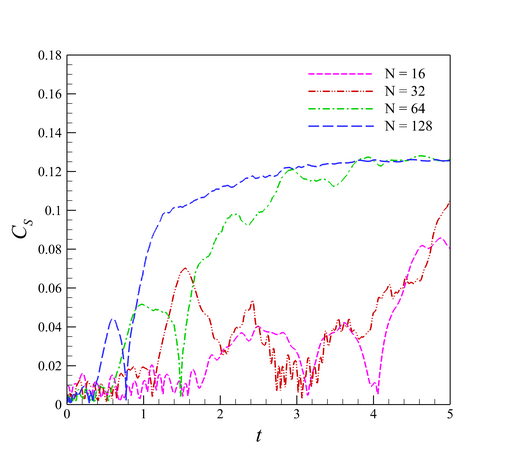}}
}\\
\mbox{
\subfigure[Dynamic Smagorinsky (Global)]{\includegraphics[width=0.48\textwidth]{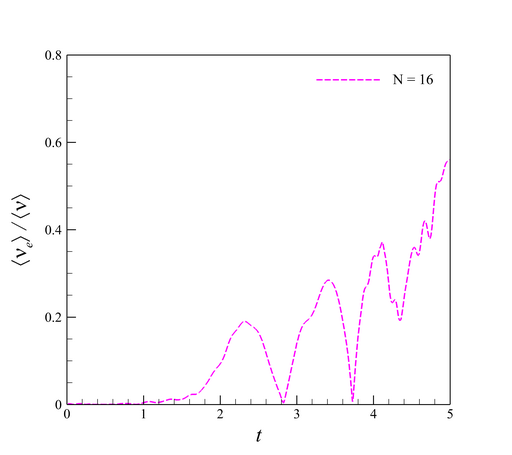}}
\subfigure[Localized Dynamic Smagorinsky]{\includegraphics[width=0.48\textwidth]{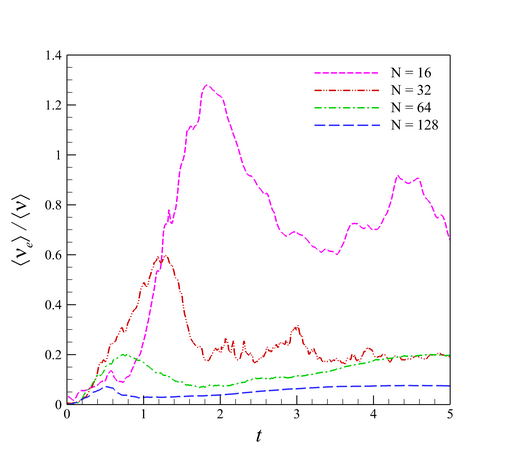}}
}
\caption{Evolution of Smagorinsky constant, $C_s$ and eddy viscosity behavior for the KHI problem ($u_{KHI}= 0.5$) with varying grid resolution, N.}
\label{fig:relative viscosity, khi3}
\end{figure}

\begin{figure}[!ht]
\centering
\mbox{
\subfigure[Dynamic Smagorinsky (Global)]{\includegraphics[width=0.48\textwidth]{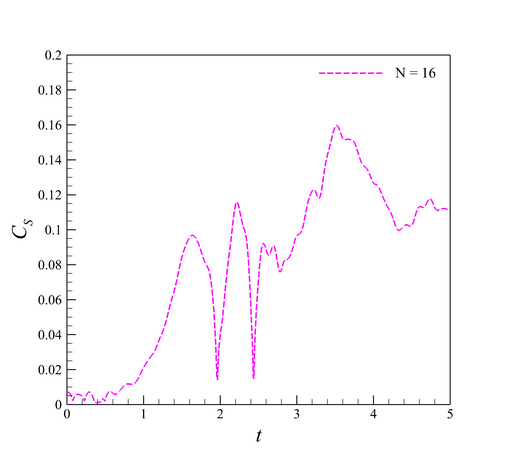}}
\subfigure[Localized Dynamic Smagorinsky]{\includegraphics[width=0.48\textwidth]{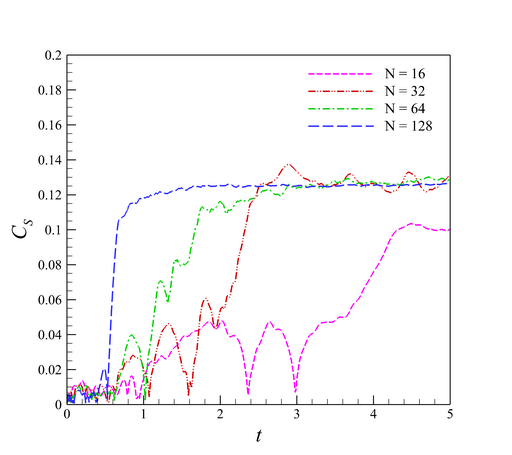}}
}\\
\mbox{
\subfigure[Dynamic Smagorinsky (Global)]{\includegraphics[width=0.48\textwidth]{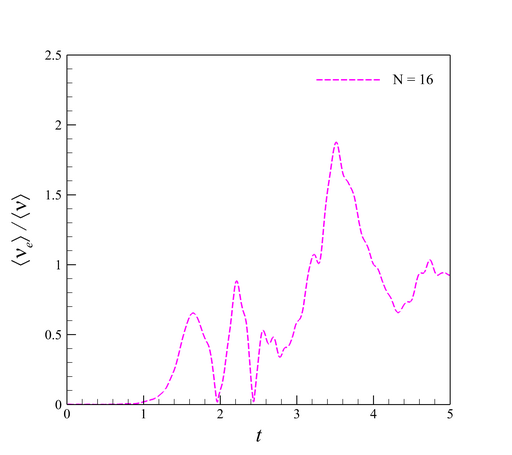}}
\subfigure[Localized Dynamic Smagorinsky]{\includegraphics[width=0.48\textwidth]{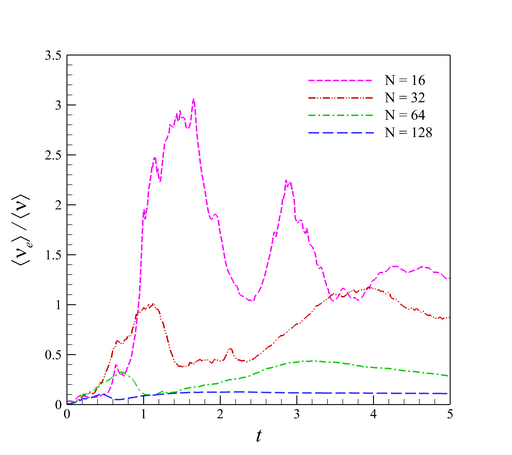}}
}
\caption{Evolution of Smagorinsky constant, $C_s$ and eddy viscosity behavior for the KHI problem ($u_{KHI}= 1.0$) with varying grid resolution, N.}
\label{fig:relative viscosity, khi4}
\end{figure}
\begin{figure}[!ht]
\centering
\mbox{
\subfigure[TGV ($Pr_t$ range: $0.1-1.5$)]{\includegraphics[width=0.48\textwidth]{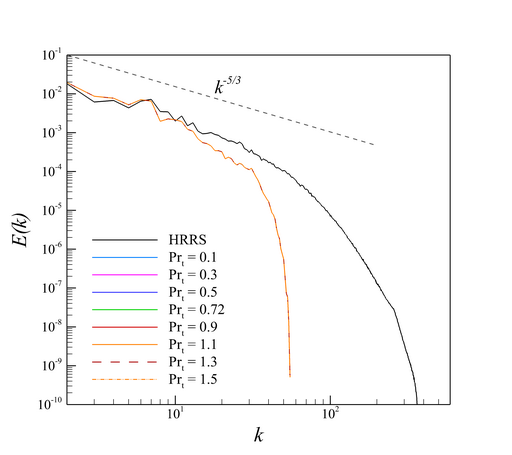}}
\subfigure[KHI ($u_{KHI} = 0.5$, $Pr_t$ range: $0.1-1.5$)]{\includegraphics[width=0.48\textwidth]{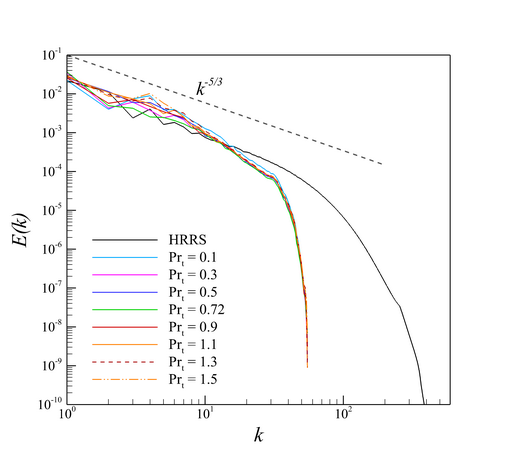}}
}\\
\mbox{
\subfigure[KHI ($u_{KHI} = 0.1$)]{\includegraphics[width=0.48\textwidth]{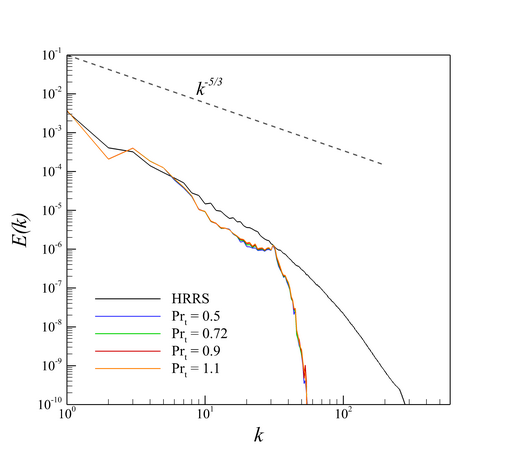}}
\subfigure[KHI ($u_{KHI} = 0.25$)]{\includegraphics[width=0.48\textwidth]{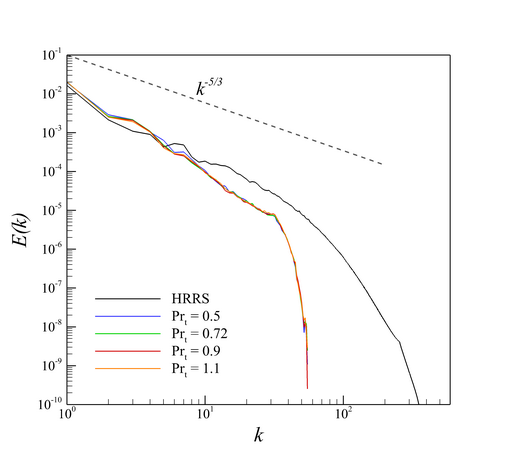}}
}\\
\mbox{
\subfigure[KHI ($u_{KHI} = 0.5$)]{\includegraphics[width=0.48\textwidth]{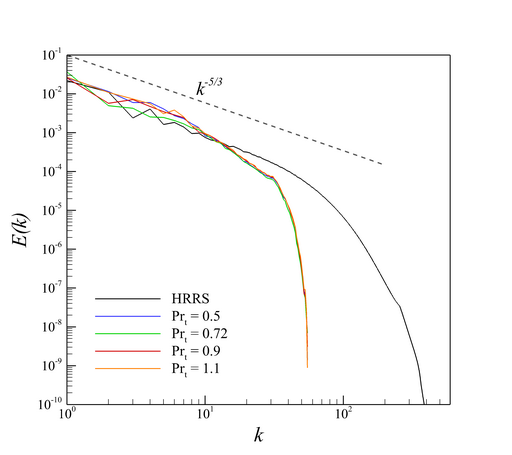}}
\subfigure[KHI ($u_{KHI} = 1.0$)]{\includegraphics[width=0.48\textwidth]{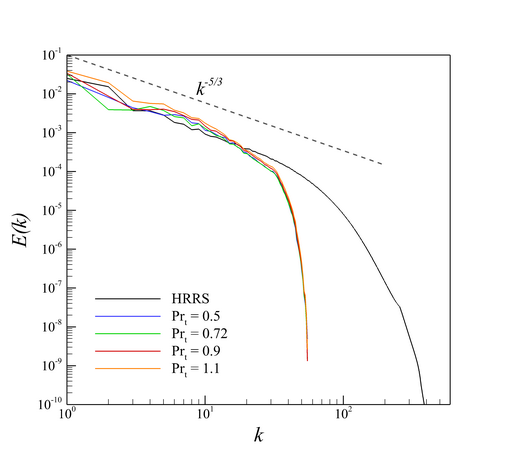}}
}
\caption{Angle-averaged kinetic energy spectra for TGV and KHI problem with varying Turbulent Prandtl number obtained by localized dynamic eddy viscosity model on $64^3$ grid resolutions. HRRS: High Resolution Reference Simulation.}
\label{fig:Prt}
\end{figure}

To show the robustness of the localized dynamic model, we perform an analysis on turbulent Prandtl number ($Pr_t$). $Pr_t$ is a dimensionless number which is accounted for the turbulent transport of heat, or for the accurate predication of the interaction or the heat flux exchange across any interface. Generally, $Pr_t$ is defined as the ratio between the momentum and heat transfer diffusivity which can be expressed as $Pr_t = K_m/K_h$. Here $K_m$ and $K_h$ are turbulent or eddy diffusivities for momentum and heat, respectively. Based on the assumption that the turbulent transport of heat or other scalars is same as the transport of momentum, it is used in Reynolds-averaged Navier-Stokes closure schemes as well as LES modeling where eddy viscosities or diffusivity are required \citep{bou2008scale,li2015revisiting}. As a result, $Pr_t$ has no meaning for ILES and $CS+RF$ models since there is no eddy viscosity or diffusivity present in those cases \citep{kays1994turbulent}. Since the exchanges of momentum, heat or other scalars are usually generated by shear, buoyancy force or any unstable stratification which is accounted by the $Pr_t$ \citep{businger1988note,stensrud2009parameterization,brutsaert2005hydrology}, $Pr_t$ is widely used in LES closures \citep{moin1991dynamic,lilly1992proposed,bou2008scale} and has the approximate value of range from $0.7$ to $0.92$ in the literature. There is another branch of studies on the relationship among $Pr_t$, instabilities and the exchange of momentum and heat flux \citep{li2011coherent,li2015revisiting} which is not inside of the scope of this present study. Fig.~(\ref{fig:Prt}) shows the averaged kinetic energy spectra plots for both the TGV and KHI flow problems using the localized dynamic eddy viscosity model on $64^3$ resolutions. As shown in Fig.~(\ref{fig:Prt})(a), the averaged kinetic energy spectra results for the TGV case are not influenced by the $Pr_t$ since TGV is an incompressible flow phenomenon (i.e., $M_0=0.08$). On the other hand, Fig.~(\ref{fig:Prt})(b) clearly shows that there are variations in the results at low wave number at $u_{KHI}=0.5$ for the shear layer turbulence test case.  To illustrate the effect of $Pr_t$ for KHI case more clearly, we can observe Fig.~(\ref{fig:Prt})(c) to Fig.~(\ref{fig:Prt})(f). When the value of $u_{KHI}$ is $0.1$ and $0.25$ that means the flow is less compressible, the spectra for $Pr_t$ above $0.5$ show almost constant results for a range up to $1.5$ (spectra up to $Pr_t$ value $1.1$ are shown in the figures). Yet the more compressible flows with $u_{KHI} = 0.5$ or $1.0$ gives fluctuations in low wave number spectra. Overall, it can be said that for all the cases of shear layer problem, the value of $Pr_t$ can be selected over a range of $0.5$ to $1.1$ without affecting the solution. We can conclude from the above observations that the localized dynamic model shows a good agreement with the existing literature on the value of $Pr_t$ for both incompressible and compressible cases.

\section{Summary and Conclusions}
\label{sec:5}
We have proposed a localized dynamic approach capable of solving Euler system of equations effectively and accurately to capture a wide range of Euler turbulence flow physics. We have also investigated our proposed model along with some of the most widely used numerical schemes in the case of incompressible decaying turbulence and compressible shear layer turbulence problems. The mean features of the proposed framework are: (i) providing a consistent filter definition for LES studies, (ii) designing a computational framework particularly for numerical schemes with non-dissipative characteristics, and (iii) implementing such schemes locally in a dynamic way. Specifically, we use a sixth-order central reconstruction for minimizing numerical dissipation and utilize an optimized Gaussian filter which has full attenuation property as well as consistent control over the numerical dissipation. Indeed, our model converts the Euler equations to the Navier-Stokes equations using a local and consistent test filtering procedure dynamically.

In order to assess the performance of different schemes,  we have considered a high-resolution WENO-Roe ILES simulation as reference for both the TGV and KHI problems. We have used statistical quantities such as averaged kinetic energy spectra and total kinetic energy to quantify the models, and density or $Q$ criterion field variables to visualize the flow physics. To further illustrate the strength of the proposed localized dynamic model, we have shown that the results produced by the model are in accordance with the theory and literature. Overall, the results suggest the robustness of the localized dynamic model to capture a wide range of scales for both compressible and incompressible flows. In future work, we plan to implement the proposed localized dynamic model to solve system of Navier-Stokes or Euler equations in simulating complex flow phenomena.

\section*{Acknowledgements}
The computing for this project was performed at the High Performance Computing Center at Oklahoma State University.

\section*{Disclosure Statement}
No potential conflict of interest was reported by the authors.

\section*{ORCID}
O. San http://orcid.org/0000-0002-2241-4648

\bibliographystyle{gCFD}
\bibliography{references}

\end{document}